\documentclass[useAMS]{mn2e} 
 
\usepackage{graphicx} 
\usepackage{txfonts} 
\usepackage{longtable,lscape} 
 
\newcommand\aj{AJ} 
\newcommand\apj{ApJ} 
\newcommand\apjs{ApJS}       
 
\newcommand\aap{A\&A} 
\newcommand\mnras{MNRAS} 
\newcommand\apjl{ApJ} 
\newcommand\pasp{PASP} 
\newcommand\nat{Nature} 
 
\newcommand\araa{ARA\&A}

\title[]{The {\it Hubble Space Telescope} UV Legacy Survey of Galactic Globular Clusters. XVI. The helium abundance of multiple populations}  
\author[A.\,P.\, Milone et al.] 
       {A.\,P.\,Milone$^{1}$,
         A.\,F.\,Marino$^{2}$,
         A.\,Renzini$^{3}$, 
         F.\,D'Antona$^{4}$,
         J.\,Anderson$^{5}$, \newauthor
         B.\,Barbuy$^{6}$,
         L.\,R.\,Bedin$^{3}$,         
         A.\,Bellini$^{5}$,
         T.\,M.\,Brown$^{5}$, 
         S.\,Cassisi$^{7}$, 
         G.\,Cordoni$^{1}$,   \newauthor
         E.\,P.\,Lagioia$^{1}$,
         D.\,Nardiello$^{1,3}$,
         S.\,Ortolani$^{1}$,
         G.\,Piotto$^{1,3}$,
         A.\,Sarajedini$^{8}$, \newauthor 
         M.\,Tailo$^{1}$,
         R.\,P.\,van der Marel$^{5,9}$,
         E.\,Vesperini$^{10}$
\\        
$^{1}$Dipartimento di Fisica e Astronomia ``Galileo Galilei'', Univ. di Padova, Vicolo dell'Osservatorio 3, Padova, IT-35122\\
$^{2}$Research School of Astronomy \& Astrophysics, Australian National University, Canberra, ACT 2611, Australia \\
$^{3}$Istituto Nazionale di Astrofisica - Osservatorio Astronomico di Padova, Vicolo dell'Osservatorio 5, Padova, IT-35122\\
$^{4}$Istituto Nazionale di Astrofisica - Osservatorio Astronomico di Roma, Via Frascati 33, I-00040 Monteporzio Catone, Roma, Italy\\
$^{5}$Space Telescope Science Institute, 3800 San Martin Drive, Baltimore,  MD 21218, USA\\
$^{6}$ Universidade de Sao Paulo, IAG, Rua de Matao 1226, Ciudade Universitaria. Sao Paulo 05508-900, Brazil\\
$^{7}$Istituto Nazionale di Astrofisica - Osservatorio Astronomico di Teramo, Via Mentore  Maggini s.n.c., I-64100 Teramo, Italy\\
$^{8}$Department of Astronomy, University of Florida, 211 Bryant Space Science Center, Gainesville, FL 32611, USA\\
$^{9}$Center for Astrophysical Sciences, Department of Physics \& Astronomy, Johns Hopkins University, Baltimore, MD 21218, USA\\
$^{10}$Department of Astronomy, Indiana University, Bloomington, IN 47405, USA\\
} 
\begin{document} 
%\date{Accepted 2016 March 10. Received 2016 March 8; in original form 2016 January 7} 
%\date{Draft Version Jun, 21, 2017} 
 
\pagerange{\pageref{firstpage}--\pageref{lastpage}} \pubyear{2017}  
\maketitle 
\label{firstpage}

\begin{abstract}
  Recent work, based on data from the {\it Hubble Space Telescope} ({\it HST\,}) UV Legacy Survey of Galactic Globular Clusters (GCs), has revealed that all the analyzed clusters host two groups of first- (1G) and second-generation (2G) stars. In most GCs, both 1G and 2G stars host sub-stellar populations with different chemical composition.
  
  We compare multi-wavelength {\it HST} photometry with synthetic spectra to determine for the first time the average helium difference between the 2G and 1G stars in a large sample of 57 GCs and the maximum helium variation within each of them.  We find that in all clusters 2G stars are consistent with being enhanced in helium with respect to 1G.
   The maximum helium variation ranges from less than 0.01 to more than 0.10 in helium mass fraction and correlates with both the cluster mass and the color extension of the horizontal branch (HB). These findings demonstrate that the internal helium variation is one of the main (second) parameters governing the HB morphology.
\end{abstract} 

\begin{keywords} 
  globular clusters: general, stars: population II, stars: abundances, techniques: photometry.
\end{keywords} 
 
\section{Introduction}\label{sec:intro}
 Although helium is the second most-abundant element in stars and in the Universe, we have little direct information on the relative helium content of stellar populations in Globular Clusters (GCs).
 The main challenge to infer the helium content from spectroscopy is that the helium line can be detected and used to derive reliable abundances in the spectra of stars that span a small interval of effective temperature, 8,000\,K $\lesssim T_{\rm eff} \lesssim$ 11,500\,K, and these conditions are present in only a small number of clusters and stars (e.g.\,Villanova et al.\,2009; Marino et al.\,2014). As an alternative, the helium abundance of GC stars can be inferred from chromospheric spectral lines, but these pioneering studies have been performed in just a few stars of three GCs (e.g.\,Pasquini et al.\,2011; Dupree et al.\,2011). 

 Thus, given these limitations of a direct, spectroscopic method, other ways of estimating the helium of GC stars have been adopted using its effect on stellar structure as predicted by stellar evolution theory. The first attempt was built on the fact that the time spent by stars on the red giant branch (RGB) decreases with increasing helium whereas that spent on the horizontal branch increases, so their ratio is a strong function of helium (Iben 1968, see also Buzzoni et al. 1983, Iben \& Renzini 1984). With this method it was proven that even the most metal poor GCs have a helium abundance as high as $Y=0.23\pm 0.03$, thus confirming the prediction of the Big Bang nucleosynthesis.

 Another stellar evolution feature that is sensitive to helium is the so called RGB Bump (e.g.\,Cassisi \& Salaris 1997, Lagioia et al.\,2018, hereafter paper XII, and references therein), whose luminosity and strength  can offer a consistency check to the helium abundance obtained with other method, at least if samples are statistically significant.
% Star-to-star variation of helium in GCs were then suggested to account for the multiple population of HB stars in the GC NGC 2808 (D'Antona et al.\,2002, 2005), an anticipation then confirmed by the discovery of multiple main sequences (MS) in this cluster (Piotto et al.\,2007). On the other hand, multiple MSs in $\omega$ Cen  could be accounted for only by appealing to distinct helium abundances (e.g.\,Bedin et al.\,2004; Norris 2004; Piotto et al.\,2005).  Another stellar evolution feature that is sensitive to helium is the so called RGB Bump (e.g.\, Cassisi \& Salaris 1997, Lagioia et al.\,2018, hereafter paper XII, and references therein), whose luminosity and strength  can offer a consistency check to the helium abundance obtained with other method, at least if samples are statistically significant.

 The discovery that the color-magnitude diagrams (CMDs) of nearly all the GCs host multiple MSs and RGBs (e.g.\,Piotto et al.\,2015 hereafter, Paper\,I) has provided a new window to infer the relative helium abundance of the distinct stellar populations. Indeed, the color separation between the distinct sequences is closely connected with their helium abundance and provides strong information on the relative helium content of the distinct stellar populations (e.g.\,D'Antona et al.\,2002, 2005;  Bedin et al.\,2004; Norris 2004; Piotto et al.\,2005; 2007; Milone 2015 and references therein).
% that can be followed continuously along the different evolutionary states (e.g.\,Milone et al.\,2012a) has allowed to infer the relative helium abundance of the distinct stellar populations (e.g.\,Bedin et al.\,2004; Norris 2004; Piotto et al.\,2005, 2007; D'Antona et al.\,2005). These studies have shattered the traditional idea that GC stars have all the same helium abundance as earlier suggested by D'Antona et al.\,(2002).

 Specifically, papers based on multi-wavelength {\it Hubble Space Telescope} ({\it HST\,}) photometry have demonstrated that it is possible to infer the relative helium abundance of multiple populations with a precision better than 0.01 in helium mass fraction (e.g.\,Milone et al.\,2015a, hereafter Paper III, and Paper XII). % and have shown that the internal helium variation in GCs ranges from $\Delta Y \sim$0.01 up to $\sim$0.10 or more (Milone et al.\,2012b, 2015a). 
 
In this context, the {\it HST} UV legacy survey of Galactic GCs  (Paper\,I), provides an optimal dataset to infer the helium content of stars in a large number of GCs. As shown in Paper\,I, the CMDs of all the 57 studied GCs are consistent with multiple populations.
 In addition, we have introduced the pseudo two-color diagram or `chromosome map' that maximizes the separation between the stellar populations along the MS and the RGB by using appropriate combination of photometry in the F275W, F336W, F438W, and F814W bands (Milone et al.\,2015b, 2017, hereafter Papers II and IX).

In this work, we exploit the chromosome maps of RGB stars derived in Paper IX  to investigate the helium abundance of multiple stellar populations in 57 GCs. The paper is organized as follows. In Section~\ref{sec:data} we describe the dataset, and define the main stellar generations of each cluster. Section~\ref{sec:Teo} describes the impact of helium and light elements on the magnitudes of GC stars,  while the method used to infer the relative helium abundance is described in Section~\ref{sec:He}.  Results are provided in Section~\ref{sec:results} and the relations between the helium abundance and the cluster parameters are discussed in Section~\ref{sec:correlations}. Finally, a summary and discussion is provided in Section~\ref{sec:discussion}.

\section{Data and data analysis} \label{sec:data} 
To infer the relative helium abundance of multiple stellar populations we have exploited both photometric and astrometric catalogs published in previous papers and additional photometry from archive data that we have specifically analyzed for this work.

The literature material includes catalogs published in Papers I and IX of the {\it HST} UV survey of Galactic GCs, which include homogeneous astrometry and five-bands {\it HST} photometry of the central region of 57 clusters. These catalogs have been derived from images collected through the F275W, F336W, and F438W filters of the Ultraviolet and Visual channel of the Wide Field Camera 3 (UVIS/WFC3) mostly as part of the {\it HST} programs GO-11233, GO-12605, and GO-13297 (PI.\,G.\,Piotto, see Paper I) and from archive data in the same filters (see Paper IX).   This dataset includes F606W and F814W photometry from the Wide Field Channel of the Advanced Camera for Surveys (WFC/ACS) as part of GO-10775 (PI.\,A.\,Sarajedini, see Sarajedini et al.\,2007 and Anderson et al.\,2008). We refer to the work by Anderson et al.\,(2008) and to papers I and IX for details on the data and the data reduction.

 To increase the number of bandpasses and better constrain the chemical composition of multiple stellar populations in GCs, we analyzed all the UVIS/WFC3 and WFC/ACS images available from the {\it HST} archive that overlap the field of view  studied in papers I and IX and provide accurate photometry of RGB stars.
 We have excluded from the analysis other archive images collected through the filters F606W, F814W of both WFC/ACS and UVIS/WFC3 and from the F435W filter of WFC/ACS, which is very similar to the F438W band of UVIS/WFC3. 
% In the case of NGC\,5139 ($\omega$\,Cen), in addition to the data of Paper\,IX, we exploited the catalogue published by Bellini et al.\,(2017a), which includes photometry in 18 UVIS/WFC3 bands.
 We have excluded from the analysis NGC\,5897 because only F275W, F336W, F438W, and F814W photometry is available for RGB stars of this cluster.
 Since the {\it HST} archive include F275W, F336W, F438W, F606W and F814W images of the GC IC\,4499, which was not previously investigated in the context of multiple populations, we extended to this cluster the analysis from Papers I and IX and included it in our sample. 
 The main properties of the archive images are summarized in Table~\ref{tab:data}. 

All the archive images were pipeline processed to account for charge-transfer efficiency losses as described in Anderson \& Bedin (2010).
 Photometry and astrometry of the WFC/ACS data have been performed by using the program img2xym\_WFC developed by Anderson \& King (2006). Briefly, we measured stars independently in each images by using the spatially-variable 9$\times$10 array of empirical point-spread function (PSFs) from Anderson \& King (2006), plus a `perturbation PSF' that fine-tunes the fitting to account for small variation of the {\it HST} focus.  

 Similarly, the analysis of UVIS/WFC3 data has been performed on each exposure separately, by using the program img2xym\_wfc3uv, which is similar to img2xym\_WFC and img2xym\_WFI (Anderson et al.\,2006), but it is devoted to the UVIS/WFC3 images.  Details are provided in Papers I and in Soto et al.\,(2017, paper VIII).
 %Details on data reduction are provided in Paper I and in Anderson et al.\,(2008).
 The flux of saturated stars has been measured as in Gilliland (2004).
 Stellar positions have been corrected for the geometric distortion of the ACS/WFC and UVIS/WFC3 detectors by adopting the solutions provided by Anderson \& King (2006), Bellini \& Bedin (2009), and Bellini, Anderson \& Bedin (2011).

 Photometry has been calibrated to the Vega-mag system by using the zero points provided by the WFC/ACS and UVIS/WFC3 webpages and following the procedure by Bedin et al.\,(2005).  We have selected relatively isolated stars that are fitted by the PSF and have small photometric and astrometric errors and we have included in our analysis only stars that according to their proper motions are cluster members.
 
 As discussed in Papers\,I and IX, the photometry has been corrected for differential reddening by using the iterative procedure described by Milone et al.\,(2012c, see their Sect.~3). In a nutshell, we first rotated the CMD into a reference frame where the abscissa is parallel to the reddening direction, and derived the fiducial line of the MS, SGB, and RGB of cluster members. To do this, we carefully excluded by eye all the evident binaries and blue stragglers.  We selected a sample of stars that are located in the regions of the CMD where the reddening line defines a wide angle with the fiducial line and used them as reference stars to estimate the differential reddening suffered by all the stars in the field of view.  Specifically, for each star in our photometric catalog, we selected the 45 closest neighboring reference stars and calculated the color residuals from the fiducial line along the reddening direction. We assumed the median of these residuals as the best determination of the differential reddening suffered by that star.
The reference star has been excluded in the determination of its own differential reddening. This ends one iteration.   
We used our determination of differential reddening to correct the CMD, derive a more accurate selection of reference stars and calculate an improved fiducial line. We re-run the procedure to improve the determination of differential reddening. Typically the procedure converges after four iteration.
 To estimate the uncertainty on the differential-reddening correction we derived the absolute values between the 45 residuals and the corresponding median. We calculated the corresponding 68.27$^{\rm th}$ percentile ($\sigma$) and considered the quantity $1.253\cdot\sigma/45$  as the estimate of the error associated to the differential-reddening correction.
We refer to the paper by Milone et al.\,(2012c) for further details on the adopted procedure. Moreover, a forthcoming paper of this series is dedicated to the differential reddening across the field of view of the GCs of the {\it HST} UV Legacy Survey and on the public realize of the differential-reddening maps.

\begin{centering} 
\begin{figure*} 
 \includegraphics[width=7.5cm]{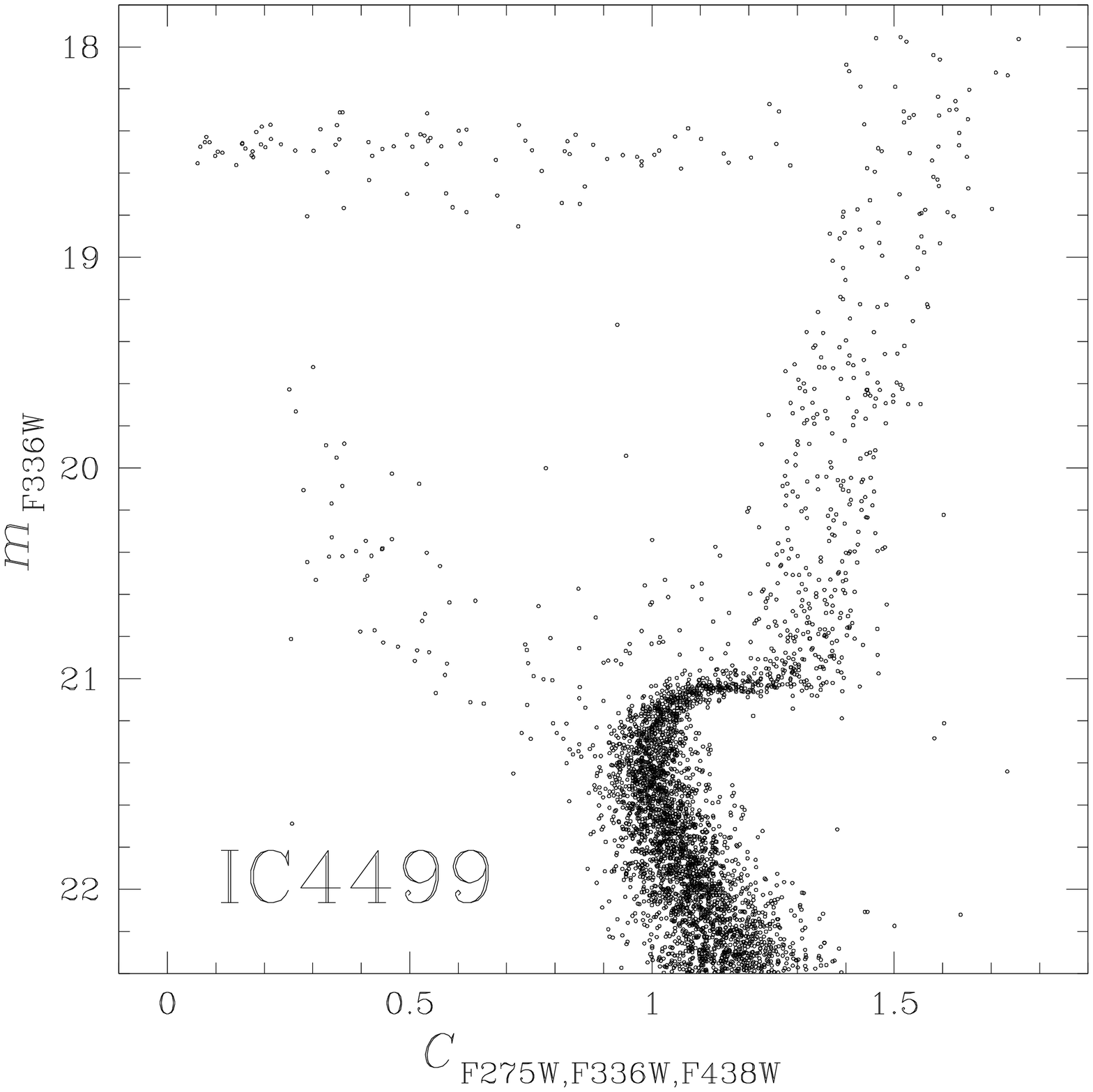} 
 \includegraphics[width=7.5cm]{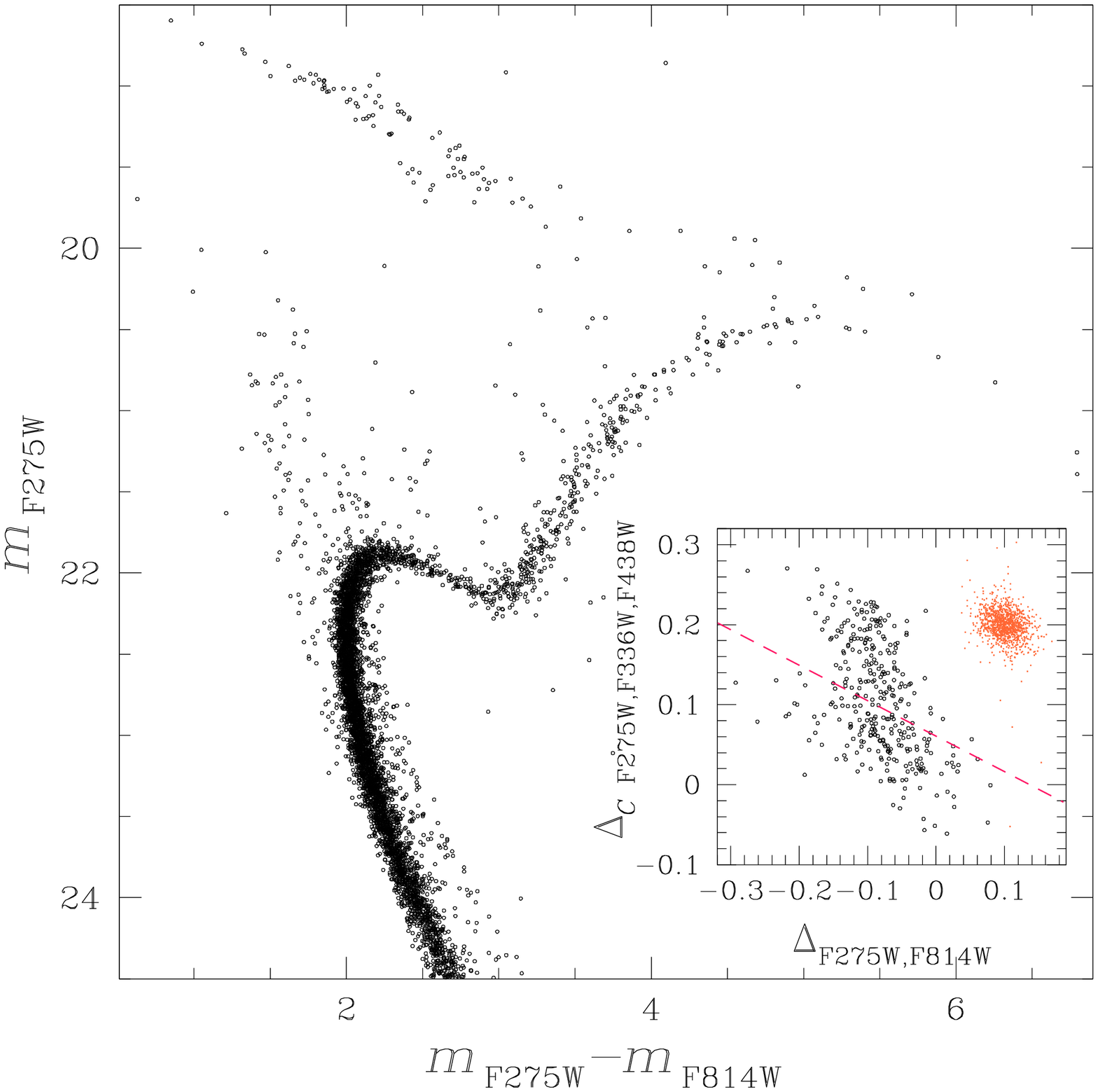} 
 %/home/milone/MW/GCs/IC4499
 \caption{$m_{\rm F336W}$ vs.\,$C_{F275W,F336W,F438W}$ (left panel) and $m_{\rm F275W}$ vs.\,$m_{\rm F275W}-m_{\rm F814W}$ (right panel) diagrams of IC\,4499, which was not investigated in our previous papers. The right-panel inset shows the chromosome map of RGB stars (black points) and the distribution of stars expected from observational errors alone. The magenta dashed line separates the selected 1G and 2G stars. See Paper IX for details.} 
 \label{fig:IC4499} 
\end{figure*} 
\end{centering}

 As an example, in the left and right panel of Fig.~\ref{fig:IC4499} we show the $m_{\rm F336W}$ vs.\,$C_{F275W,F336W,F438W}$ and the $m_{\rm F275W}$ vs.\,$m_{\rm F275W}-m_{\rm F814W}$ diagrams of IC\,4499, respectively. The chromosome map of RGB stars is shown in the inset and reveals that this cluster hosts the two main groups of 1G and 2G stars, in close analogy with all the other analyzed GCs. Specifically, by using the methods described in Paper\,IX, we find that the 1G includes the  51$\pm$5\% of the total number of RGB stars. 
%%%%%%%%%%%%%%%%%%%%%%%%%%%%%%%%%%%%%%%%%%%%%%%%%%%%%%%%%%%%%%%%%%%%%%%%%%%
%%%%%%%%%%%%%%%%%%%%%%%%%%%%%%%%%%%%%%%%%%%%%%%%%%%%%%%%%%%%%%%%%%%%%%%%%%%

\subsection{Distinguishing the main stellar generations}
The analysis of the chromosome maps of 57 GCs from Paper IX reveals that stars of most clusters (type-I GCs) separate into two main groups of first- (1G) and second-generation (2G) stars. The chromosome maps of a second group of clusters, which we named type-II GCs, exhibit a more complex pattern with seemingly split 1G and 2G. The sub-giant branch (SGB) of type-II GCs is either split or broadened also in optical colors, in contrast with what we observe in type-I GCs where the SGB splitting is visible only in CMDs based on ultraviolet bands.
Moreover, in the $m_{\rm F336W}$ vs.\,$m_{\rm F336W}-m_{\rm F814W}$ CMD, the RGB of type-II GCs splits into a blue and red component, with the red-RGB connected with the faint SGB. Spectroscopy reveals that red-RGB stars are enhanced in metallicity (i.e.\,[Fe/H]), s-process-element content and overall C$+$N$+$O abundance with respect to the blue-RGB  (Marino et al.\,2015 and references in their Table~10). Type-II GCs thus correspond to the class of anomalous GCs defined by Marino et al.\,(2011) on the basis of their chemical composition as some populations appear enhanced in iron, s-process elements, and C$+$N$+$O.

In this work we analyze all the RGB stars of type-I GCs and the blue-RGB stars of type-II GCs. We estimate the average helium difference between their 1G and 2G stars and the average helium difference between the sub-populations of 1G and 2G stars with extreme position in the chromosome map (1Ge and 2Ge).
 The two groups of 1G and 2G stars have been defined in Paper IX and their location in the chromosome maps is shown in their Figs.~3--7. We refer to the Sect.~3.3 of Paper IX for details on the method used to identify 1G and 2G stars. 

%%%%%%%%%%%%%%%%%%%%%%%%%%%%%%%%%%%%%%%%%%%%%%%%%%%%%%%%%%%%%%%%%%%%%%%%%%%
\begin{centering} 
\begin{figure*} 
 \includegraphics[width=6.5cm]{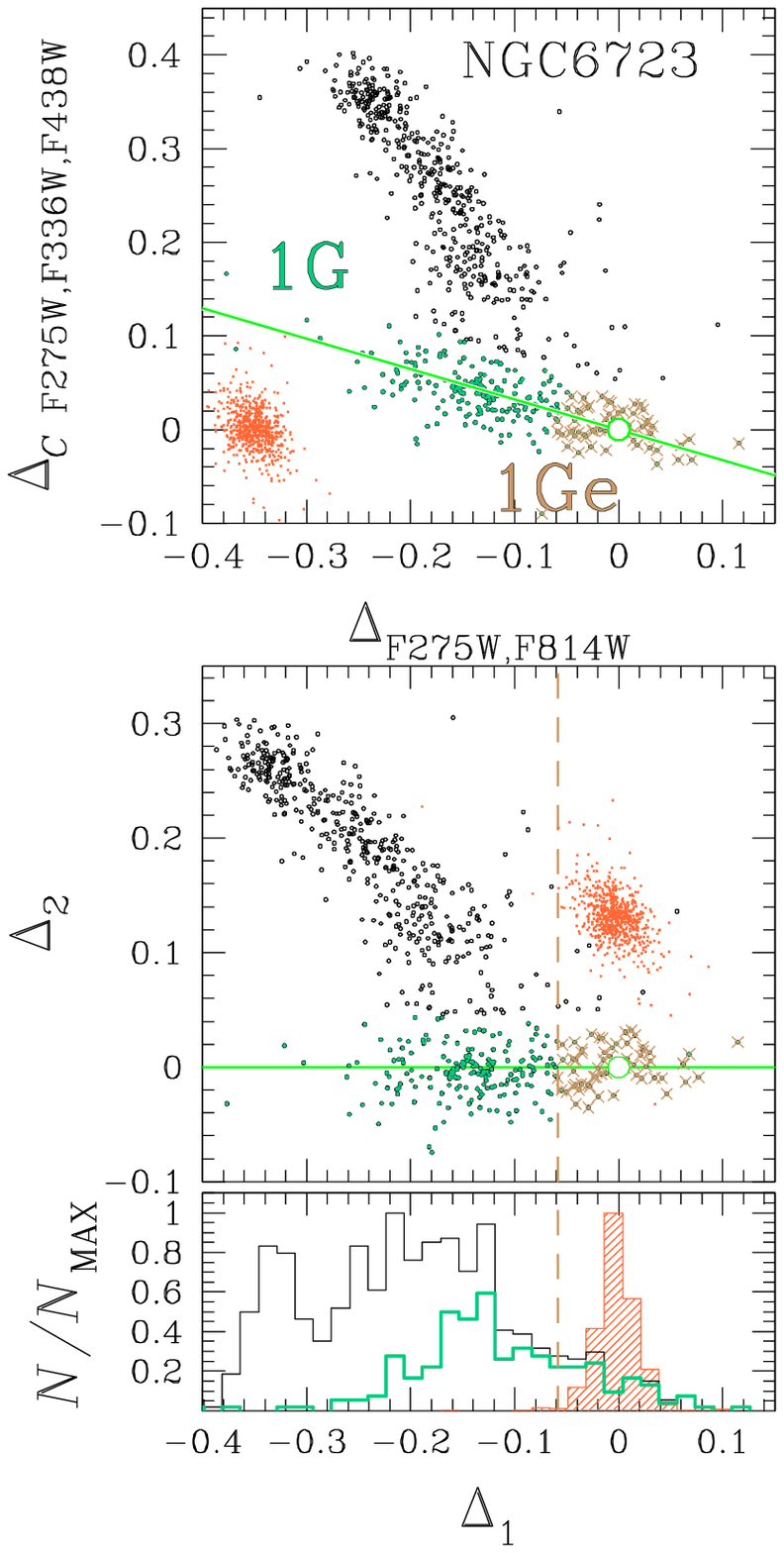} 
 \includegraphics[width=6.5cm]{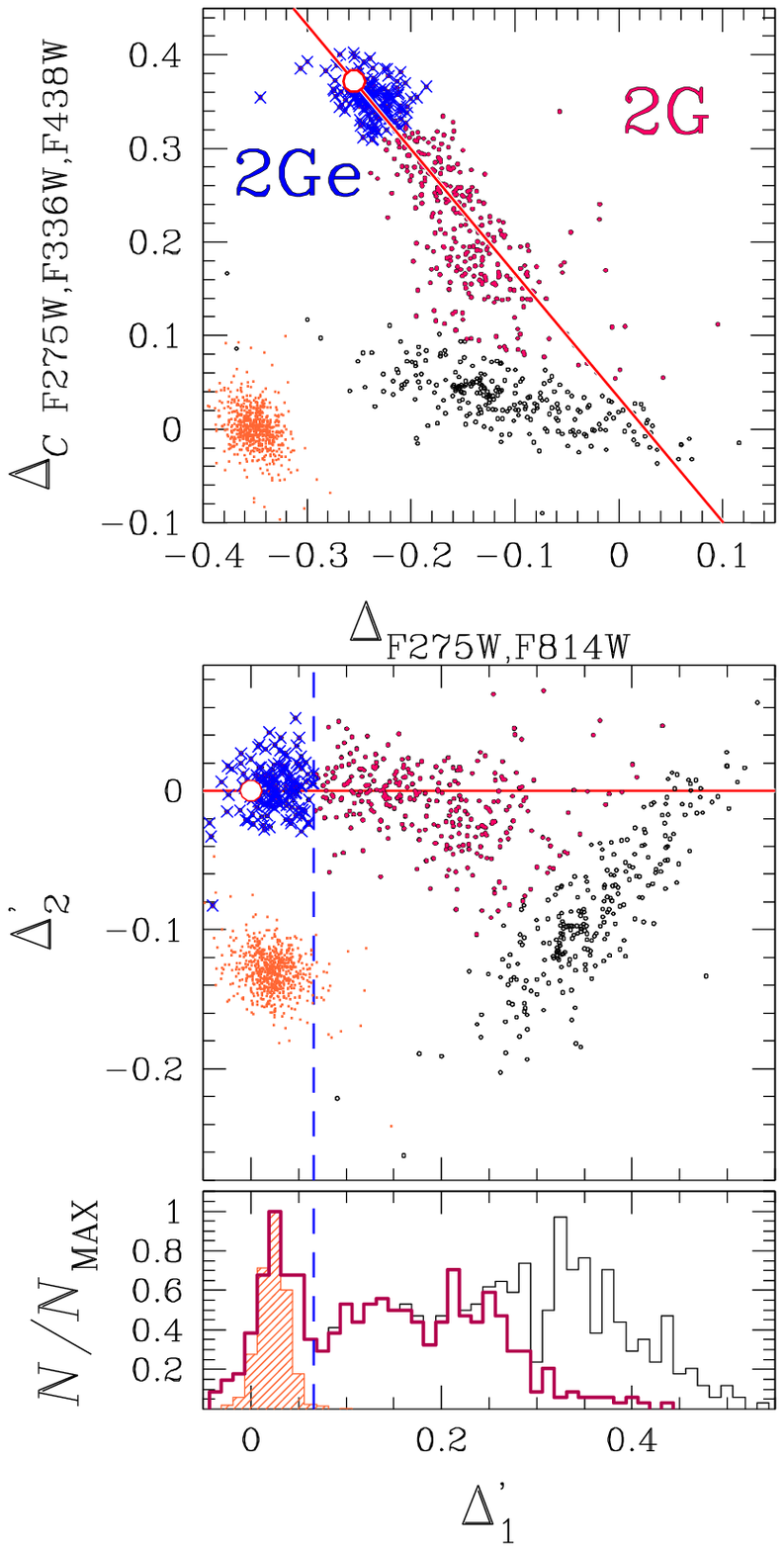} 
 %/home/milone/WORKS/treasury13/figure/mp2bin figoh figoh2
 \caption{This figure illustrates the procedure to identify the groups of 1Ge (left panels) and 2Ge stars (right panels) of NGC\,6723 with extreme $\Delta_{\rm F275W,F814W}$ and $\Delta_{C \rm F275W,F336W,F438W}$ values. \textit{Left:} The chromosome map of NGC\,6723 from Paper IX is reproduced in the upper panel where  the colored points mark 1G stars identified in that work. Middle panel shows $\Delta_{2}$ vs.\,$\Delta_{1}$ for the stars plotted in the upper panel. This diagram has been obtained by rotating counterclockwise the chromosome map in such a way the origin of the new reference frame corresponds to green circle and the abscissa to the green line defined in the upper panel. The orange points in the upper and middle panels represent the distributions that we expect from photometric errors only. Lower panel shows the $\Delta_{1}$ histogram distribution for all the stars (black), for 1G stars (aqua), and the corresponding distribution for photometric errors (orange). The vertical dashed lines in the lower and middle panel separate 1Ge stars from the remaining 1G stars. The selected 1Ge stars are marked with brown crosses in the upper and middle panel. \textit{Right:} Illustration of the procedure to identify 2Ge stars, which is similar to that shown in the left panels for 1G stars. 2Ge stars are represented with blue crosses in both the chromosome map (upper panel) and in the $\Delta^{'}_{2}$ vs.\,$\Delta^{'}_{1}$ diagram (middle panel). The blue dashed lines separate 2Ge stars from the remaining 2G stars (red points, see text for details).} 
 \label{fig:setupHe} 
\end{figure*} 
\end{centering} 
%%%%%%%%%%%%%%%%%%%%%%%%%%%%%%%%%%%%%%%%%%%%%%%%%%%%%%%%%%%%%%%%%%%%%%%%%%%

The procedure to define the sup-populations 1Ge is illustrated in the left panels of Fig.~\ref{fig:setupHe} for NGC\,6723 and is similar to the method that we used in Paper IX to identify 1G and 2G stars. The green line overimposed on the chromosome map plotted in the upper-right panel of Fig.~\ref{fig:setupHe} is the best-fit straight line for the sample of 1G stars that we have represented with colored symbols. We have rotated counterclockwise the chromosome map in such a way that the origin of the new reference frame corresponds to the green circle and the abscissa to the green line. The counter-rotated $\Delta_{2}$ vs.\,$\Delta_{1}$ diagram is shown in the middle panel of Fig.~\ref{fig:setupHe}, while the lower panel compares the $\Delta_{1}$ normalized histogram distribution of all the stars in the chromosome map (black) and the corresponding distribution of 1G stars (aqua).  

The distribution of stars in the chromosome map that we would expect from observational errors (including errors on the differential reddening corretion) only is represented with orange points in the upper and middle panels of Fig.~\ref{fig:setupHe} while the orange filled histogram plotted in the lower panel corresponds to the normalized histogram $\Delta_{1}$ distribution of the errors.

 The orange points are arbitrarily plotted in the corner of the upper-panel chromosome map and their average $\Delta_{2}$ is also chosen arbitrarily. The average $\Delta_{1}$ value, $\Delta_{1,0}$, has been determined by using the following procedure.
We calculate the 68.27$^{\rm th}$ percentile of the $\Delta_{1}$ error distribution, $\sigma$, and assume for the errors a range of $\Delta^{\rm i}_{1,0}$ values from $-1.000$ to $0.100$ in steps of 0.001. For each choice of $\Delta^{\rm i}_{1,0}$ we determine the normalized kernel-density $\Delta_{1}$ distribution of the observed 1G stars with $\Delta_{1}>\Delta^{\rm i}_{1,0}-\sigma$, $\phi^{\rm i}_{\rm obs}$, and the corresponding distribution for the error-points in the same interval of $\Delta_{1}$, $\phi^{\rm i}_{\rm err}$. We assumed as $\Delta_{1,0}$ the value of $\Delta^{\rm i}_{1,0}$ corresponding to the minimum $\chi$-squared between $\phi^{\rm i}_{\rm obs}$ and $\phi^{\rm i}_{\rm err}$. 
We consider as 1Ge, the sub-sample of 1G stars with $\Delta_{1}>\Delta_{1,0}-3\sigma$. The selected 1Ge stars are represented with brown crosses in Fig.~\ref{fig:setupHe}.

The sub-sample of 2Ge stars has been selected by using a similar procedure, which is illustrated in the right panels of Fig.~\ref{fig:setupHe}. In this case we have rotated the chromosome map clockwise in such a way that the red circle becomes into the origin of the new reference frame and the red line to its abscissa.
 The rotated $\Delta^{'}_{2}$ vs.\,$\Delta^{'}_{1}$ diagram is plotted in the middle-right panel of Fig.~\ref{fig:setupHe} and the normalized $\Delta^{'}_{1}$ distributions of all the stars, of 2G stars, and the corresponding error distribution are shown in the lower-right panel, and are represented with black, red, and orange histograms, respectively.
The average position of the error distribution in the $\Delta^{'}_{2}$ vs.\,$\Delta^{'}_{1}$ diagram has been determined by adapting the method described above to 2G stars. We consider as 2Ge stars the sub-sample of 2G stars with $\Delta^{'}_{1}<\Delta^{'}_{1,0}+3\sigma$, which we have represented with blue crosses in Fig.~\ref{fig:setupHe}.

In the Sect.~\ref{sub:ExtrTeo} we compare isochrones with the same metallicity but different abundance of He, C, N, and O, to demonstrate that 1Ge stars have primordial helium abundance (Y$\sim$0.25), while the most helium-rich stars correspond to the population 2Ge. For this reason, the average helium difference between 2Ge and 1Ge stars is indicative of the maximum internal helium variation.

\section{The impact of helium and light elements on the stellar colors}
\label{sec:Teo}

%%%%%%%%%%%%%%%%%%%%%%%%%%%%%%%%%%%%%%%%%%%%%%%%%%%%%%%%%%%%%%%%%%%%%%%%%%%
\begin{centering} 
\begin{figure*} 
  \includegraphics[width=11.0cm]{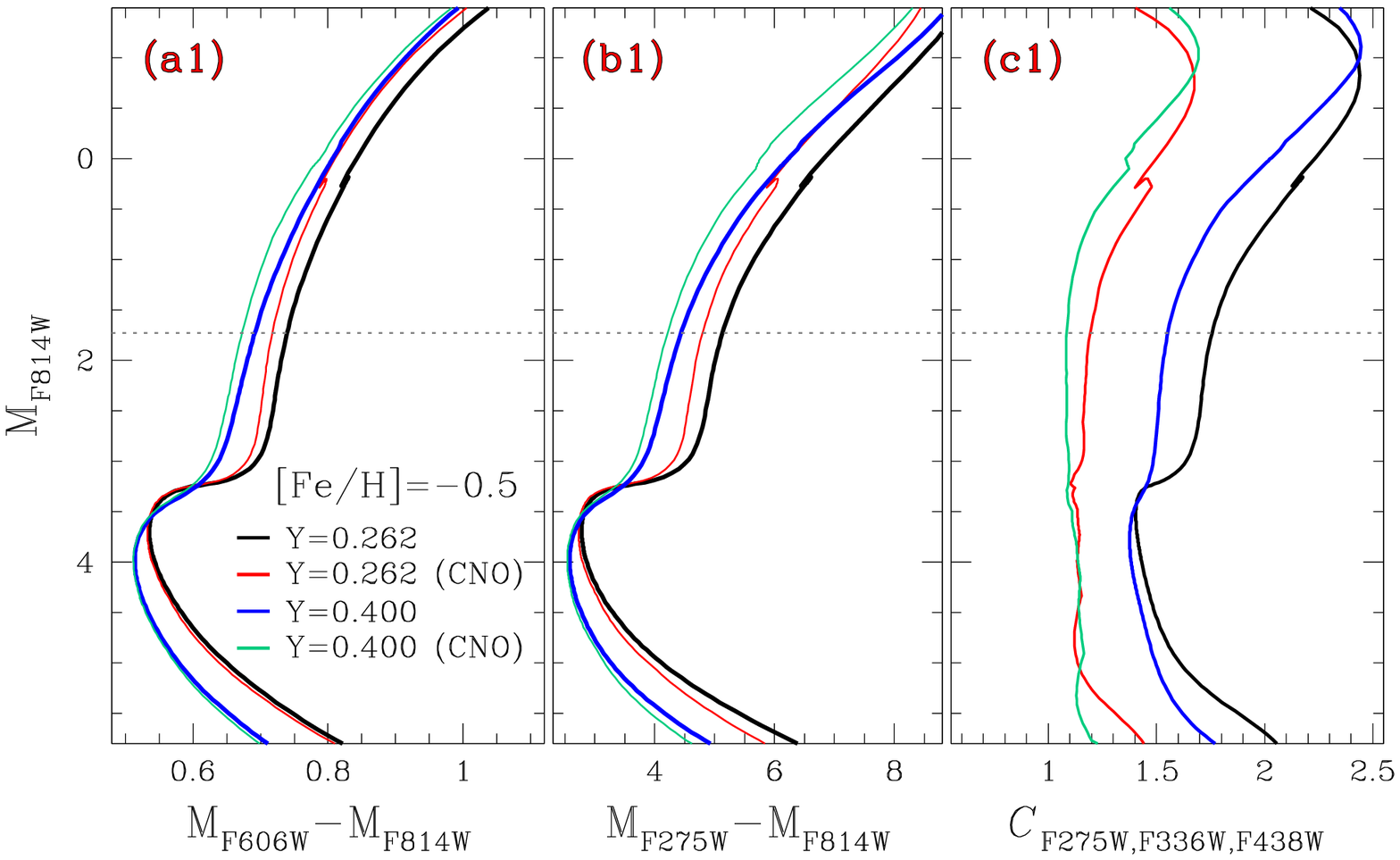}
  \includegraphics[width=11.0cm]{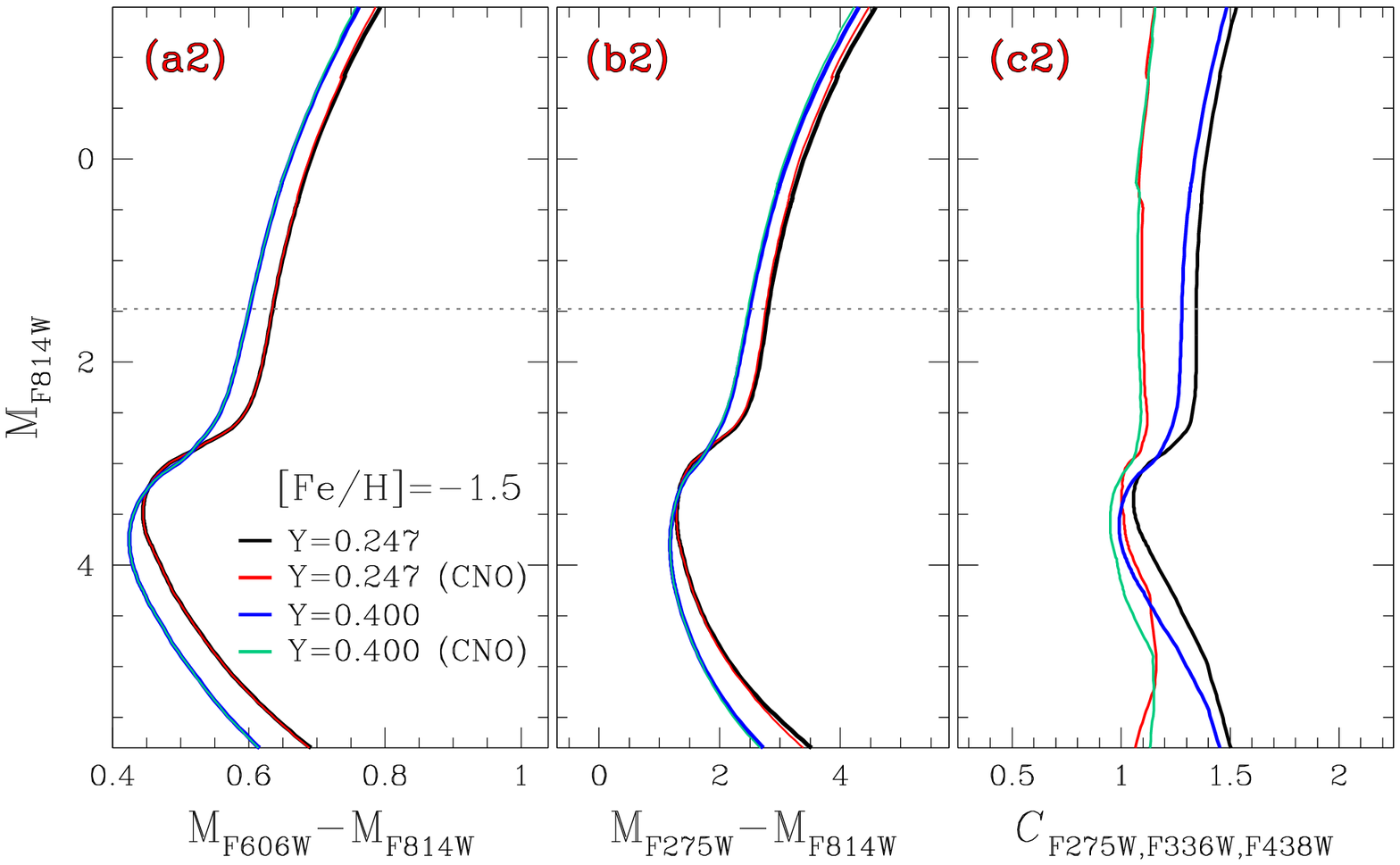} 
  \includegraphics[width=11.0cm]{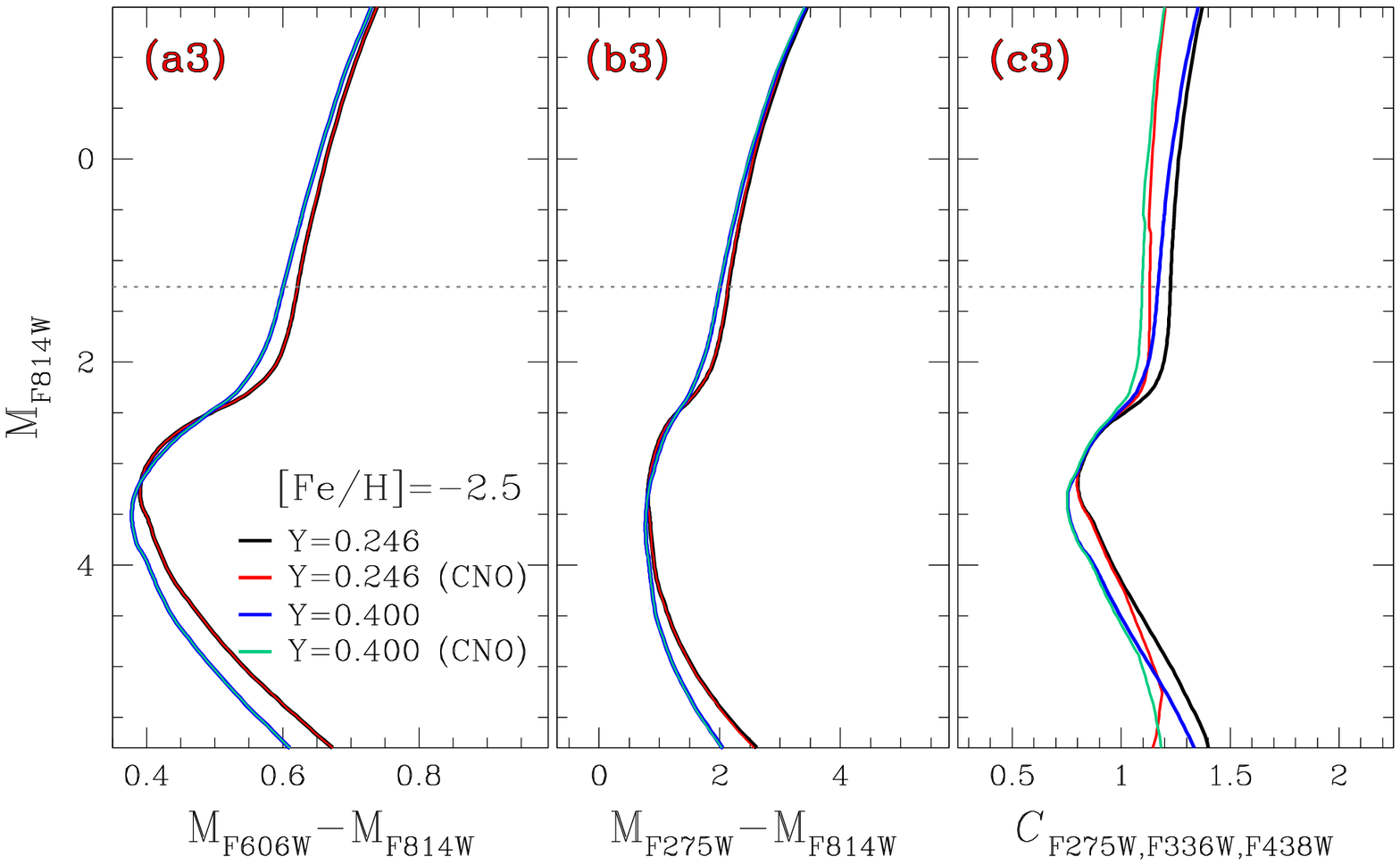}
 %/home/milone/MW/GCs/NGC5139/NIR/GO09444/CLEAN/SPETTRI/XANT/SPETTRI_SYNTH/ELIO/parametri.macro go2d
 %/home/milone/MW/GCs/NGC5139/NIR/GO09444/CLEAN/SPETTRI/XANT/SPETTRI_SYNTH/ELIOm250/parametri.macro go2d
  \caption{Isochrones with age of 13.0 Gyr and [$\alpha$/Fe]=0.3 for [Fe/H]=$-0.5$ (top), [Fe/H]=$-1.5$ (middle), and [Fe/H]=$-2.5$ (bottom) in the $M_{\rm F814W}$ vs.\,$M_{\rm F606W}-M_{\rm F814W}$ (panels a1--a3), $M_{\rm F814W}$ vs.\,$M_{\rm F275W}-M_{\rm F814W}$ (panels b1--b3) and $M_{\rm F814W}$ vs.\,$C_{\rm F275W,F336W,F438W}$ (panels c1--c3) planes.
   Black and red isochrones correspond to stellar populations with primordial helium abundance, while blue and aqua isochrone have Y=0.40. 
   In addition, both the red and the aqua isochrones are enhanced in nitrogen by 1.21 dex and depleted in both oxygen and carbon by 0.50 dex with respect to the black and blue isochrones.
   The horizontal dashed-dot lines are 2.0 magnitude brighter than the MS turn off of the black isochrone.
   To better compare the isochrones with different metallicities, we used for the x-axes of three diagrams plotted in the panels a1--a3 the same $M_{\rm F606W}-M_{\rm F814W}$ color width. Similarly, in panels b1--b3 and c1--c3 we used the same interval of $M_{\rm F275W}-M_{\rm F814W}$ and $C_{\rm F275W,F336W,F438W}$.} 
 \label{fig:ISO} 
\end{figure*} 
\end{centering} 
%%%%%%%%%%%%%%%%%%%%%%%%%%%%%%%%%%%%%%%%%%%%%%%%%%%%%%%%%%%%%%%%%%%%%%%%%%%

%%%%%%%%%%%%%%%%%%%%%%%%%%%%%%%%%%%%%%%%%%%%%%%%%%%%%%%%%%%%%%%%%%%%%%%%%%%
\begin{figure*} 
\begin{center} 
  \includegraphics[width=5.8cm]{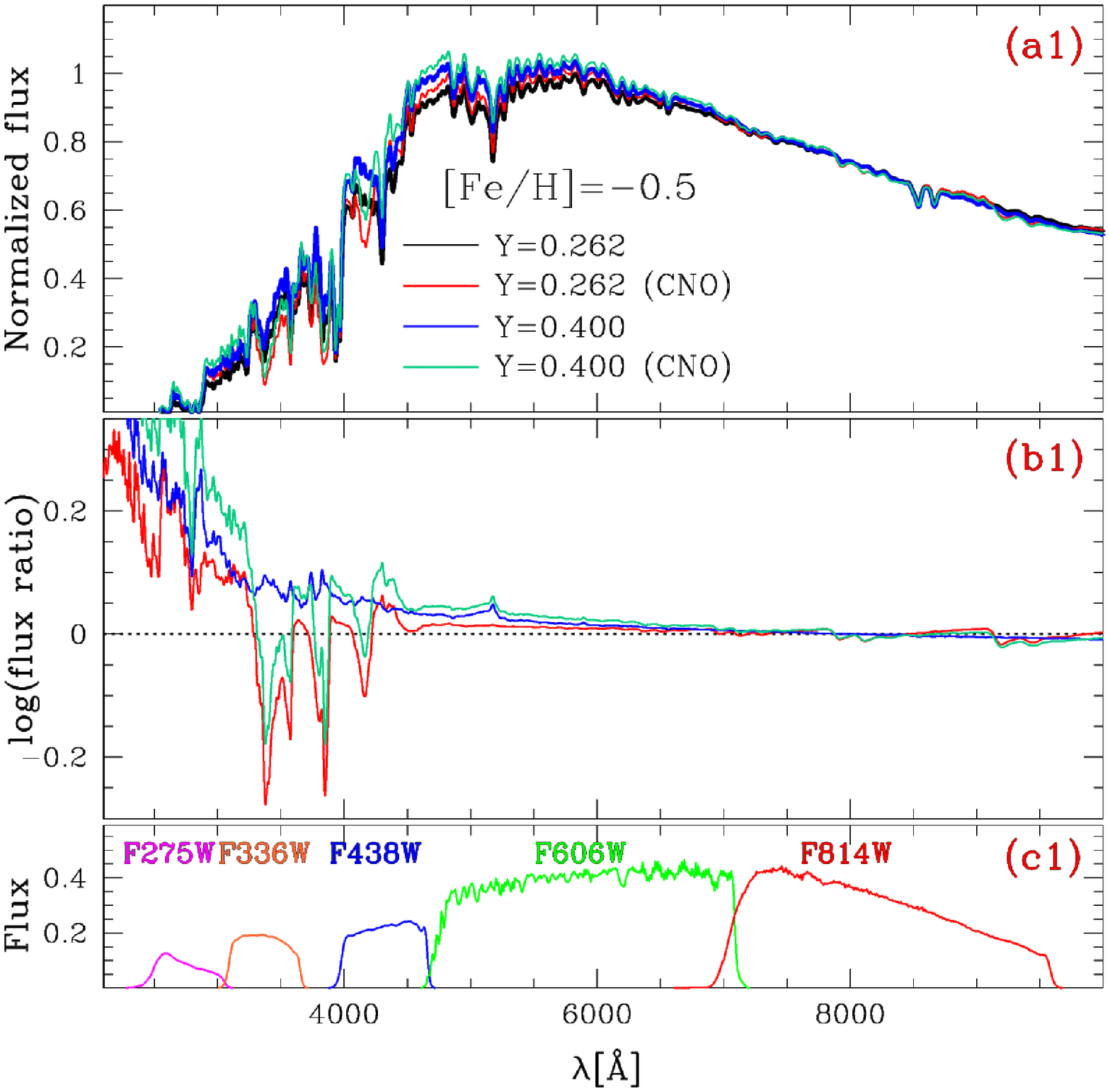}
  \includegraphics[width=5.8cm]{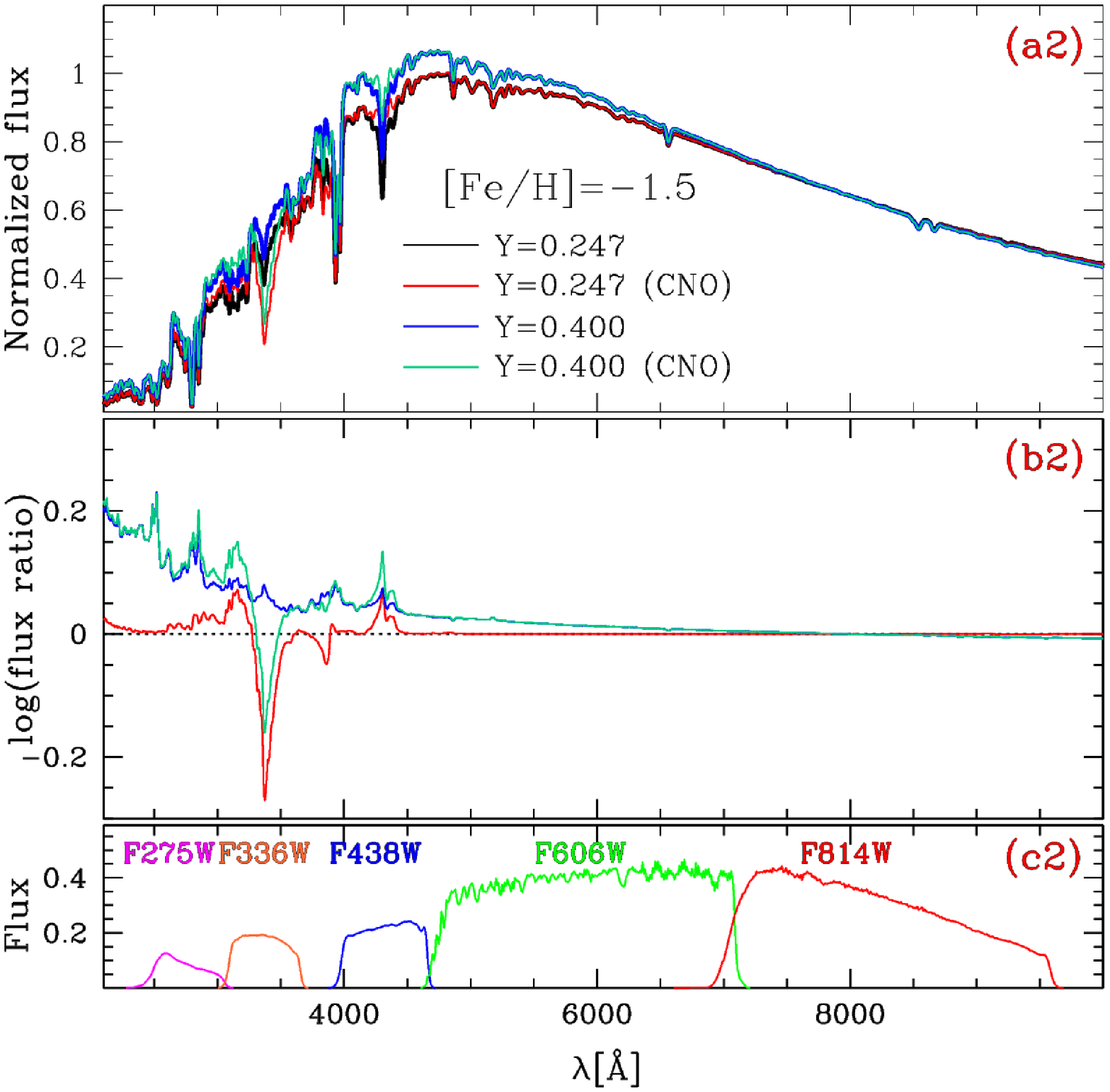} 
  \includegraphics[width=5.8cm]{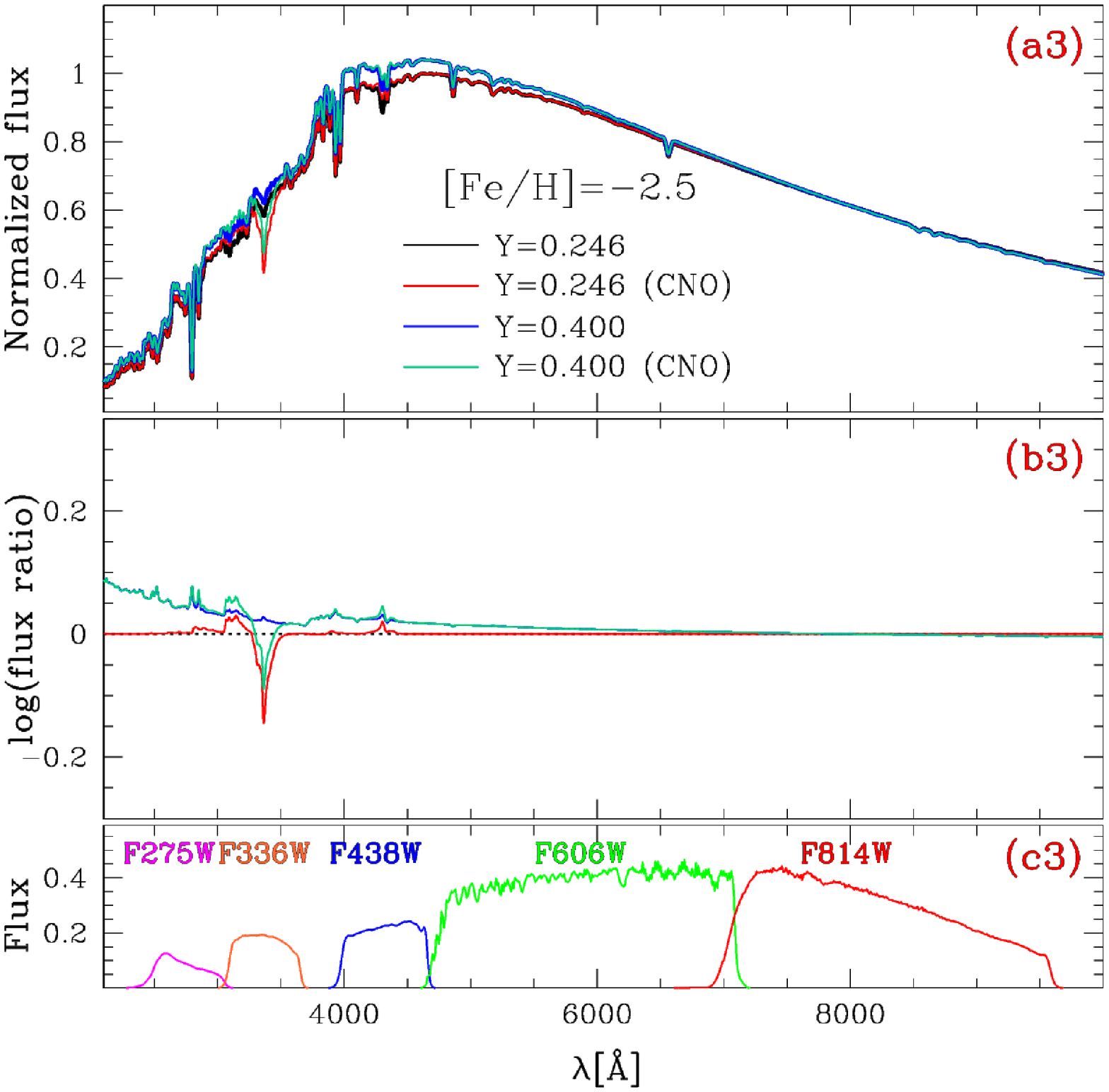}
 %/home/milone/MW/GCs/NGC5139/NIR/GO09444/CLEAN/SPETTRI/XANT/SPETTRI_SYNTH/ELIO/parametri0.macro go2
 \caption{Synthetic spectra corresponding to stars of the black, red, blue, and aqua isochrone located 2.0 F814W magnitudes above the MS turn off the black isochrone (panels a1--a3). The logarithm of the ratio between the flux of each spectrum and the flux of the black spectrum is plotted in panels b1--b3 as a function of the wavelength. Panels c1--c3 show the transmission curves of the ACS/WFC and UVIS/WFC3 filters used in this paper. See text for details.} 
 \label{fig:spettri} 
\end{center} 
\end{figure*} 

%%%%%%%%%%%%%%%%%%%%%%%%%%%%%%%%%%%%%%%%%%%%%%%%%%%%%%%%%%%%%%%%%%%%%%%%%%%

The effect of He, C, N, O variations on the stellar colors is illustrated in Fig.~\ref{fig:ISO} where we show the $M_{\rm F814W}$ vs.\,$M_{\rm F606W}-M_{\rm F814W}$ and $M_{\rm F814W}$ vs.\,$M_{\rm F275W}-M_{\rm F814W}$ CMDs and the $M_{\rm F814W}$ vs.\,$C_{\rm F275W,F336W,F438W}$ pseudo CMD for stellar populations with age of 13.0 Gyrs, [$\alpha$/Fe]=0.40, and for three distinct metallicities of [Fe/H]=$-0.50$, $-1.50$, and $-2.50$.

For each value of metallicity, [Fe/H], we show two isochrones from Dotter et al.\,(2008) with C, N, O abundance typical of 1G stars ([C/Fe]=0.0, [N/Fe]=0.0, [O/Fe]=0.4) but different helium content. Specifically, we represent in black the isochrones with primordial helium, $Y_{0}=0.245+1.5 \cdot Z$, while the isochrone with extreme helium content, Y=0.40, have been colored blue.
As widely discussed in literature, RGB and MS stars of the helium-rich isochrones have bluer colors than stars stars with primordial helium and the same luminosity (e.g.\,D'Antona et al.\,2002; Norris 2004; Sbordone et al.\,2011; Cassisi et al.\,2017). For a fixed magnitude and helium difference, the color separation increases with the color baseline. Moreover, for a fixed color baseline, the maximum color separation between the RGB and MS of helium-rich isochrones and isochrones with primordial helium increases when moving from metal-poor to metal-rich stellar populations.
 Such effects of helium variations on the CMD are illustrated in Fig.~\ref{fig:ISO} for the $M_{\rm F606W}-M_{\rm F814W}$ and $M_{\rm F275W}-M_{\rm F814W}$ colors. 

 The red and aqua isochrones plotted in each panel of Fig.~\ref{fig:ISO} correspond to stellar populations with primordial and extreme helium abundance, respectively, and are enhanced in nitrogen by 1.21 dex and depleted in both carbon and oxygen by 0.5 dex with respect to the black and blue isochrones.
  We note that the MS and RGB of the metal-poor ischrones with different C, N, O abundance are almost coincident in the optical CMDs as discussed by Sbordone et al.\,(2011) and Dotter et al.\,(2015). In contrast, light-element variations significantly affect the optical colors of metal rich RGB and MS stars, with the N-rich stars having redder $M_{\rm F606W}-M_{\rm F814W}$ colors than N-poor stars with the same luminosity and helium abundance. 

  Finally, we note that light-element and helium variations strongly affect the $C_{\rm F275W,F336W,F438W}$ pseudo-color of RGB and MS stars, thus corroborating the notion that the $M_{\rm F814W}$ vs.\,$C_{\rm F275W,F336W,F438W}$ pseudo-CMD is a powerful tool to identify the distinct stellar populations of GCs (Milone et al.\,2013). It is worth noting that, for a fixed variation of C, N, and O the maximum separation between N-rich and N-poor stars increases with the cluster metallicity.

  The red and aqua isochrones are derived by combining information from the isochrones from Dotter et al.\,(2008) and from synthetic spectra.
 We identified 15 points along the black and the blue isochrone and extracted the corresponding values of temperature, $T_{\rm eff}$, and gravity, $\log{g}$. For each pair of $T_{\rm eff}$ and $\log{g}$ we calculated a reference synthetic spectrum with solar nitrogen and carbon abundance and with [O/Fe]=0.40 and a comparison spectrum with [C/Fe]=$-0.50$, [N/Fe]=$1.21$, [O/Fe]=$-0.10$. The adopted abundances roughly resemble the chemical composition of 1G and 2G stars derived in GCs from high-resolution spectroscopy (e.g.\,Yong et al.\,2015).

 Synthetic spectra have been calculated over the wavelength range between 1,800 and 20,000 \AA\, by using the ATLAS12 and SYNTHE codes (Castelli 2005; Kurucz 2005; Sbordone et al.\,2007). As an example, in the panels b1--b3 and in panels a1--a3 of Fig.~\ref{fig:spettri} we show the reference and the comparison spectra with primordial helium abundance and with $Y=0.40$, respectively. When we changed the helium abundance we accounted for the variation in effective temperature and gravity predicted by the isochrones by Dotter and collaborators.
  The atmospheric parameters of each spectrum correspond to an RGB star located 2.0 F814W magnitudes above the MS turn off. The corresponding flux ratios are shown in the panels b1--d3 as a function of the wavelength, while in the panels c1--c3 we provide the bandpasses of the {\it HST} filters used in this paper. 

Each spectrum has been integrated over the bandpasses of the five filters used in this paper to derive synthetic magnitudes. These magnitudes are used to calculate the magnitude difference, $\delta m_{\rm X}$, between the comparison and the reference spectrum. The red and aqua isochrones have been determined by adding to the black and the blue isochrones the corresponding values of $\delta m_{\rm X}$. 

%%%%%%%%%%%%%%%%%%%%%%%%%%%%%%%%%%%%%%%%%%%%%%%%%%%%%%%%%%%%%%%%%%%%%%%%%%%
\begin{figure*} 
\begin{center} 
  \includegraphics[width=11cm]{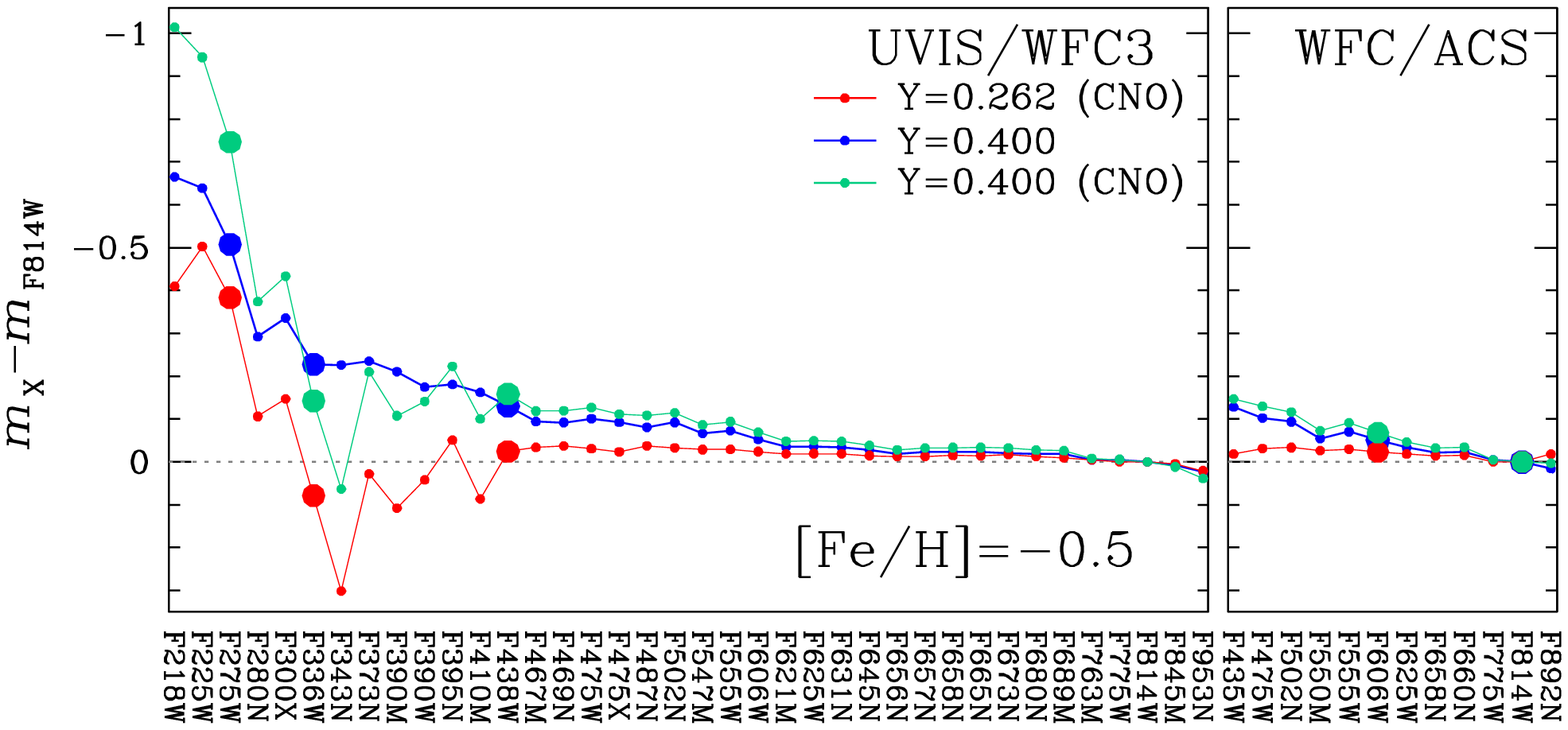}
  \includegraphics[width=11cm]{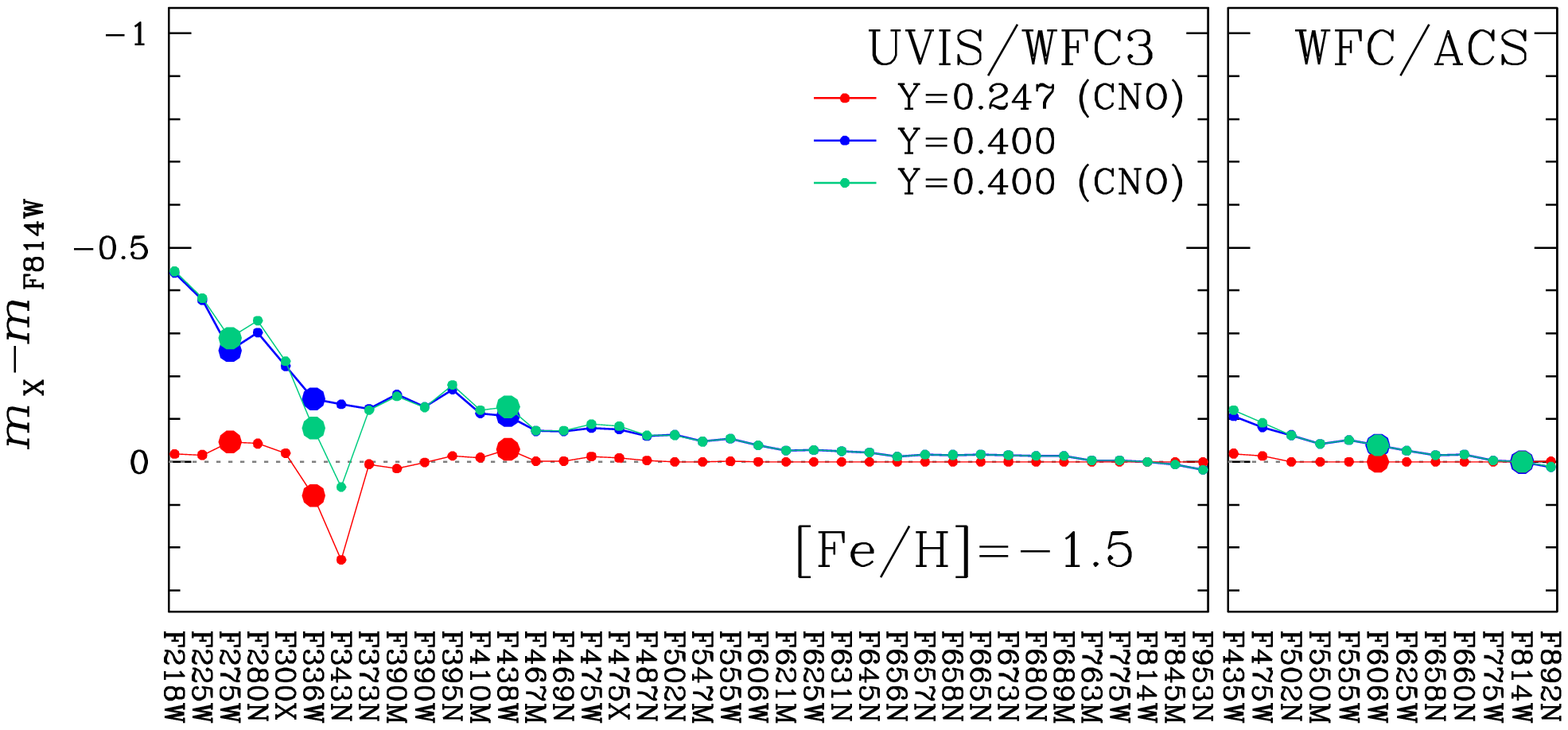} 
  \includegraphics[width=11cm]{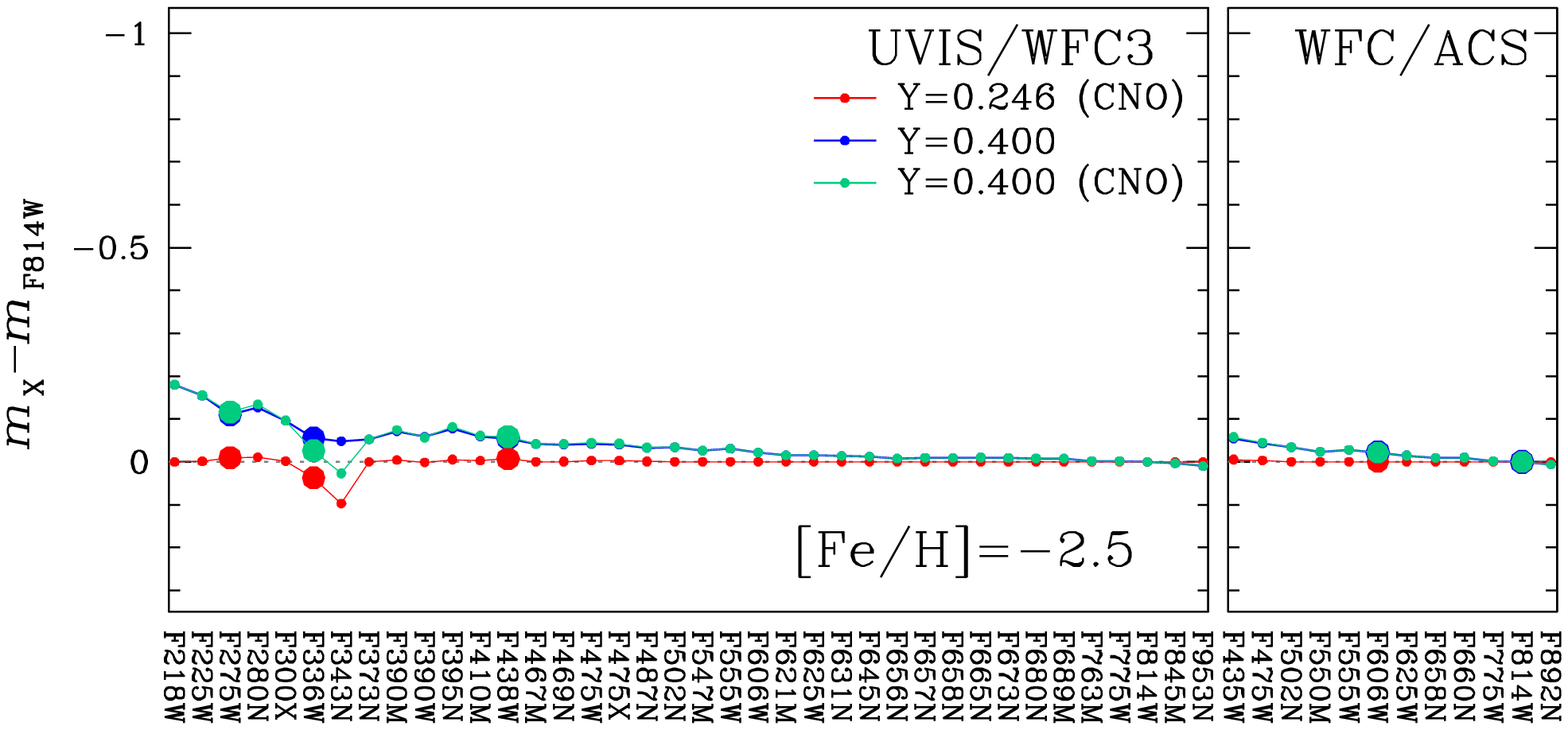}
  \caption{Color difference between each isochrone of Fig.~\ref{fig:ISO} and the black isochrone for UVIS/WFC3 (left) and WFC/ACS filters (right).  The color difference corresponding to the five X filters available for all the clusters of this paper, namely F275W, F336W, F438W, F606W and F814W, are marked with large dots. Upper, middle, and lower panel correspond to [Fe/H]=$-0.5$, $-1.5$, and $-2.5$, respectively. The color differences have been calculated at the level of the dashed-dot lines plotted in Fig.~\ref{fig:ISO}, 2.0 F814W magnitudes above the MS turn off of the black isochrone.} 
 \label{fig:DCOL} 
\end{center} 
\end{figure*} 
%%%%%%%%%%%%%%%%%%%%%%%%%%%%%%%%%%%%%%%%%%%%%%%%%%%%%%%%%%%%%%%%%%%%%%%%%%%

To illustrate the effect of changing He, C, N, and O on the color of the isochrones of Fig.~\ref{fig:ISO}, we plot in Fig.~\ref{fig:DCOL} the $m_{\rm X}-m_{F814W}$ color difference, $\Delta$($m_{\rm X}-m_{F814W}$), between each isochrone and the black isochrone for RGB stars located 2.0 F814W magnitudes above the MS turn off. 
The blue points show that, in the case of a variation in helium only, the $m_{\rm X}-m_{F814W}$ color difference decreases almost steadily when moving from red towards UV filters. Moreover, for a fixed X filter the absolute value of $\Delta$($m_{\rm X}-m_{F814W}$) increases from metal-poor to metal-rich isochrones.

A variation in C, N, and O only (red dots in Figure~\ref{fig:DCOL}) is responsible for positive values of $\Delta$($m_{\rm X}-m_{F814W}$) values for X=F336W and X=F343N and negative color differences for X filters bluer than F336W, like F275W.
 In the case of [Fe/H]=$-$0.5, the $m_{\rm X}-m_{F814W}$ color difference is significantly lower than zero for optical X filters, but the corresponding $\Delta$($m_{\rm X}-m_{F814W}$) values become negligible in metal-poor isochrones.
 Noticeably, the $m_{\rm F438W}-m_{F814W}$ color difference is smaller than zero for all the metallicities.

 Qualitativelly, the behaviour of the aqua dots plotted in Fig.~\ref{fig:DCOL}, which correspond to variations in helium, carbon, nitrogen, and oxygen, sums up the effects described above for the blue and red dots, separately.
By and large, the color separation becomes more negative from red to blue colors (i.e., colors get bluer with increasing helium), with the exception of  $\Delta$($m_{\rm F343N}-m_{\rm F814W}$) and to a lesser extent $\Delta$($m_{\rm F336W}-m_{\rm F814W}$), which are positive.
 In first approximation, the color separation decreases from red to blue color but a deviation towards positive $\Delta$($m_{\rm X}-m_{F814W}$) values is present for X=F343N and, to less extent for X=F336W.

\subsection{Stellar populations with extreme helium abundance in the chromosome map.}
\label{sub:ExtrTeo}
\begin{figure*} 
\begin{center} 
  \includegraphics[width=13.8cm]{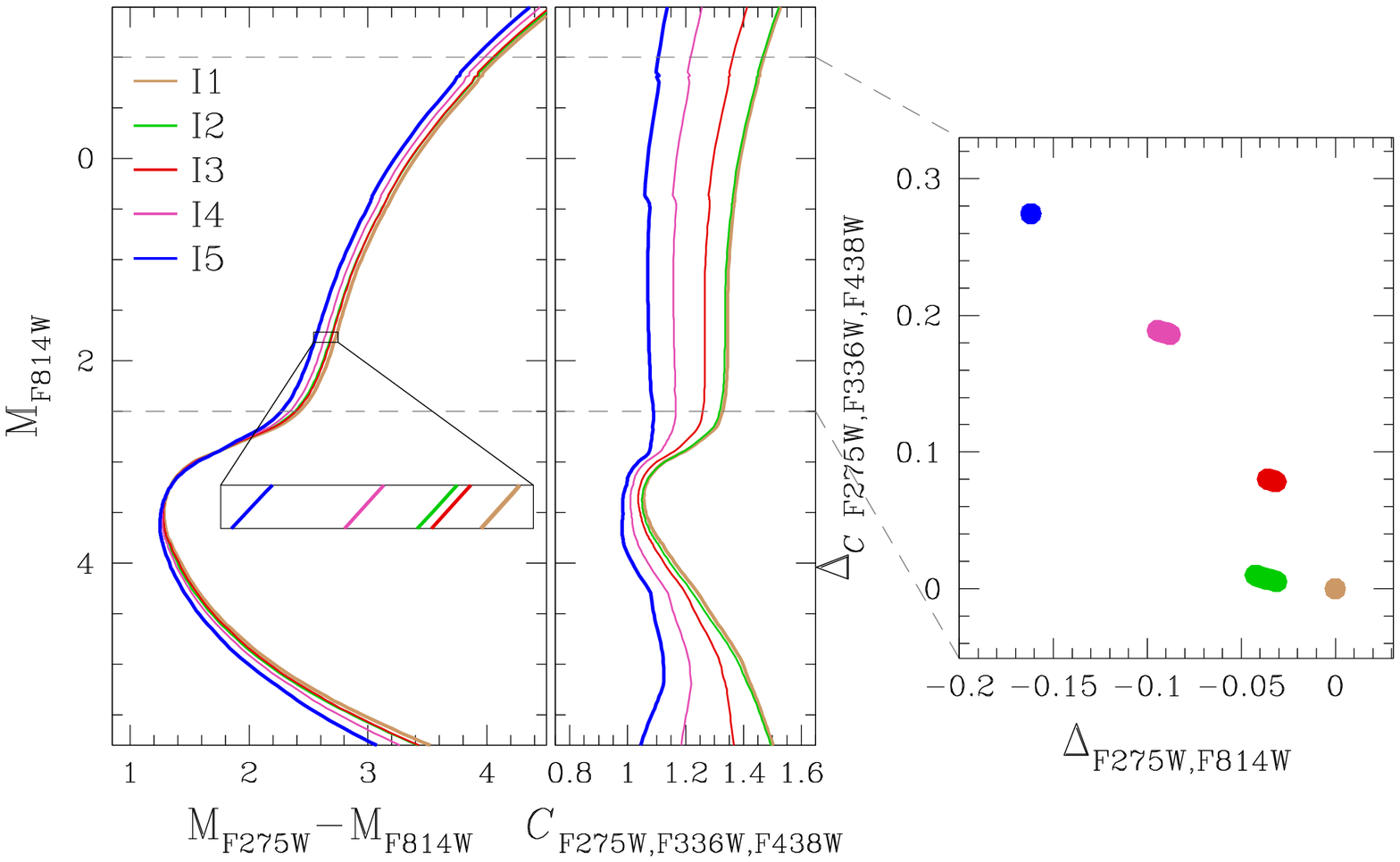}
%/home/milone/MW/GCs/NGC5139/NIR/GO09444/CLEAN/SPETTRI/XANT/SPETTRI_SYNTH/ELIOm150/parametri.macro go3
 \caption{Left and middle panels show isochrones with [Fe/H]=$-1.5$ and different abundance of He, C, N, and O in the $m_{\rm F814W}$ vs.\,$m_{\rm F275W}-m_{\rm F814W}$ and $m_{\rm F814W}$ vs.\,$C_{\rm F275W,F336W,F438W}$ diagrams. The two dashed lines delimit the RGB interval used to derive the chromosome map plotted in the right panel. See text for details.} 
 \label{fig:CMteo} 
\end{center} 
\end{figure*}

One of the main objectives of this paper is to constrain the maximum internal helium variations in GCs. 
In this subsection, we use isochrones with different abundances of He, C, N, and O to show that the most-helium-rich and the most-helium-poor stellar populations are located on the upper-left and lower-right side of the chromosome map, respectively.
 To do this, we plotted in Fig.~\ref{fig:CMteo} five isochrones with the same metallicity ([Fe/H]=$-$1.5) but different content of He, C, N, O in the $m_{\rm F814W}$ vs.\,$m_{\rm F275W}-m_{\rm F814W}$ and $m_{\rm F814W}$ vs.\,$C_{\rm F275W,F336W,F438W}$ planes. Specifically, we assumed that the brown and the blue isochrones have extreme helium values of Y=0.247 and 0.297, while the green, red, and magenta isochrones have intermediate helium abundances and are enhanced in helium by $\Delta$Y=0.02, 0.02, and 0.03, respectively, with respect to the brown isochrone.  %Further details on the adopted chemical compositions are provided in Table~\ref{}.
 
These isochrones are used to derive the chromosome map shown in the right panel of Fig.~\ref{fig:CMteo}, which corresponds to the RGB segment between the dashed lines plotted in the left and middle panels. This figure clearly shows that stars with primordial helium are clustered around the origin of the chromosome map, while the most helium-rich stellar population is located on the upper-left extreme of the chromosome map. These two populations with minimum and maximum helium abundance clearly correspond to the populations 1Ge and 2Ge selected in the observed chromosome map as illustrated in Fig.~\ref{fig:setupHe}.  
%%%%%%%%%%%%%%%%%%%%%%%%%%%%%%%%%%%%%%%%%%%%%%%%%%%%%%%%%%%%%%%%%%%%%%%%%%%

%%%%%%%%%%%%%%%%%%%%%%%%%%%%%%%%%%%%%%%%%%%%%%%%%%%%%%%%%%%%%%%%%%%%%%%%%%%

\subsection{The effect of Mg, Al, and Si variations on the colors of RGB stars}
\label{sec:mgteo}
In addition to the internal C, N, O variations, 2G stars of some GCs are depleted in Mg and enhanced in Al and Si. 

To investigate the effect of Mg, Al, and Si variations on the colors of RGB stars we  adopted the procedure described in Section~\ref{sec:Teo} to calculate the synthetic spectra of RGB stars located 2.0 F814W magnitudes above the turn off with the same chemical composition but different [Mg/Fe].
For all the stars we used [C/Fe]=$-0.50$, [N/Fe]=$1.21$, [O/Fe]=$-0.10$ and primordial helium content, which are the same abundances adopted for the red spectra of Fig.~\ref{fig:spettri} and are representative of 2G stars.

\begin{figure*} 
\begin{center} 
  \includegraphics[width=8.5cm]{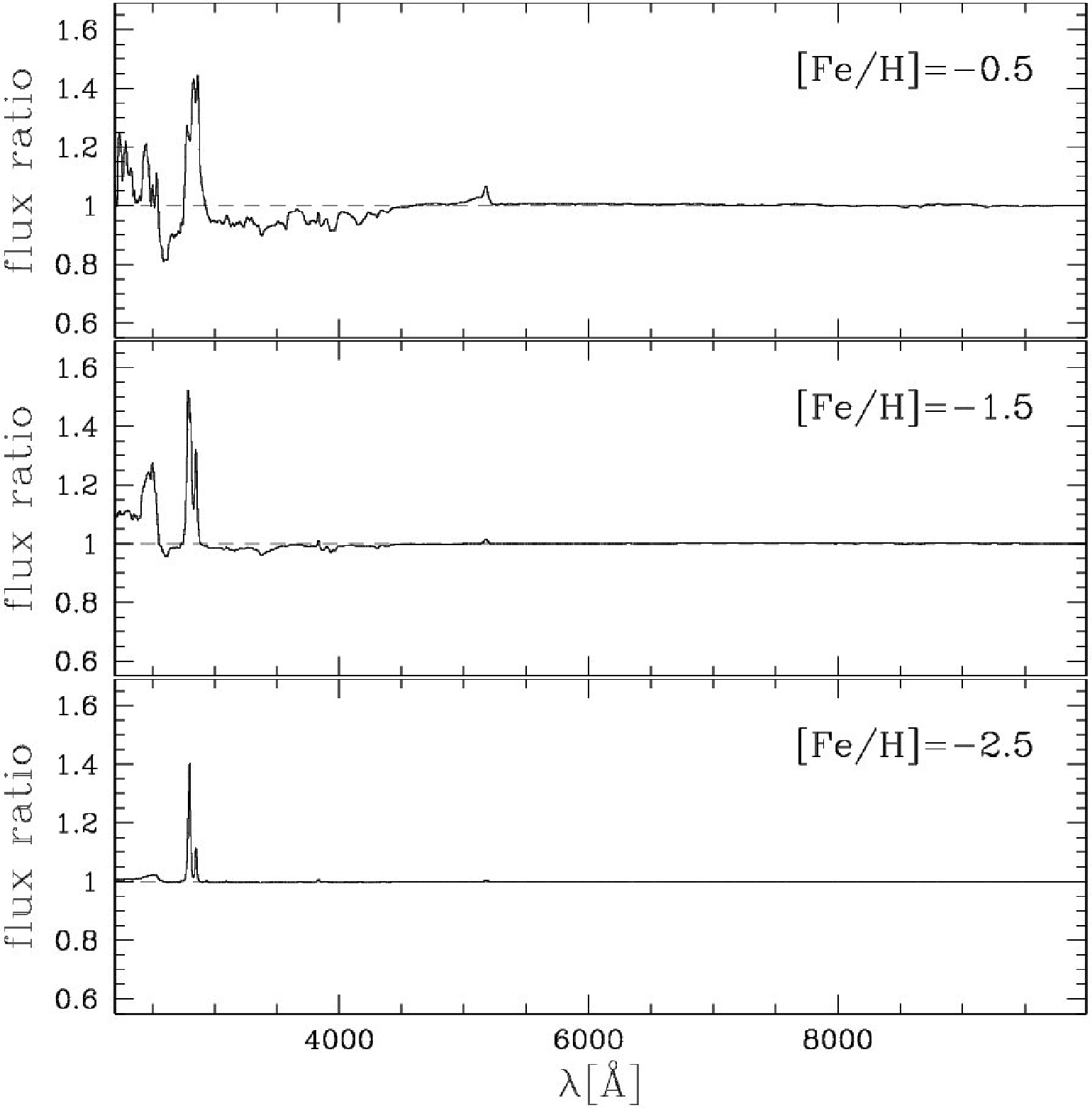}
  \includegraphics[width=8.5cm]{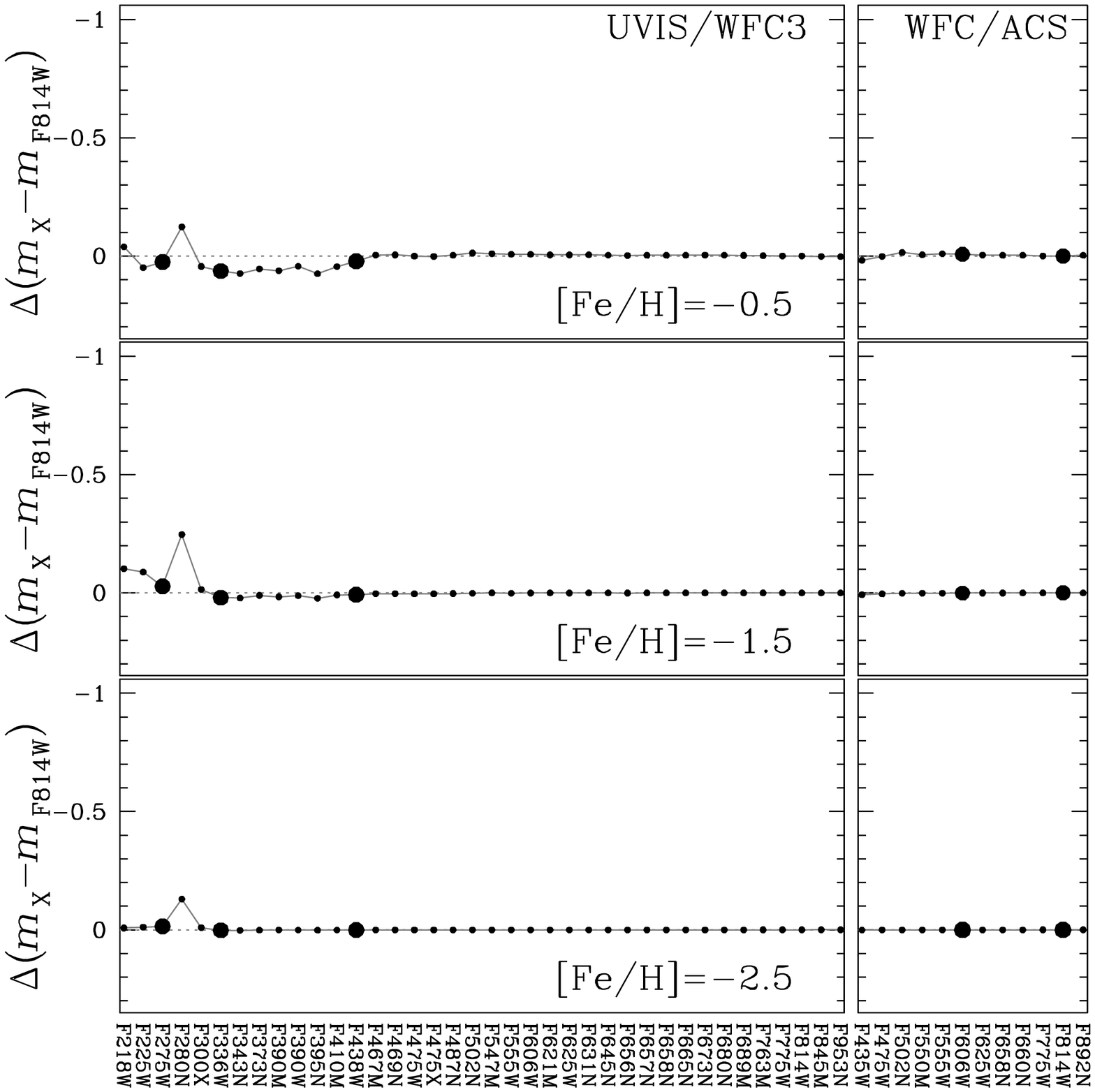} 
  \caption{{\it Left panels.} Logarithm of the ratio between the fluxes of synthetic spectra of RGB stars with different  Mg, Al and Si. All the spectra have the same content of He, C, N, O as the red spectra of Fig.\ref{fig:spettri} but different [Mg/Fe], [Al/Fe], and [Si/Fe].
    The spectra belong to stars located 2.0 F814W magnitudes above the MS turn off.
    Upper, middle and lower panels correspond to different metallicities of [Fe/H]=$-0.5$, $-1.5$, and $-2.5$, respectively. 
    {\it Right panels.} $m_{\rm X}-m_{\rm F814W}$ color differences between the spectra of the Mg-rich and the Mg-poor stars inferred from spectra plotted in the right panels. The color differences corresponding to the five X filters that are available for all the clusters are marked with large dots. 
   } 
 \label{fig:Mg} 
\end{center} 
\end{figure*}

We compared synthetic spectra for two stars, one with [Mg/Fe]=0.4, [Al/Fe]=0.0, and [Si/Fe]=0.4  and the other with [Mg/Fe]=0.0, [Al/Fe]=1.0, and [Si/Fe]=0.3 and show the flux ratio in the left panels of Fig.~\ref{fig:Mg} for three different metallicities. The fluxes of the two spectra have been then convolved with the transmission curves of the UVIS/WFC3 and WFC/ACS filters to derive the corresponding colors. The right panels of Fig.~\ref{fig:Mg} show the difference between the color of the Mg-poor and the Mg-rich star.

%We confirm previous findings by Cassisi et al.\,(2013) that magnesium variations have a negligible impact on the optical colors of metal poor stars.
We find that variations in Mg, Al, and Si have a negligible impact on the optical colors of metal poor stars thus confirming previous results by Cassisi et al.\,(2013) based on spectra with different [Mg/Fe] and [Al/Fe].
Moreover, we find that the adopted differences in Mg, Al, and Si between the two spectra, significantly affect the ultraviolet spectral region with $\lambda < 3000 \AA$ of all the spectra and the region with $\lambda < 4500 \AA$ of metal rich stars.

The flux difference is maximum around $\lambda \sim 2800 \AA$ due to the presence of strong Mg-II lines and corresponds to a F280N magnitude difference of $\sim$0.25 for [Fe/H]=$-1.5$. The impact of the adopted Mg, Al and Si difference is smaller in the F275W band and corresponds to $\sim$0.02 mag.
In the case of [Fe/H]=$-1.5$, we note a flux variation of $\sim$0.05 mag in F218W and F225W, while at high metallicity ([Fe/H]=$-$0.5) the adopted Mg, Al and Si variation affects the filters with central wavelength between 300 and 410\,nm a the level of $\sim$0.05 mag.

%%%%%%%%%%%%%%%%%%%%%%%%%%%%%%%%%%%%%%%%%%%%%%%%%%%%%%%%%%%%%%%%%%%%%%%%%%%
%%%%%%%%%%%%%%%%%%%%%%%%%%%%%%%%%%%%%%%%%%%%%%%%%%%%%%%%%%%%%%%%%%%%%%%%%%%

\section{How to read chromosome maps}\label{sec:read}
%Assumo che la descrizione di come si sono calcolate le frecce sia stata fatta prima e quindi mi riferisco ad una figura X che mostra la CM %di 2808 nel pannello a sinistra e un plot con le "frecce" nella stessa scala nel pannello di destra.

%For the physical interpretation of the CMs we take the example of NGC~2808 shown in the left panel of Fig. X and compare it  to the plot in the right panel where vectors represent the expected correlated changes of $\Delta_{\rm C}$ and $\Delta_{\rm F275W,F814W}$ when the abundance of the element C, N, O and He are changed, one a at a time.

\begin{figure*} 
\begin{center} 
  \includegraphics[width=12.5cm]{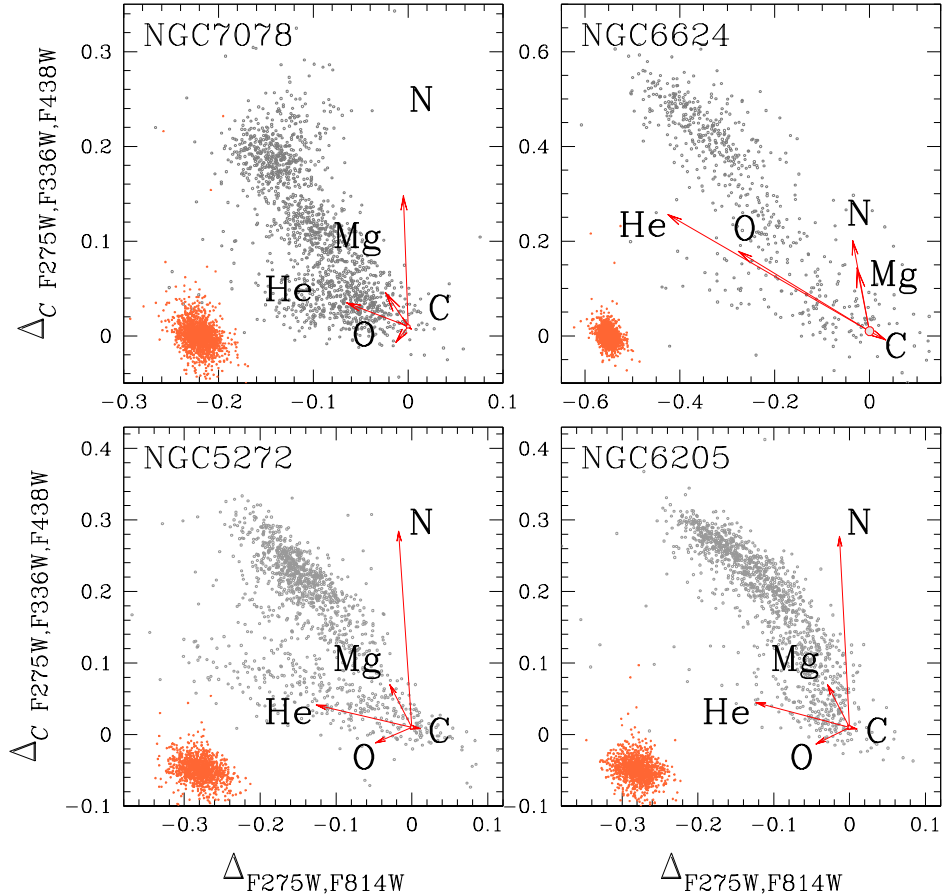}
 \caption{Reproduction of the chromosome maps of four GCs from Paper\,IX. Upper panels show the diagrams of the most metal poor (NGC\,7078) and the most metal rich (NGC\,6624) clusters in our sample. In the lower panels we plot the diagrams of second-parameter pair NGC\,5272 and NGC\,6205, which have similar metallicity ([Fe/H]$\sim-$1.50, Harris\,1996, updated as in 2010) but different extension. The orange points indicate the observation-error distribution (see Paper\,IX for details). The arrows indicate the effect of changing He, C, N, Mg, and O,  one a at a time, on $\Delta_{\rm C, F275W,F336W,F438W}$ and $\Delta_{\rm F275W,F814W}$.} 
 \label{fig:ICMs} 
\end{center} 
\end{figure*}

For the physical interpretation of the chromosome maps we used the four GCs shown in Fig.~\ref{fig:ICMs} as an example. Specifically, we plotted in the upper panels of Fig.~\ref{fig:ICMs} the chromosome maps of NGC\,7078 and NGC\,6626, which according to Harris (1996, updated in 2010) have metallicities of [Fe/H]=$-$2.37 and [Fe/H]=$-$0.44, respectively, and are the most metal poor and the most metal rich GCs in our sample.
In the lower panels of Fig.~\ref{fig:ICMs} we show the chromosome maps of NGC\,5272 and NGC\,6205, which according to the same scale have similar metallicity ([Fe/H]$\sim-$1.50) but have very different HB morphology making them classical {\it second parameter} pair of clusters.
The  vectors  overimposed on the chromosome maps of each cluster represent the expected correlated changes of $\Delta_{\rm C, F275W,F336W,F438W}$ and $\Delta_{\rm F275W,F814W}$ when the abundance of the element C, N, O, Mg and He are changed, one at a time. We assumed abundance variations of $\Delta$[C/Fe]=$-0.50$, $\Delta$[N/Fe]=$1.21$, $\Delta$[O/Fe]=$-0.50$, $\Delta$[Mg/Fe]=$-0.40$ and $\Delta Y=0.08$.

Thus, we see that the nitrogen vector is almost vertical, meaning that an increase of [N/Fe] has a strong effect on $\Delta_{\rm C, F275W,F336W,F438W}$ but a negligible one on $\Delta_{\rm F275W,F814W}$, whereas helium has the opposite effect, producing a strong change in $\Delta_{\rm F275W,F814W}$ but a small one in $\Delta_{\rm C, F275W,F336W,F438W}$. Decreasing the carbon abundance has a completely negligible direct effect, whereas decreasing oxygen and magnesium has a less-pronounced effect on both indices.
 
Of course, in real stars the variations of C, N, O and Mg are not independent of each other, as a decrease in oxygen and/or carbon inevitably
products an increase in nitrogen. Moreover, since the cosmic C:N:O proportions are roughly 4:1:10, even a modest decrease in oxygen and/or carbon produce a sizable increase in Nitrogen. 

A visual inspection at Fig.~\ref{fig:ICMs} immediately reveals that the 2G sequence in the chromosomic map can be reproduced mostly by a combination of the nitrogen and helium vectors, as indeed expected for material having undergone various degrees of CNO processing, leading to the depletion of carbon and oxygen and the enhancement of nitrogen, accompanied by helium enrichment.

Thus, we can see the 2G as a {\it CNO-cycle Sequence}.
But what about the 1G sequence? Clearly, the helium vector runs almost perfectly parallel to the 1G sequence and one is tempted to ascribe to a pure helium spread the inhomogeneity of the so-called first generation. Indeed, a physically sound combination of oxygen depletion accompanied by the nitrogen enhancement cannot combine to give a vector parallel to the 1G sequence. In fact, one would need a large  depletion in oxygen to produce the $\Delta_{\rm F275W,F814W}$ spread of the 1G, but this would come with a roughly factor of $\sim 10$ increase in nitrogen, with the combination of the two vectors falling somewhere on the 2G sequence, not on the 1G!

So, apparently the 1G spread seems produced almost exclusively by a helium spread, as already suggested in Milone et al.\,(2015), an idea revisited  in Lardo et al.\,(2018).
The helium variation needed to reproduce the 1G sequence would dramatically change from one cluster to another.
Reading from Fig. \ref{fig:DCOL}, we estimate $d(m_{\rm F275W}-m_{\rm F814W})/dY\simeq$ --3.6, --2.0 and --0.65 for [Fe/H]=--0.5, --1.5 and --2.5, respectively. 

As an example, NGC\,5272 exhibits a very-extend 1G sequence in its chromosome maps, which is consistent with an extreme helium variation of $\Delta Y^{\rm 1G} \gtrsim 0.10$, whereas NGC\,6205, which has similar metallicity as NGC\,5272, show a less-extended 1G sequence.
 
But, how can one have a sizable helium enrichment without it being accompanied by appreciable CNO processing? In principle, pure {\it pp}-chain reactions would do just that and we are tantalized by the idea of calling 1G the {\it pp-chain Sequence} as opposed to the 2G {\it CNO-Sequence}.
But how, concretely could almost pure {\it pp}-chain products  pollute 1G stars and do it in different star-by-star degree? We shall explore some speculative options in Sect.~\ref{sec:He1G}.
%%%%%%%%%%%%%%%%%%%%%%%%%%%%%%%%%%%%%%%%%%%%%%%%%%%%%%%%%%%%%%%%%%%%%%%%%%%

\section{Determination of the helium abundance of the multiple populations}
\label{sec:He}
\begin{centering} 
\begin{figure*} 
 \includegraphics[width=11.0cm]{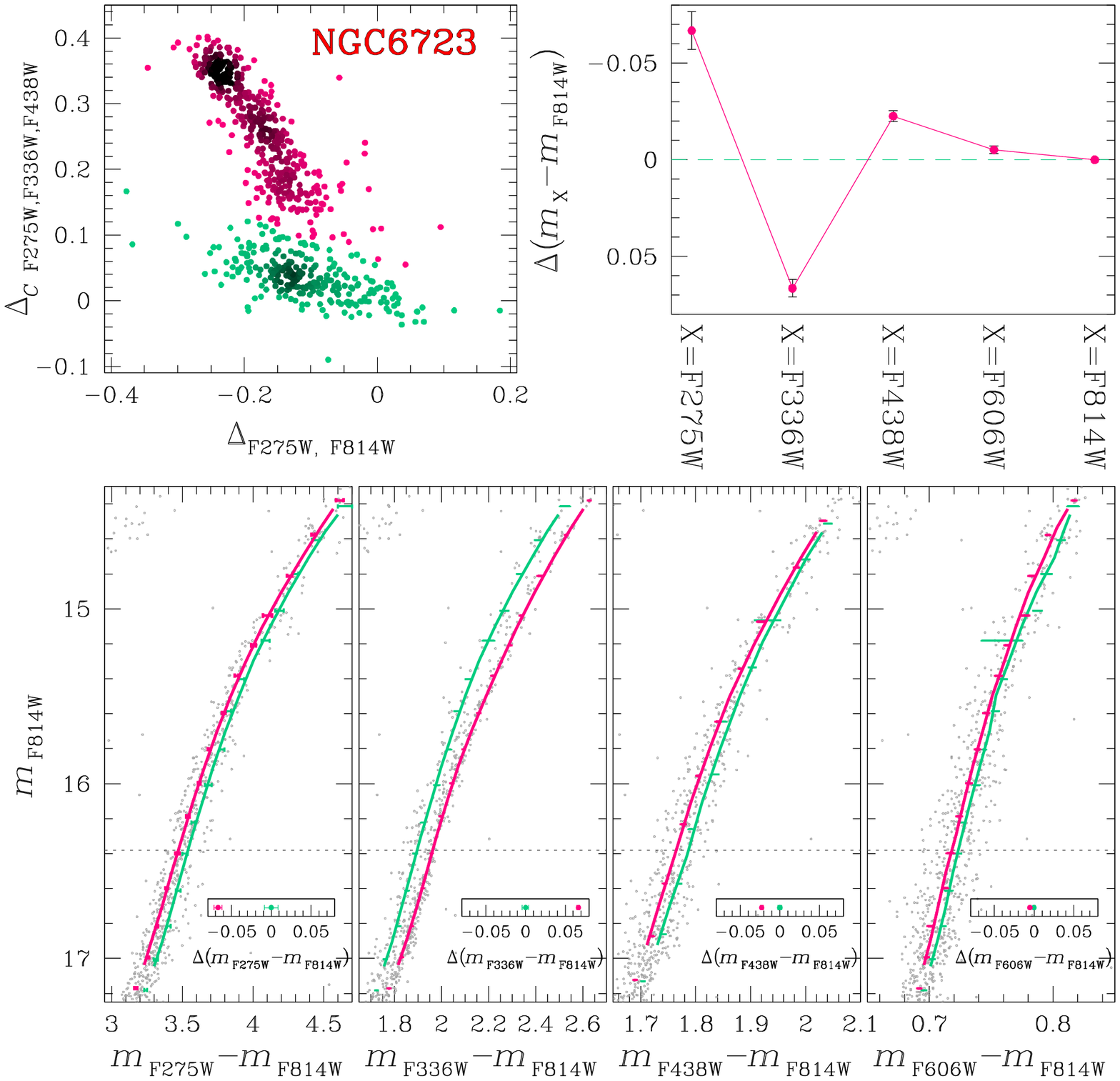} 
 %/home/milone/WORKS/treasury13/NGC6723/MATCH/popolazioni.macro fig
 \caption{This figure illustrates the procedure to derive the color differences between 1G and 2G RGB stars of NGC\,6723. The upper-left panel reproduces the chromosome map plotted in Fig.~\ref{fig:setupHe}, and the aqua and magenta colors  represent 1G and 2G stars, respectively. The lower panels show the fiducial lines of the corresponding RGBs in the $m_{\rm F814W}$ vs.\,$m_{\rm X}-m_{\rm F814W}$ plane, where X=F275W, F336W, F438W, F606W, and F814W. The color separation, $\Delta(m_{\rm X}-m_{\rm F814W})$, between RGB2 and RGB1 is determined at the luminosity level $m_{\rm F814W}=16.38$, indicated by the horizontal dotted lines and is highlighted in each inset. The upper left panel shows $\Delta(m_{\rm X}-m_{\rm F814W})$ as a function of the central wavelength. } 
 \label{fig:fiducials6723} 
\end{figure*} 
\end{centering}

\begin{centering} 
\begin{figure*} 
\includegraphics[width=11.0cm]{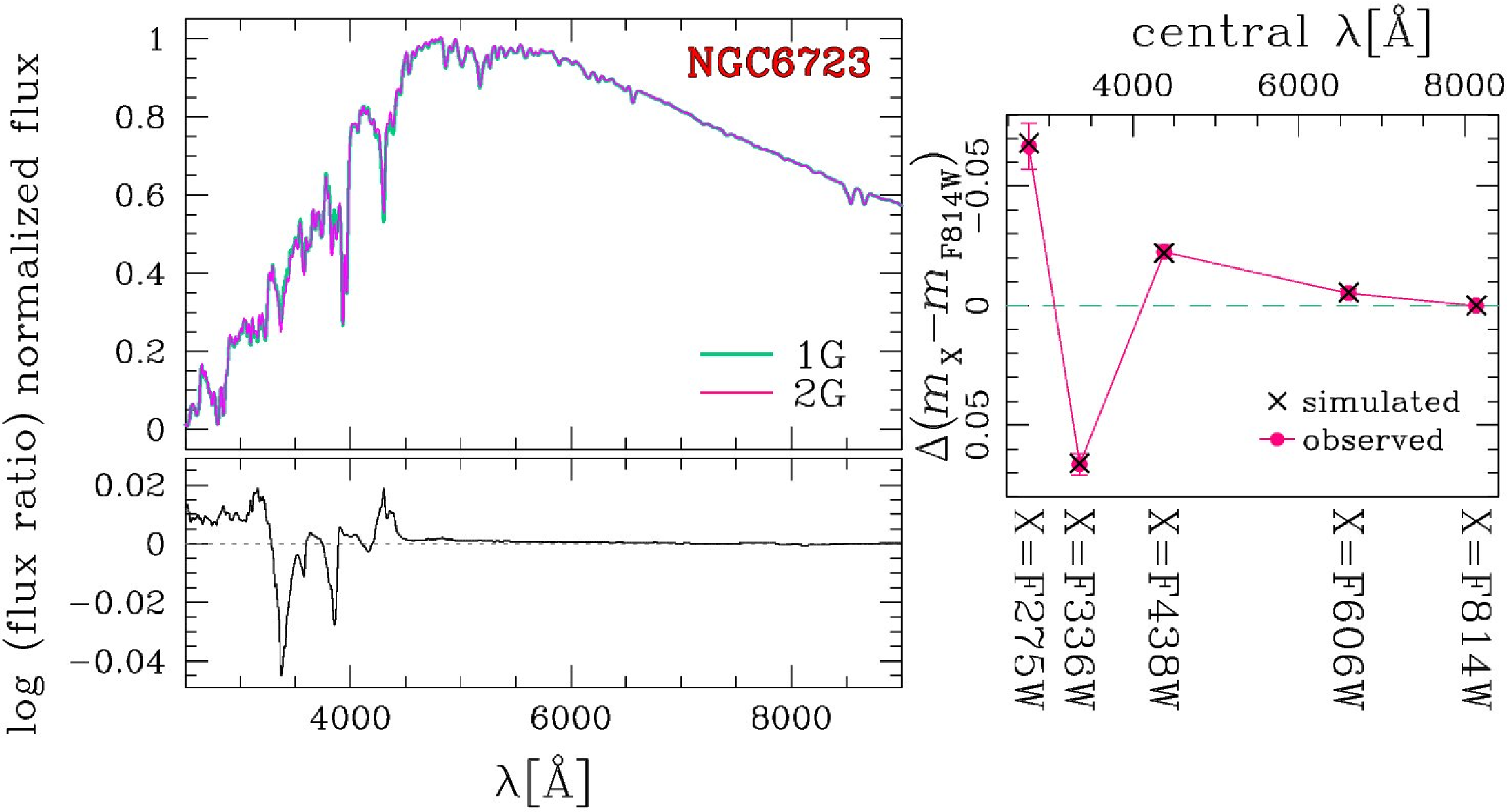} 
 %/home/milone/WORKS/treasury13/NGC6723/MATCH/popolazioni.macro fig2
 \caption{The aqua and magenta synthetic spectra plotted in the upper-left panel correspond to 1G and 2G stars of NGC\,6723 with $m_{\rm F814W}=16.38$, respectively. Lower-left panel shows the logarithm of the flux ratio of 2G and 1G spectra as a function of wavelength. Right panel reproduces the plot of Fig.~\ref{fig:fiducials6723} where we show with magenta points the $\Delta(m_{\rm X}-m_{\rm F814W})$ color difference between 2G and 1G fiducials of NGC\,6723 with $m_{\rm F814W}=16.38$ against the central wavelength of the X filter. The black crosses overimposed on this plot are the corresponding color difference derived from synthetic spectra.} 
 \label{fig:spettri6723} 
\end{figure*} 
\end{centering}

%NGC7089, NGC6779, NGC6101, NGC5466, NGC6397, NGC5024, NGC4590, NGC5053, NGC7099, NGC6341, NGC7078, NGC6624, NGC6304, NGC6496, NGC6441, NGC5927, NGC6388, NGC6366, NGC6637, NGC6352, NGC104, NGC6838, NGC6652, NGC6362, NGC6723, NGC2808, NGC6121, NGC6717, NGC1851, NGC362, NGC1261, NGC5904, NGC288, NGC6934, NGC6715, NGC6218, NGC6981, NGC5272, NGC6584, NGC6205, NGC6752, NGC6254, NGC5986, NGC3201, NGC6681, NGC5139, NGC5286, NGC6656, NGC6535, NGC6093, NGC6144, NGC6541, NGC4833, NGC2298, NGC6809
To infer the relative helium abundance between 2G and 1G stars, $\delta Y_{\rm 2G,1G}$, we adopted the following procedure, which is based on the method introduced by Milone et al.\,(2012) and is illustrated in the Figures~\ref{fig:fiducials6723} and~\ref{fig:spettri6723} for NGC\,6723. The same approach has been used to derive the helium difference, $\delta Y_{\rm max}$,  between 2Ge and 1Ge populations, which is indicative of the maximum helium variation within each cluster.

 We first analyzed the $m_{\rm F814W}$ vs.\,$m_{\rm X}-m_{\rm F814W}$ CMDs, where X=F275W, F336W, F438W, and F606W, and derived the fiducial line for each group of 1G and 2G RGB stars selected in paper\,IX.  To do this, we divided the RGB into F814W magnitude intervals of size $\delta m$ which are defined over a grid of points spaced by magnitude bins of size $s=\delta m/3$. For each bin we calculated the median F814W magnitude and X-F814W color and smoothed these median points by  boxcar averaging, where each point has been replaced by the average of the three adjacent points.
  The fiducial line of 1G and 2G stars is derived by linearly interpolating the resulting points and are represented with aqua and magenta lines, respectively, in the $m_{\rm F814W}$ vs.\,$m_{\rm X}-m_{\rm F814W}$ CMDs of Fig.~\ref{fig:fiducials6723}.

  Then we defined a list of $N$ points along the RGB, which are regularly spaced in F814W magnitude by intervals of size $2 \delta m$ and  are represented with filled circles in the lower panels of Fig.~\ref{fig:fiducials6723}. For each point, $i$, we calculated the $\Delta (m_{\rm X}-m_{\rm F814W})$ color difference between the fiducial of 2G and 1G stars. The $\Delta (m_{\rm X}-m_{\rm F814W})$ values derived for $m_{\rm F814W, i}=16.38$ in NGC\,6723 are plotted as a function of the central wavelength of the X filter in the upper-right panel of Fig.~\ref{fig:fiducials6723}. 
 
 We estimated the effective temperature and gravity corresponding to each point, $i$, by using the isochrones by Dotter et al.\,(2008) and assuming the same age, reddening, distance modulus, and metallicity derived by Dotter et al.\,(2010). For those clusters, namely NGC\,1851, NGC\,2808, NGC\,6388, NGC\,6441, NGC\,6656, and NGC\,6715, which are not investigated by Dotter et al.\,(2010), we adopted the values of age, reddening, distance modulus, and metallicity derived by Milone et al.\,(2014) by using the same recipes from Dotter and collaborators.

 The helium difference between 2G and 1G stars corresponding to each point, $i$, has been derived by using the following iterative procedure.
 We first computed a reference synthetic spectrum, corresponding to  a star with the effective temperature, gravity, and metallicity, $Z$, inferred from the best-fit isochrone, helium, $Y=0.245+1.5 \cdot Z$, solar abundances of carbon and nitrogen, and [O/Fe]=0.40.
 Moreover, we derived a sample of comparison spectra with the same atmospheric parameters and chemical composition as the reference spectrum but different abundance of either He, C, N, or O only. Specifically, we simulated two spectra enhanced in [N/Fe] by 0.5 and 1.5 dex, two spectra depleted in [O/Fe] by $-0.2$ and $-0.5$ dex, two spectra depleted in [C/Fe] by $-0.2$ and $-0.5$ dex, and one helium-rich spectrum with Y=0.33. When we change helium we also change $T_{\rm eff}$ and $\log{g}$ according to the isochrones by Dotter et al.\,(2008). We verified that the dependence of the relative helium abundances inferred from this procedure from the C, N, O abundances of the reference spectra is smaller than 0.001 in helium mass fraction.  

 Each spectrum has been integrated over the transmission curves of the {\it HST} filters used in this paper to derive the corresponding $\Delta (m_{\rm X}-m_{\rm F814W, i})^{\rm synth}$ color difference with the reference spectrum.
 We estimated the color difference due to a given variation in nitrogen, $\Delta (m_{\rm X}-m_{\rm F814W, i})^{\rm synth}$($\Delta$[N/Fe]),  by linearly interpolating the values of $\Delta (m_{\rm X}-m_{\rm F814W, i})_{\rm synth}$ derived from the nitrogen-rich comparison spectra and the adopted nitrogen variations of $\Delta$[N/Fe]=0.5 and 1.5. 
 We estimate the color variations, $\Delta (m_{\rm X}-m_{\rm F814W, i})_{\rm synth}$($\Delta$[Y]), $\Delta (m_{\rm X}-m_{\rm F814W, i})_{\rm synth}$($\Delta$[C/Fe]), $\Delta (m_{\rm X}-m_{\rm F814W, i})_{\rm synth}$($\Delta$[O/Fe]), due to variations in helium, carbon, and oxygen similarly. 

At this stage, we assumed as a first-guess abundance of He, C, N, O of 2G stars the values of these elements, Y$^{*,i}$, C$^{\rm *,i}$, N$^{\rm *,i}$, O$^{\rm *,i}$, that provide the best match between the observed $\Delta (m_{\rm X}-m_{\rm F814W, i})$ values and the sum $\Delta (m_{\rm X}-m_{\rm F814W, i})_{\rm synth, He, C, N, O}$=$\Delta (m_{\rm X}-m_{\rm F814W, i})_{\rm synth}$($\Delta$[Y])+$\Delta (m_{\rm X}-m_{\rm F814W, i})_{\rm synth}$($\Delta$[C/Fe])+$\Delta (m_{\rm X}-m_{\rm F814W, i})_{\rm synth}$($\Delta$[N/Fe])+$\Delta (m_{\rm X}-m_{\rm F814W, i})_{\rm synth}$($\Delta$[O/Fe]).

The derived abundances are used to simulate a new sample of comparison spectra.  The content of He, C, N, and O of the main comparison spectrum corresponds to the first-guess abundances previously defined (Y$^{\rm *, i}$, C$^{\rm *, i}$, N$^{\rm *, i}$, and O$^{\rm *, i}$).  Similarly to what we have done before, the other comparison spectra all have the same chemical composition but different abundance of either He, C, N, or O only. Specifically, both [C/Fe], [N/Fe], and [O/Fe] have been changed by $\pm$0.2 dex and $Y$ by $\pm$0.02. 
These synthetic spectra are used to derived improved estimates of Y$^{\rm *, i}$, C$^{\rm *, i}$, N$^{\rm *, i}$, and O$^{\rm *, i}$. This procedure was was repeated, changing the C, N, and O abundance by 0.1 dex and the helium abundance by 0.01.  This procedure typically converged after three or four iterations.
%repeated until convergence. At subsequent iterations we changed the C, N, and O abundances by 0.1 dex and the helium abundance by 0.01. The convergence typically requires either three or four iterations. 

%%%%%%%%%%%%%%%%%%%%%%%%%%%%%%%%%%%%%%%%%%%%%%%%%%%%%%%%%%%%%%%%%%%%%%%%%%%
%%%%%%%%%%%%%%%%%%%%%%%%%%%%%%%%%%%%%%%%%%%%%%%%%%%%%%%%%%%%%%%%%%%%%%%%%%%

%%%%%%%%%%% ADDED Mg
As discussed in Section~\ref{sec:mgteo}, in addition to C, N, O variations, which are common features of all the GCs, in some GCs the content of magnesium, aluminum and silicon is also not uniform.
% In Section~\ref{sec:mgteo} we investigated the effect of these elements on the colors of RGB stars. We find that the star-to-star variations in Mg, Al, and Si that are typically observed in some massive GCs do not provide any significant flux variation in optical filters (central wavelength, $\lambda > 4,000 \AA$) of all the GCs and in ultraviolet filters with $\lambda > 3,000 \AA$ of metal-poor GCs. 
%
%Mg, Al and Si variations are responsible for significant flux variation (at the level of $\sim$0.02 mag) in ultraviolet filters with $\lambda$ smaller than $\sim 3,000 \AA$ in all the clusters and in filters with $\lambda < 4,000 \AA$ in the metal-rich GCs. As an exception, the flux variation can be as large as $\sim$0.3 mag in the F280N band. Magnesium is the element which is mostly responsible for such large flux variations. 

To account for the effects of Mg, Al, and Si on the observed colors of RGB stars, we assumed the abundances of these elements inferred from spectroscopy. For GCs with no spectroscopic determination of these elements we provide two determinations of $\delta Y_{\rm 2G,1G}$ and $\delta Y_{\rm max}$.
We first determined both helium variations by using constant [Mg/Fe]=0.40, [Al/Fe]=0.0, and [Si/Fe]=0.4.
Then we estimated $\delta Y_{\rm max}$ by assuming that 2G stars are enhanced in [Al/Fe] and [Si/Fe] by 0.8 and 0.07 dex, respectively and depleted in [Mg/Fe] by 0.1 dex, with respect to 1G stars. 
 Similarly, we obtained $\delta Y_{\rm max}$ by assuming that 2Ge stars are enhanced in [Al/Fe] and [Si/Fe] by 1.0 and 0.15 dex, respectively and depleted in [Mg/Fe] by 0.3 dex, with respect to 1Ge stars. These values resemble the abundance difference between 2G and 1G stars of NGC\,6752 and the maximum abundance variations among NGC\,6752 stars derived by Yong et al.\,(2015).

We find that the adopted choices for the Mg, Al, and Si abundances have a negligible impact on the inferred values of $\delta Y_{\rm max}$ (less than 0.002).
 This finding is consistent with the fact that these elements do not affect the  stellar flux in optical bands.

 In contrast, the adopted abundances for Mg, Al, and Si do strongly affect the best-fit values of oxygen and nitrogen in metal-poor GCs. This fact is not unexpected, since O and N are largely constrained by the F275W band.
 The flux variations in this filter, which is mostly due to magnesium and (to a less extent) to Si and Al, are generally small ($\sim 0.02$ mag) in all GCs. Nevertheless, in metal-poor clusters they are comparable to those due to oxygen and nitrogen.
The adopted values for the relative content of C, N, O, Mg, Al, and Si in 1G and 1G stars and in 2Ge and 1Ge stars are listed in Table~\ref{tab:CNO}.
%%%%%%%%%%%

As an example of the procedure to infer $\delta Y_{\rm 2G,1G}$, in Fig.~\ref{fig:spettri6723} we illustrate the results from the analysis of 1G and 2G stars of NGC\,6723 with $m_{\rm F814W, i}=16.38$, i.e.\, 2.0 magnitudes above the turn off. The upper-left panel shows the reference synthetic spectrum (aqua), which is representative of 1G stars, and the comparison spectrum (magenta), which provides the best fit with the observed $\Delta(m_{\rm X}-m_{\rm F814W, i})$ color differences. The corresponding flux ratio is provided in the lower-left panel, while in the right panel we compare the observed color differences and those obtained from the best-fit synthetic spectrum. We assumed as the best estimate of the relative helium abundance between 2G and 1G stars, $\delta Y_{\rm 2G,1G}$, the difference between the mean value of the $N$ determinations of $Y^{\rm *,i}$ and the helium abundance that we assumed for 1G stars ($Y=0.245+1.5 \cdot Z$).

%%%%%%%%%%%%%%%%%%%%%%%%%%%%%%%%%%%%%%%%%%%%%%%%%%%%%%%%%%%%%%%%%%%%%%%%%%%
%%%%%%%%%%%%%%%%%%%%%%%%%%%%%%%%%%%%%%%%%%%%%%%%%%%%%%%%%%%%%%%%%%%%%%%%%%%

\begin{centering} 
\begin{figure*} 
  \includegraphics[width=13.15cm]{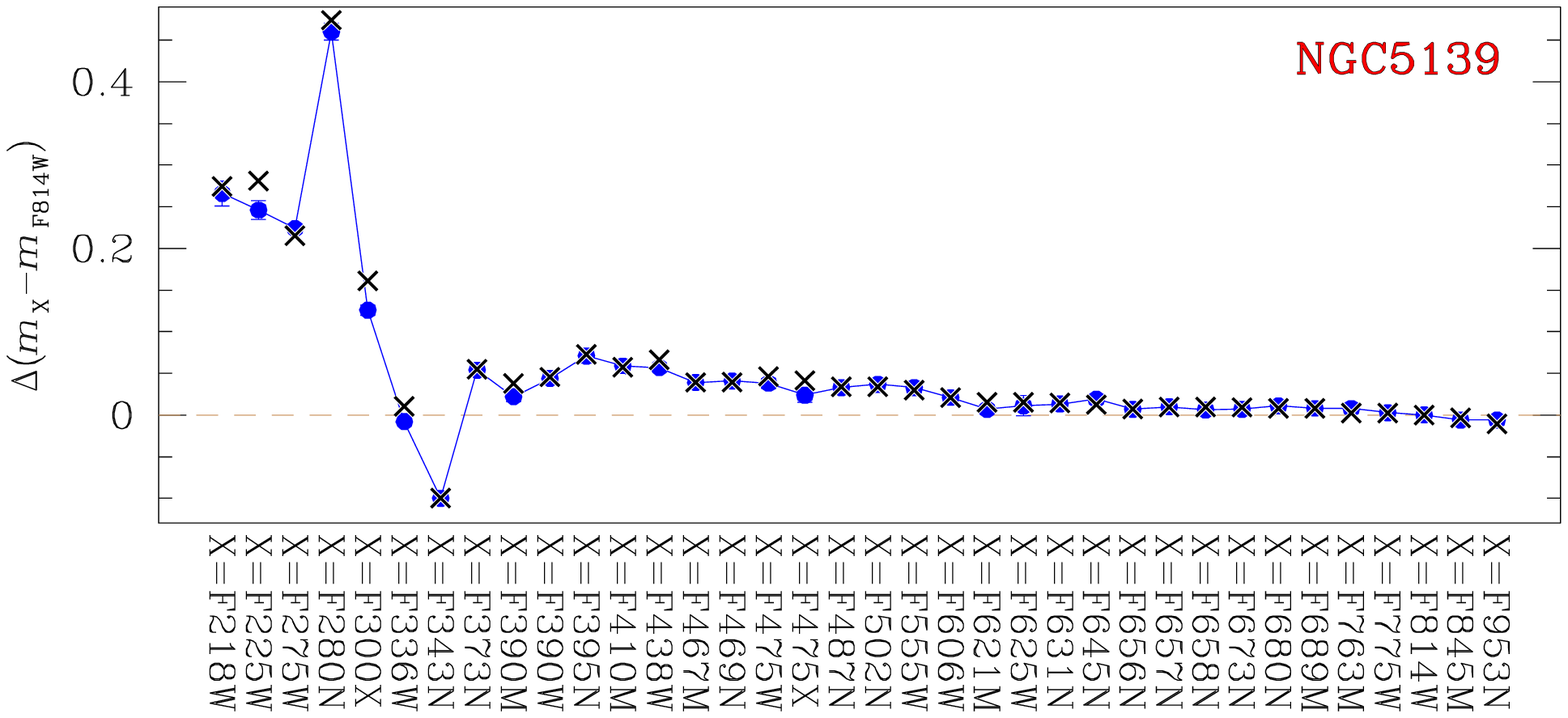}  
  \includegraphics[width=6.5cm]{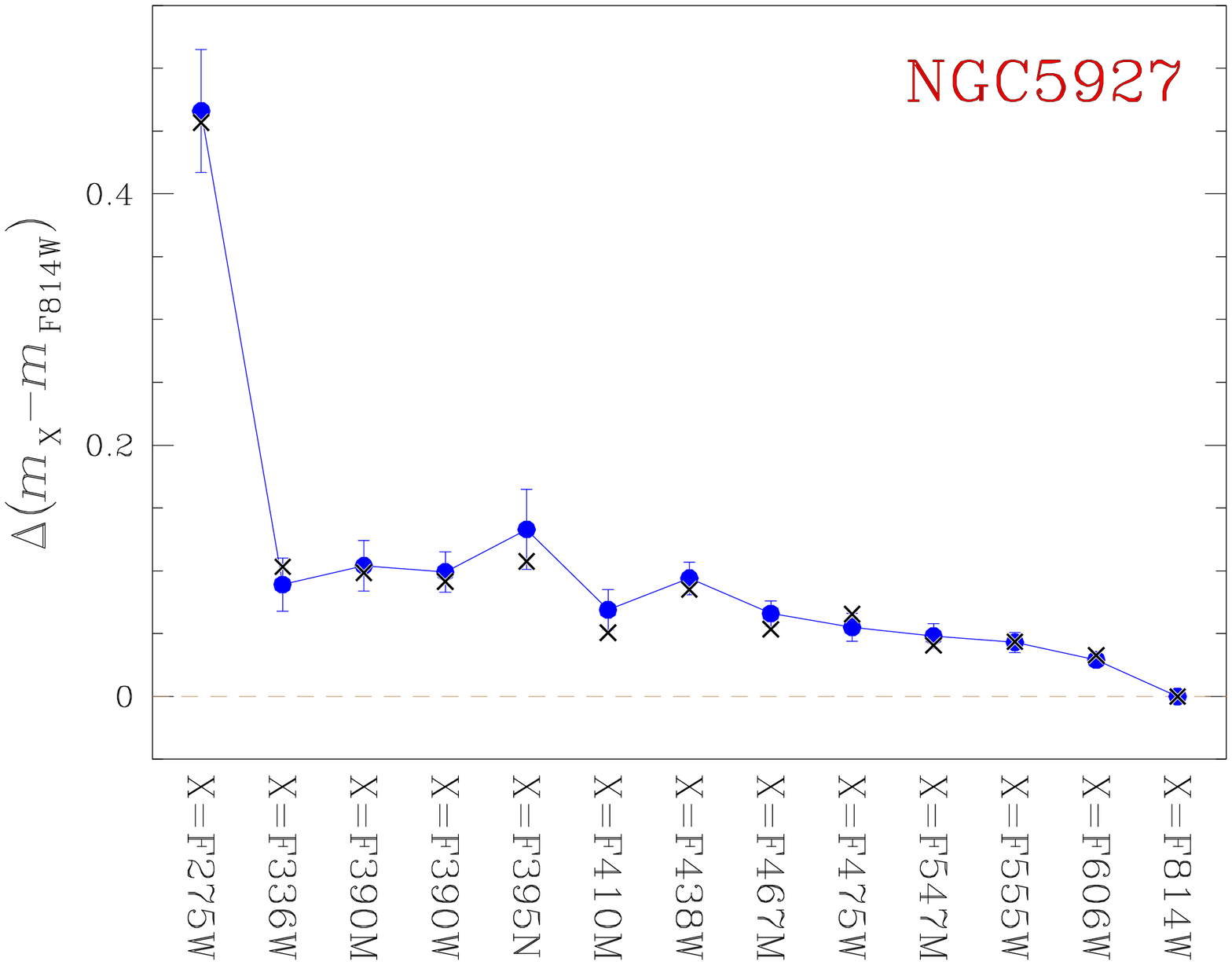}
  \includegraphics[width=6.5cm]{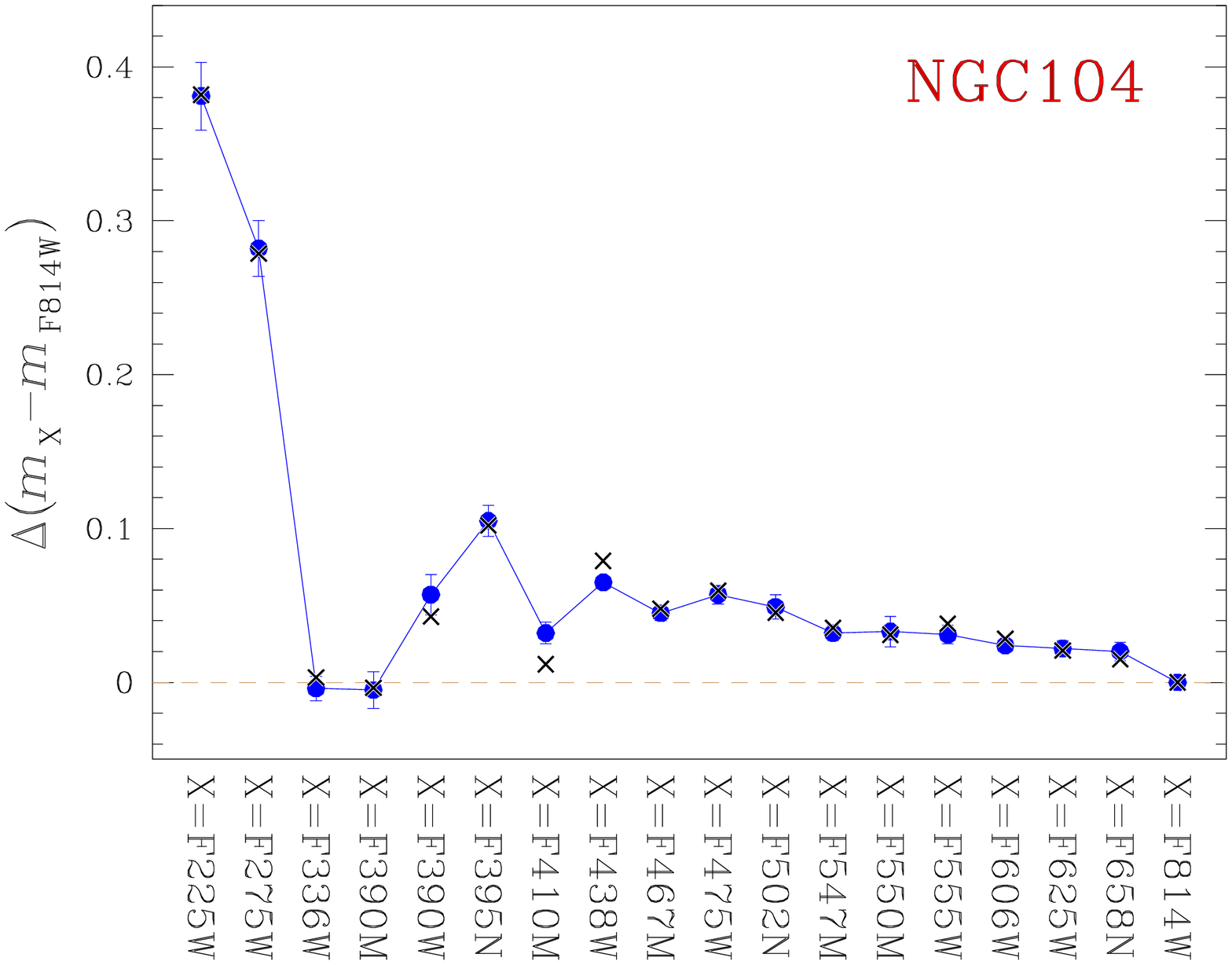}
  \includegraphics[width=6.5cm]{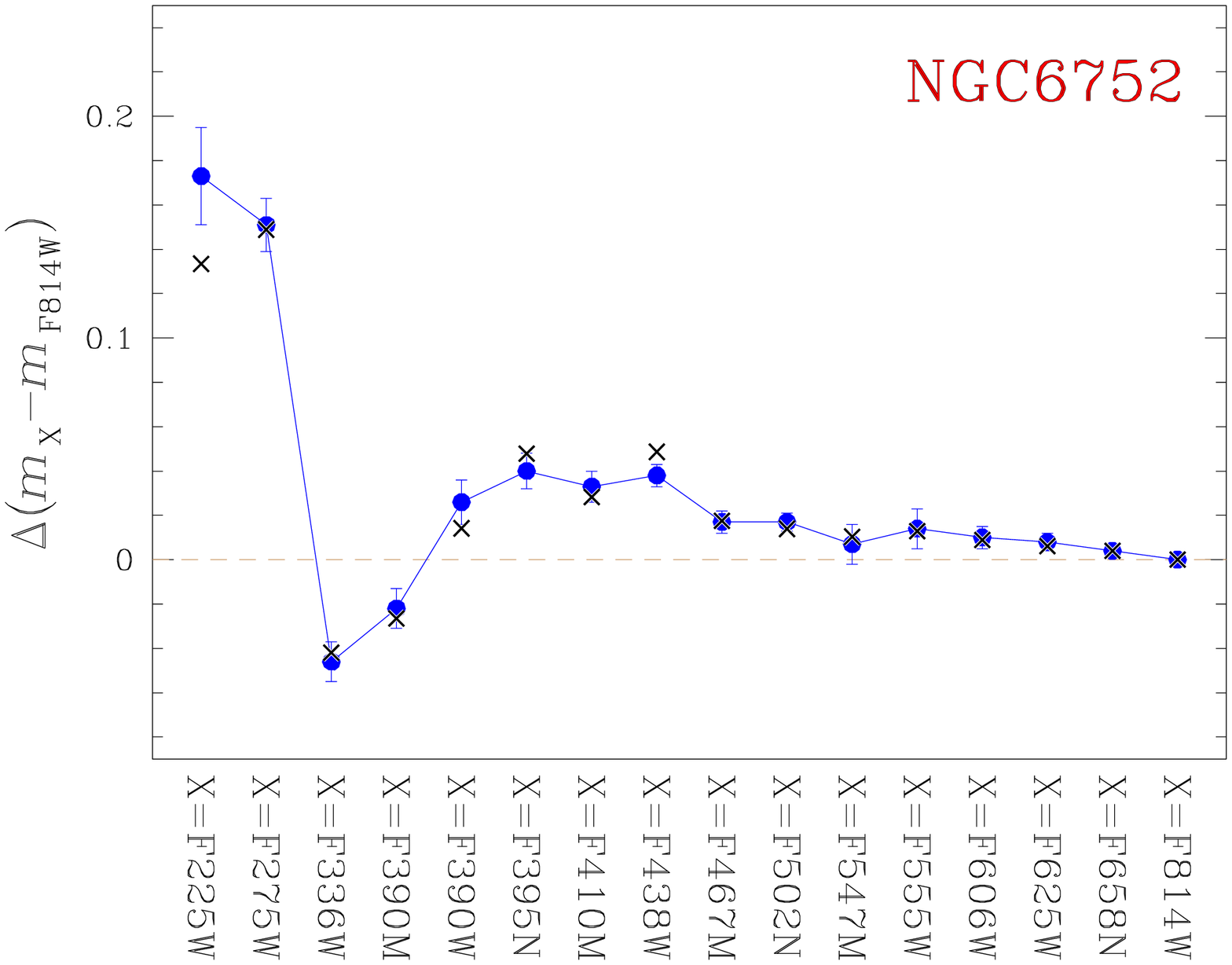}
  \includegraphics[width=6.5cm]{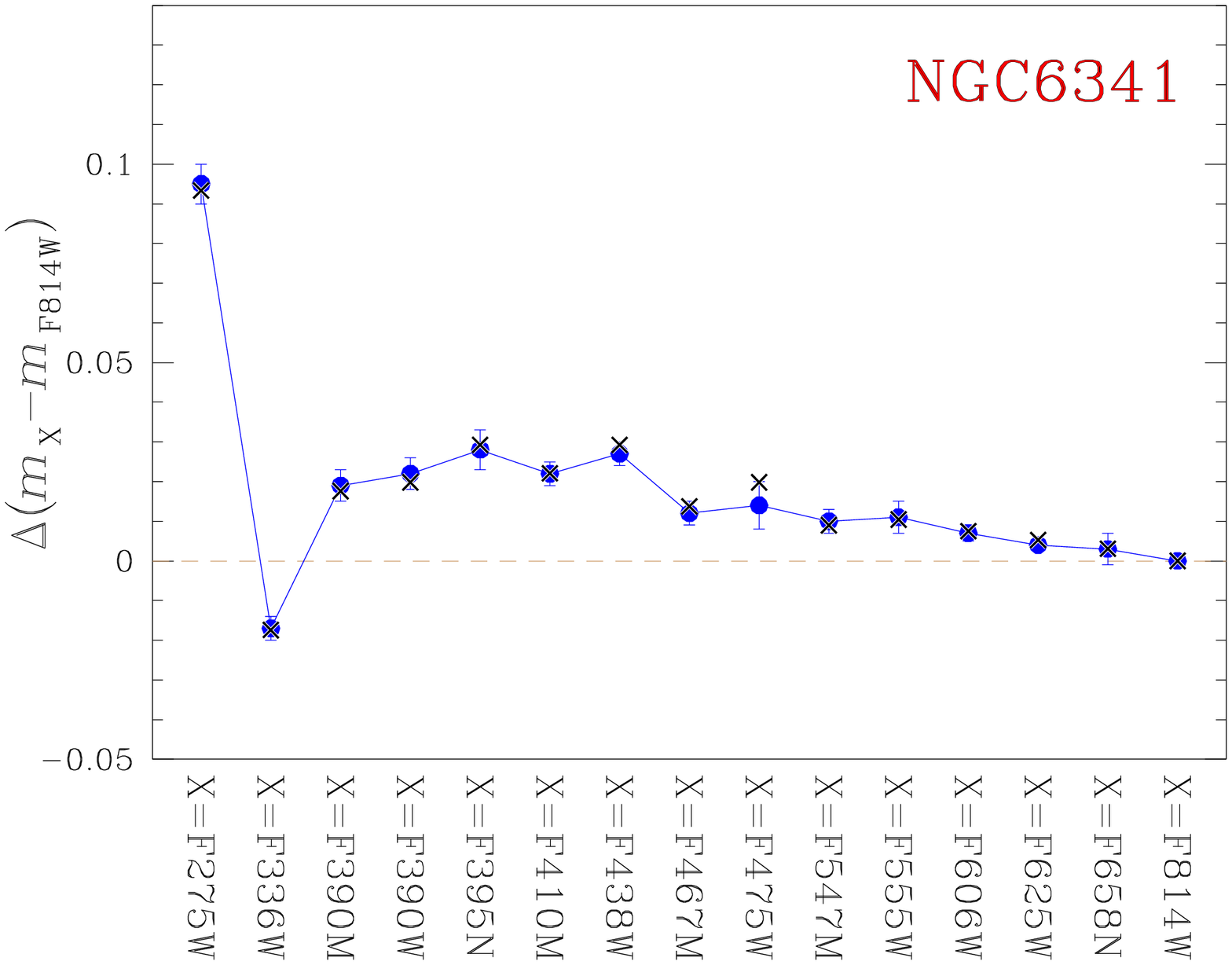} 
  \includegraphics[width=6.5cm]{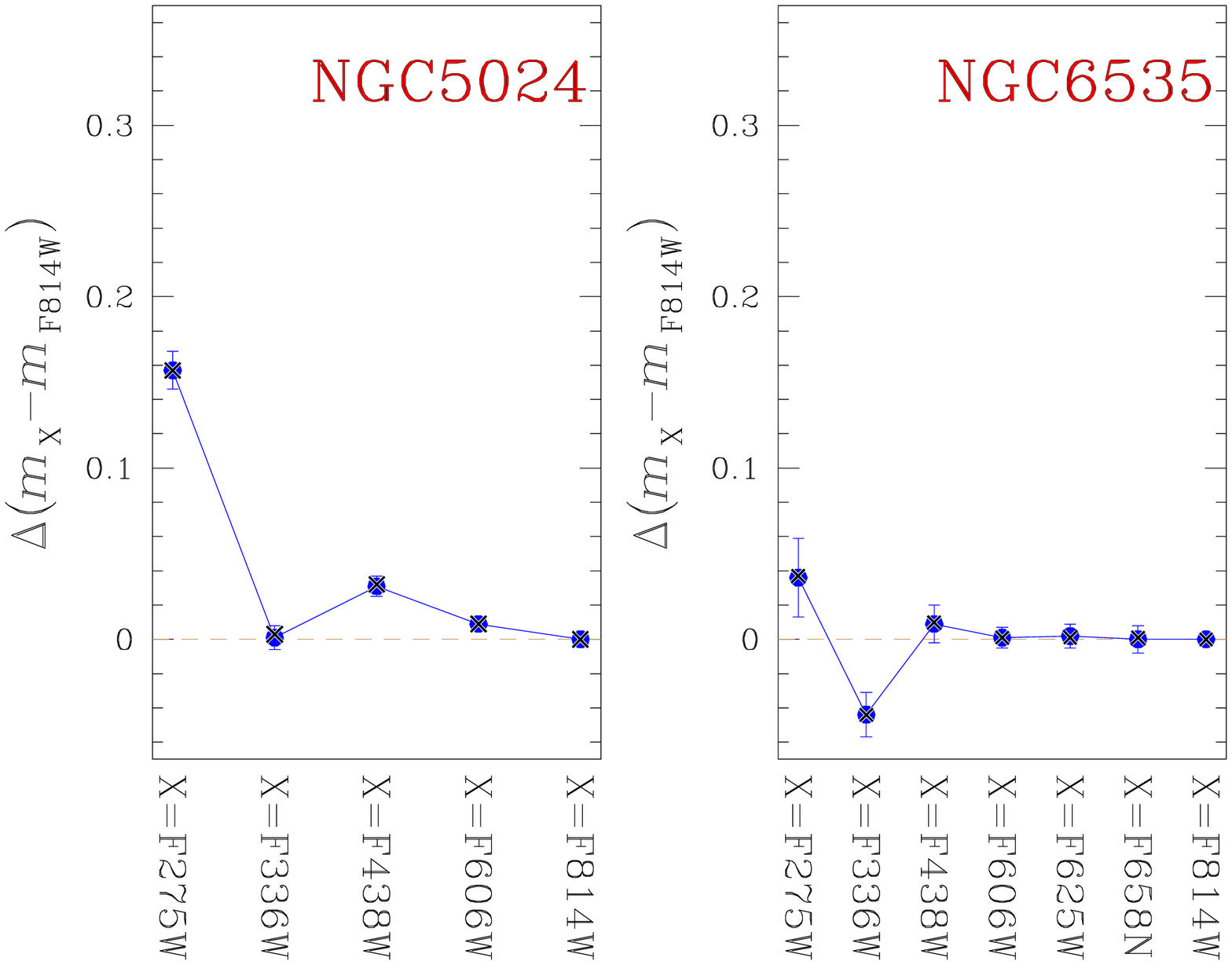}
  \includegraphics[width=6.5cm]{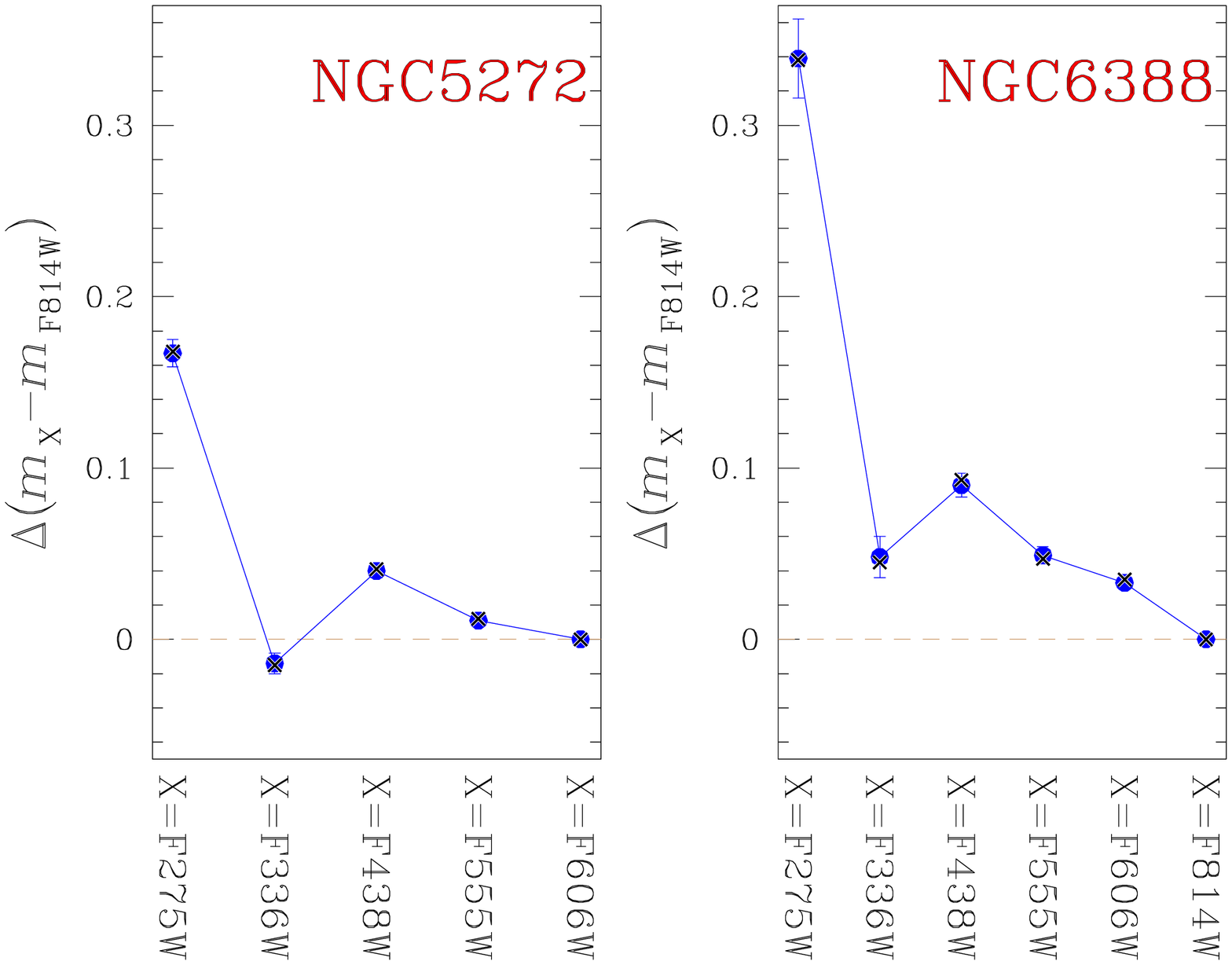}
 \caption{Color difference between 2Ge and 1Ge stars for nine GCs  against the central wavelength of the X filter. Blue dots with error bars indicate the observed color differences and the corresponding uncertainties, while black crosses correspond to the best-fit model. The color differences have been estimated for stars that are 2.0 F814W magnitudes brighter than the turn off.} 
 \label{fig:DCOLobs} 
\end{figure*} 
\end{centering}

As an example, we show in Fig.~\ref{fig:DCOLobs} the $\Delta(m_{\rm X}-m_{\rm F814W})$ color difference between 2Ge and 1Ge stars corresponding to various X filters for nine analyzed GCs with different metallicities.
 We included NGC\,5139, NGC\,5927, NGC\,104, NGC\,6752, and NGC\,6341, for which photometry in a large number of 13--35 filters  is available, and NGC\,5024, NGC\,6535, NGC\,5272, and NGC\,6388 where the number of available filters ranges from five to seven.

%%%%%%%%%%%%%%%%%%%%%%%%%%%%%%%%%%%%%%%%%%%%%%%%%%%%%%%%%%%%%%%%%%%%%%%%%%%
%%%%%%%%%%%%%%%%%%%%%%%%%%%%%%%%%%%%%%%%%%%%%%%%%%%%%%%%%%%%%%%%%%%%%%%%%%%

 \section{Results}\label{sec:results}
\begin{centering} 
\begin{figure} 
\includegraphics[width=8.0cm]{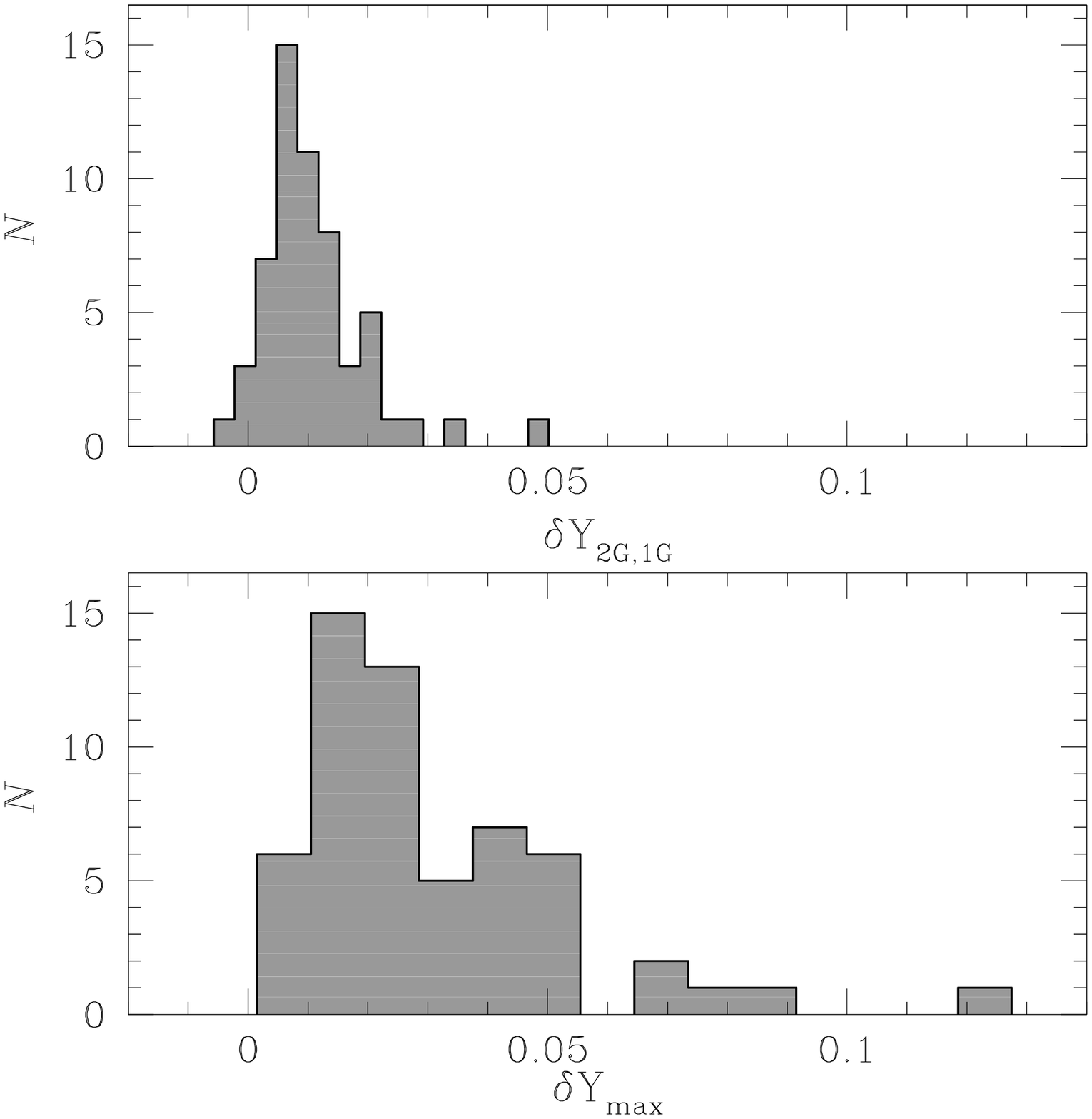} 
 %/home/milone/WORKS/treasury13/ELIO/test.macro go2h
 \caption{Histogram distribution of the derived values for the helium differences between 2G and 1G stars (top) and the maximum helium variation within each GC (bottom). We remark that in type II GCs, only blue-RGB stars have been included in the analysis.} 
 \label{fig:HeHisto} 
\end{figure} 
\end{centering}

The relative helium abundances derived in the previous Section are listed in Table~\ref{tab:He} for all the analyzed clusters, while the histogram distributions of $\delta Y_{\rm 2G,1G}$, and $\delta Y_{\rm max}$ are plotted in Fig.~\ref{fig:HeHisto}.

%%%%%%%%%%%%%%%%%%%%%%%%%%%%%%%%%%%%%%%%%%%%%%%%%%%%%%%%%%%%%%%%%%%%%%%%%%%
%%%%%%%%%%%%%%%%%%%%%%%%%%%%%%%%%%%%%%%%%%%%%%%%%%%%%%%%%%%%%%%%%%%%%%%%%%%

We find that 2G stars are typically more helium-rich than 1G stars. The median helium enhancement for the 2G stars in the 57 analyzed GCs is $0.009$ in mass fraction  and the 68.27$^{\rm th}$ percentile of the measurements corresponds to $\sigma=0.007$.
Noticeably, the two main populations of several clusters are consistent, within measurement errors, with constant helium content but in none of the analyzed objects is the first generation more He-rich than the second. The average helium enhancement of 2G stars never exceeds $\sim 0.05$. 

% Marino et al.\,(2017) has determined the chemical composition of 1G and 2G stars defined in paper IX for XX GCs by using literature elemental abundance derived from high-resolution spectroscopy. They find that in most GCs 1G and 2G stars have on average different abundance of some light-elements, including nitrogen, sodium, oxygen, and in some cases magnesium, aluminum, and silicon. In contrast, 1G and 2G stars in stars of both type I GCs and blue-RGB stars of type-II GCs show no evidence for star-to-star variations in  $\alpha$ and heavy elements.
% Figure~\ref{} shows $\delta Y_{\rm 2G,1G}$ against the average 2G-1G abundance difference of some elements ($\delta [el/Fe]_{\rm 2G,1G}$) and against the average difference of iron abundance ($\delta [Fe/H]_{\rm 2G,1G}$) calculated by Marino et al.\,(2017).
% We find that the average helium difference between 2G and 1G stars correlates with the corresponding nitrogen, sodium, and alluminum and anticorrelates with the abundance of oxygen and magnesium. 

 The maximum internal variation of helium significantly changes from one cluster to another and ranges from $\sim$0.00 in small mass clusters like NGC\,5053,  NGC\,5466, NGC\,6362, NGC\,6535, NGC\,6717 to $\sim$0.12 in NGC\,2808.
  The median value corresponds to 0.027 ($\sigma=0.018$).
 The fact that our determinations of $\delta Y_{\rm max}$ are typically larger than the corresponding values of $\delta Y_{\rm 2G,1G}$ clearly demonstrates that at least 2G stars (and possibly the 1G population as discussed in Sect.~4) host sub-populations of stars with different helium content.

% \section{The helium abundance and the cluster parameters}
 \subsection{Univariate relations between the helium abundance and the global cluster parameters}
\label{sec:correlations}
In the following we investigate the relation between the helium abundance and the main parameters of the host GCs, in close analogy with what we did in Paper\,IX for the RGB width and the fraction of 1G stars with respect to the total number of cluster stars.

Our analysis involves some global parameters from the Harris\,(1996, updated as in 2010), including metallicity ([Fe/H]), absolute visual magnitude ($M_{\rm V}$), central velocity dispersion ($\sigma_{\rm V}$), ellipticity ($\epsilon$), central concentration ($c$), core relaxation time ($\tau_{\rm c}$), half-mass relaxation time($\tau_{\rm hm}$), central stellar density ($\rho_{0}$), central surface brightness ($\mu_{\rm V}$), reddening ($E(B-V)$), and Galactocentric distance ($R_{\rm GC}$).  Moreover, we used cluster masses from McLaughlin \& van der Marel\,(2005), and ages from Mar{\'i}n Franch et al.\,(2009, MF09), Dotter et al.\,(2010, D10), and Vandenberg et al.\,(2013, V13). 
The binary fraction determined by Milone et al.\,(2012) within the cluster core ($f_{\rm bin}^{\rm C}$),  in the region between the core and the half-mass radius ($f_{\rm bin}^{\rm C-HM}$), and beyond  the half-mass radius ($f_{\rm bin}^{\rm oHM}$).
%We also used the RGB widths in $C_{\rm F275W,F336W,F438W}$ and $m_{\rm F275W}-m_{\rm F814W}$ ($W_{\rm C~ F275W, F336W, F438W}$ and $W_{\rm F275W, F814W}$) and the 1G fraction ($N_{1}/N_{\rm TOT}$) calculated in Paper\,IX and the HB F606W$-$F814W and F275W$-$F814W HB extension ($L_{\rm F606W-F814W}$ and $L_{\rm F275W-F814W}$) derived by Milone et al.\,(2014) and in Sect.~\ref{sub:HB}.

For each pair of analyzed quantities we calculated Spearman's rank correlation coefficient, $r$, and estimated the corresponding uncertainty, ($\sigma_{\rm r}$), by  bootstrapping statistics as in Milone et al.\,(2014).
The results are listed in Table~\ref{tab:rel}.

% When we analyze the HB color extension we considered both the entire sample of GCs and the three sub-groups of $g1$, $g2$, and $g3$ GCs defined by Milone et al.\,(2014).  Results are listed in Table~\ref{}. 

%%%%%%%%%%%%%%%%%%%%%%%%%%%%%%%%%%%%%%%%%%%%%%%%%%%%%%%%%%%%%%%%%%%%%%%%%%%
\begin{centering} 
\begin{figure*} 
  \includegraphics[width=6.0cm]{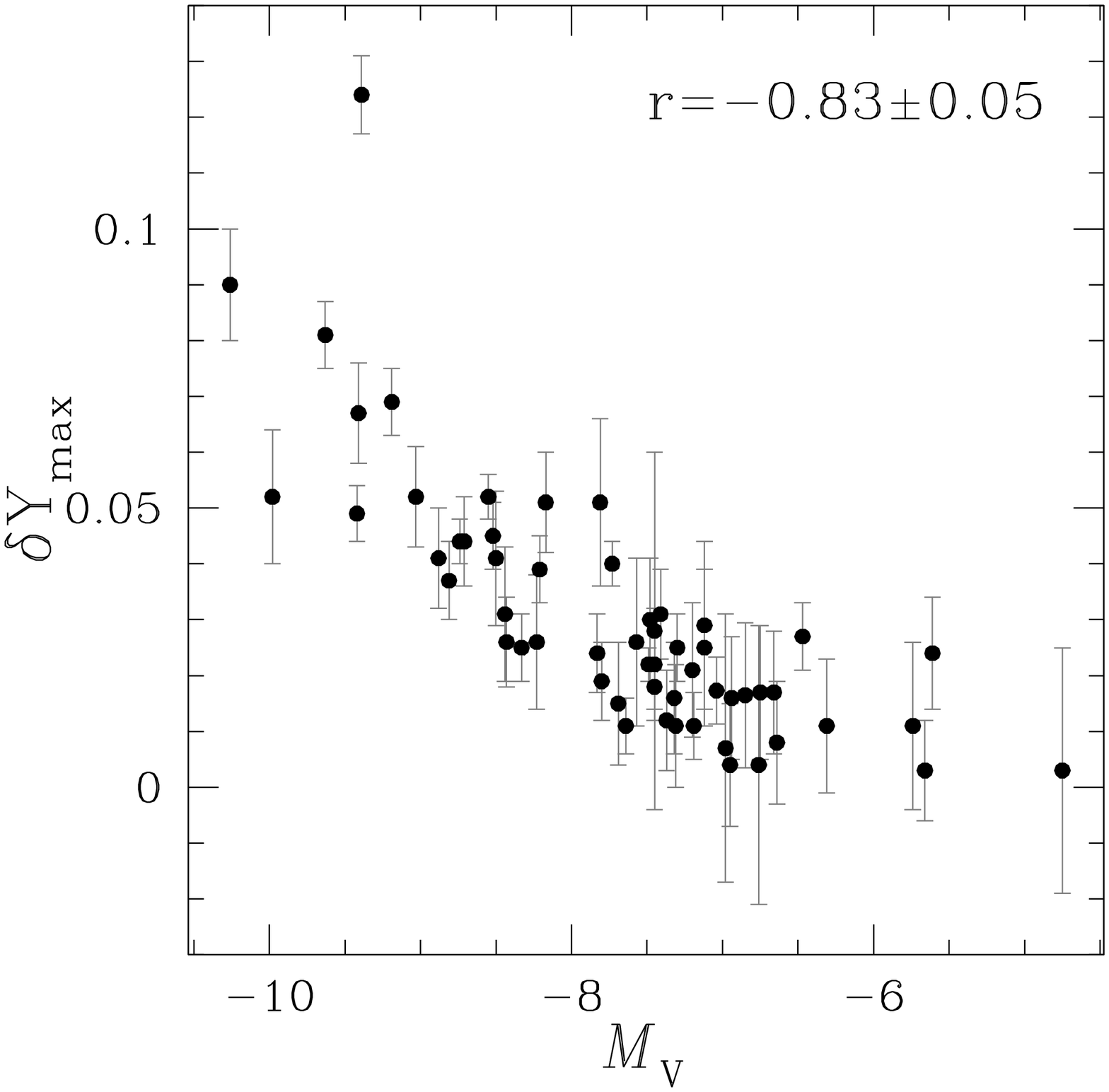}
  \includegraphics[width=6.0cm]{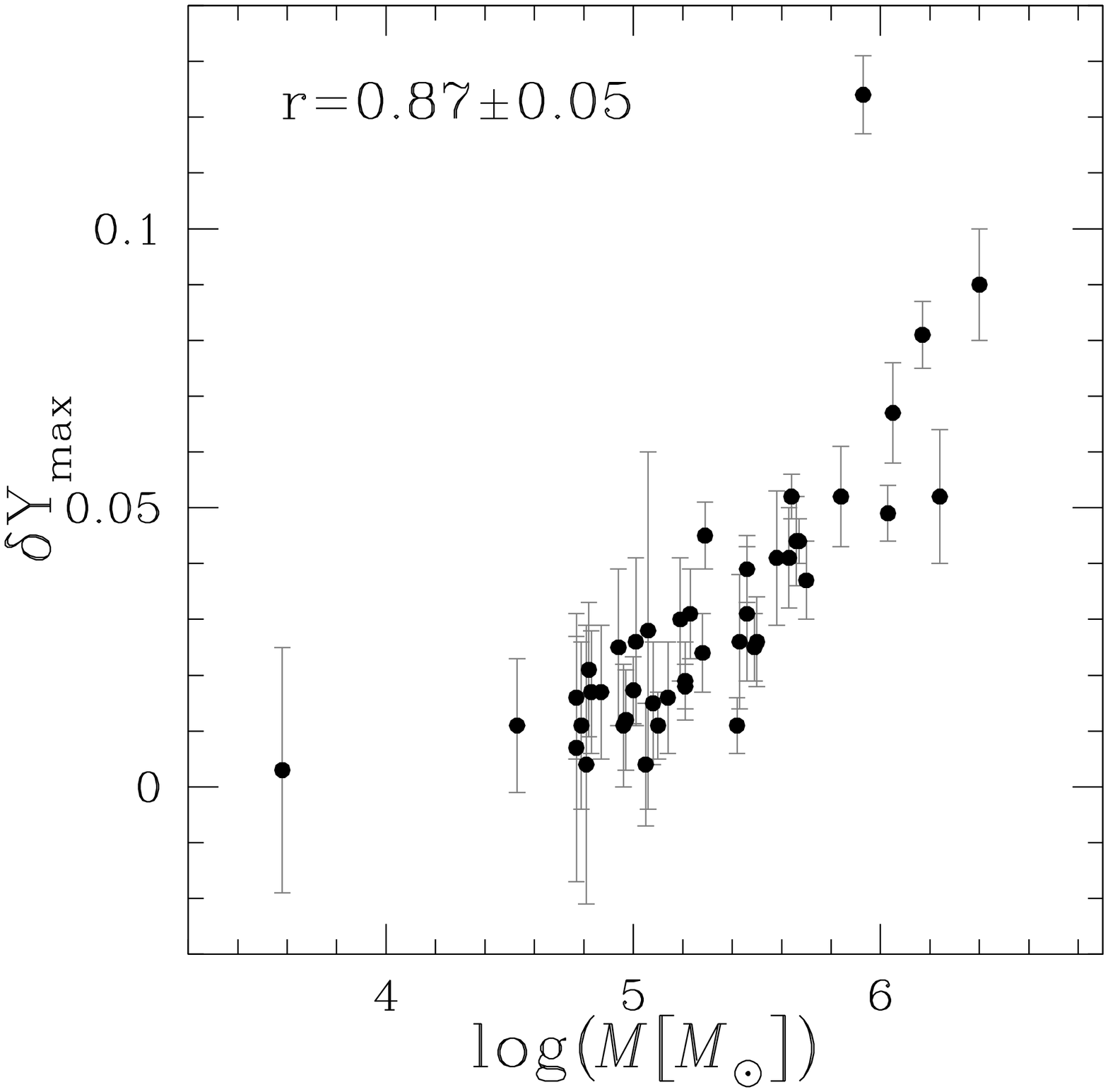}
 \caption{Maximum internal helium variation, $\delta Y_{\rm max}$ (left) as a function of the absolute magnitude (left) and the mass (right) of the host cluster. In each panel we indicate the Spearman's rank correlation coefficient and the corresponding uncertainty.} 
 \label{fig:figHe} 
\end{figure*} 
\end{centering} 
%%%%%%%%%%%%%%%%%%%%%%%%%%%%%%%%%%%%%%%%%%%%%%%%%%%%%%%%%%%%%%%%%%%%%%%%%%%

%%%%%%%%%%%%%%%%%%%%%%%%%%%%%%%%%%%%%%%%%%%%%%%%%%%%%%%%%%%%%%%%%%%%%%%%%%%
\begin{centering} 
\begin{figure*} 
  \includegraphics[width=6.0cm]{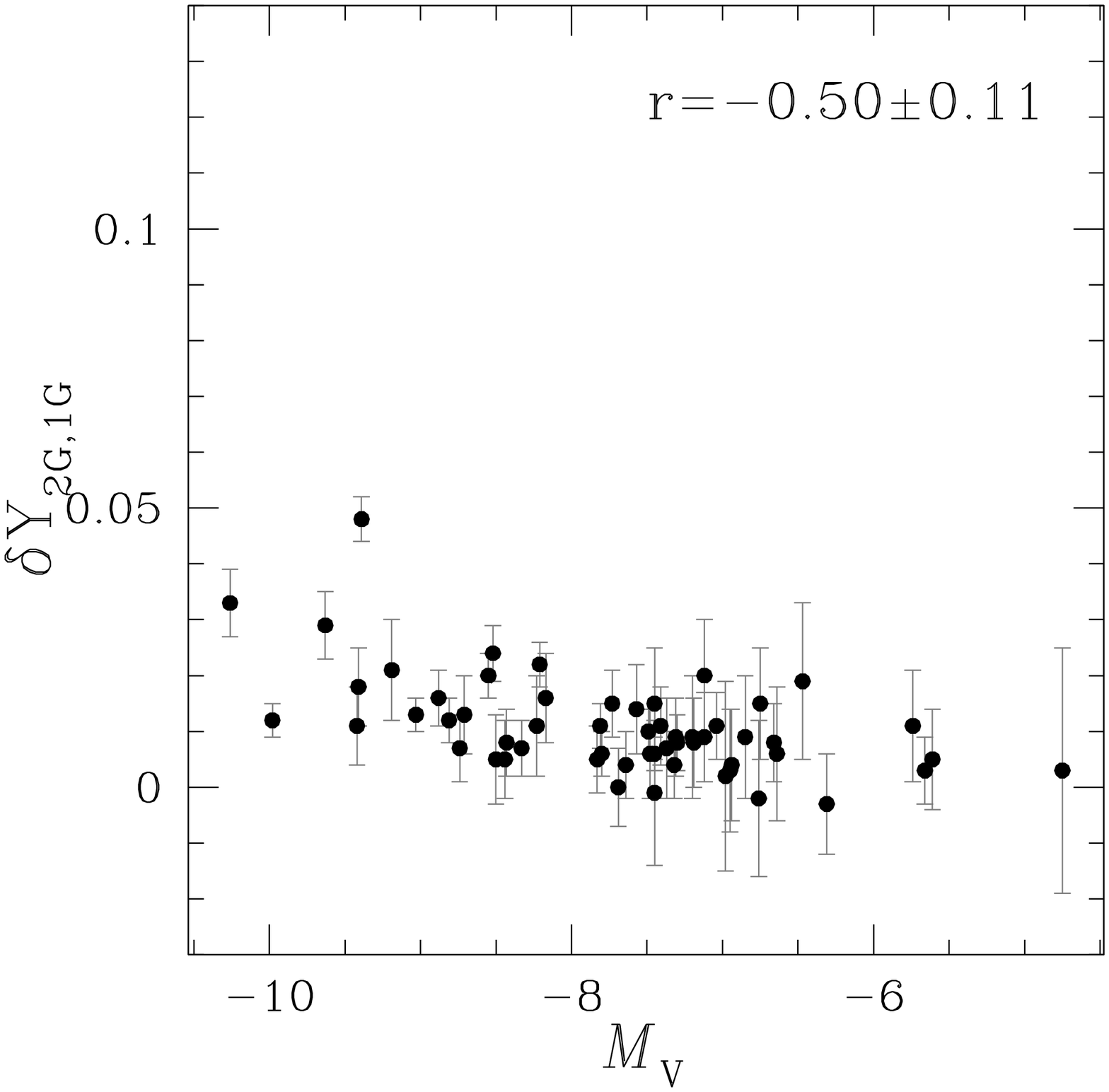} 
   \includegraphics[width=6.0cm]{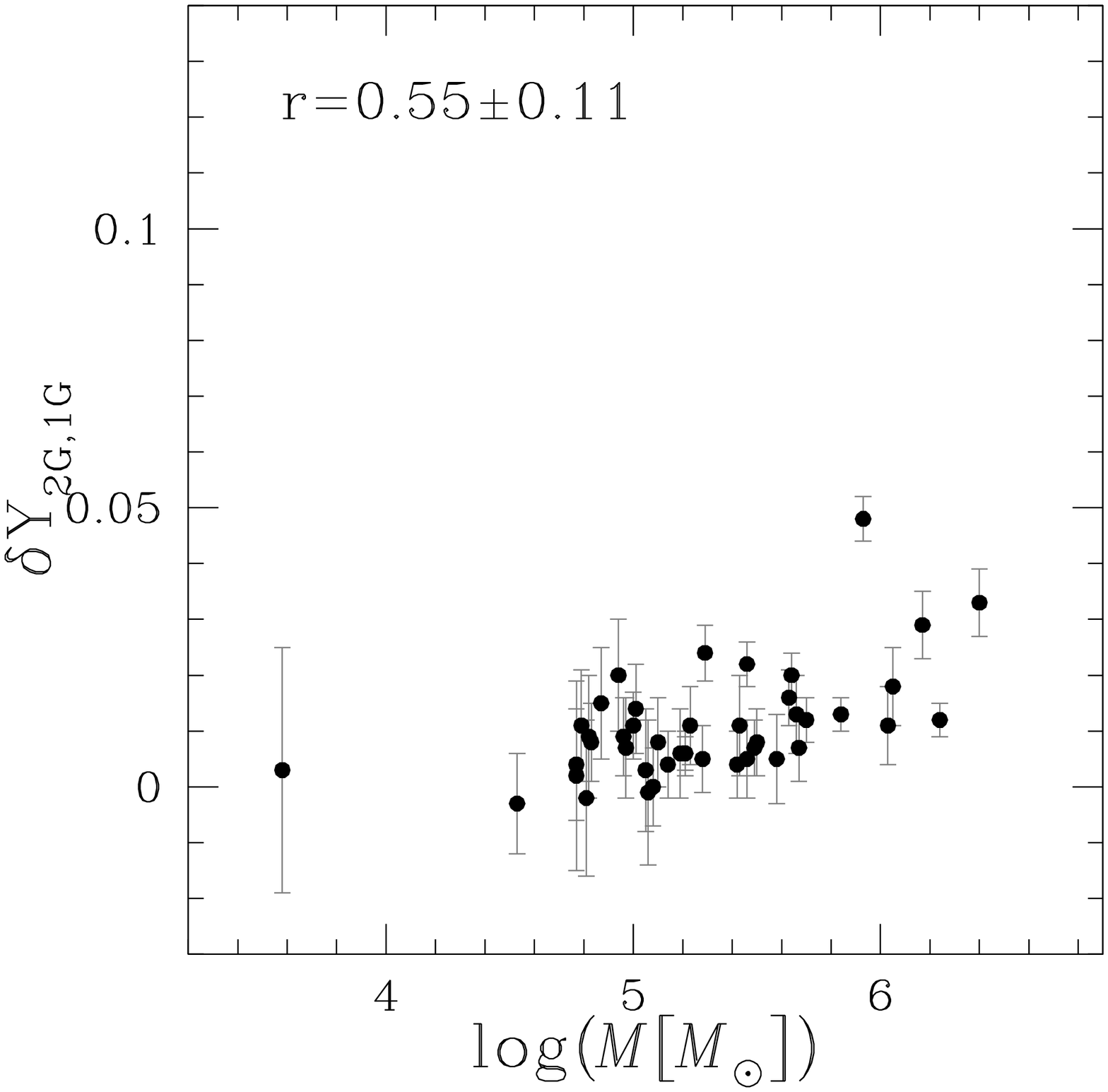} 
 \caption{Average helium difference between 2G and 1G stars, $\delta Y_{\rm 2G,1G}$, as a function of the absolute magnitude (left) and the mass (right) of the host cluster. In each panel we indicate the Spearman's rank correlation coefficient and the corresponding uncertainty. Note that in this plot we adopted the same limits as in Fig.~\ref{fig:figHe}. } 
 \label{fig:figHe2} 
\end{figure*} 
\end{centering} 
%%%%%%%%%%%%%%%%%%%%%%%%%%%%%%%%%%%%%%%%%%%%%%%%%%%%%%%%%%%%%%%%%%%%%%%%%%%

%\subsection{The maximum helium variation and the global cluster parameters}
As illustrated in Fig.~\ref{fig:figHe}, low-mass clusters clearly show smaller helium variations than massive ones.
We find a strong anticorrelation between the maximum internal helium variation ($r=-0.83\pm 0.05$) and the absolute magnitude of the host GC and a strong correlation with the logarithm of cluster mass ($r=0.86\pm 0.05$). These findings confirm a previous result by Milone (2015) based on a smaller data set.
 There is some mild correlation between the maximum helium variation and the central velocity dispersion ($r=0.53\pm 0.12$) and an anticorrelation with the central surface brightness. These results are not surprising because these quantities are related with the cluster mass (see e.g. Djorgovski \& Meylan 1994). There is no evidence for significant correlation and anticorrelations with the other parameters analyzed in this work. 
%As expected, the maximum helium variation correlates or anticorrelates with those quantities which %, including.

%\subsection{The helium difference between 2G and 1G stars the global cluster parameters}
 The relation between $\delta Y_{\rm 2G,1G}$ and the absolute luminosity and the mass of the host cluster are shown in Fig.~\ref{fig:figHe2}. In this case, we find only a mild anticorrelation with $M_{\rm V}$ ($r=-0.50\pm 0.11$) and some hints of a correlation with the logarithm of the cluster mass ($r=0.54\pm 0.11$).

 %\subsection{The F275W-F814W color extension of the horizontal-branch}
 \subsection{Relations with the F275W-F814W color extension of the horizontal branch}
\label{sub:HB}
%%%%%%%%%%%%%%%%%%%%%%%%%%%%%%%%%%%%%%%%%%%%%%%%%%%%%%%%%%%%%%%%%%%%%%%%%%%
\begin{centering} 
\begin{figure} 
 \includegraphics[width=9.0cm]{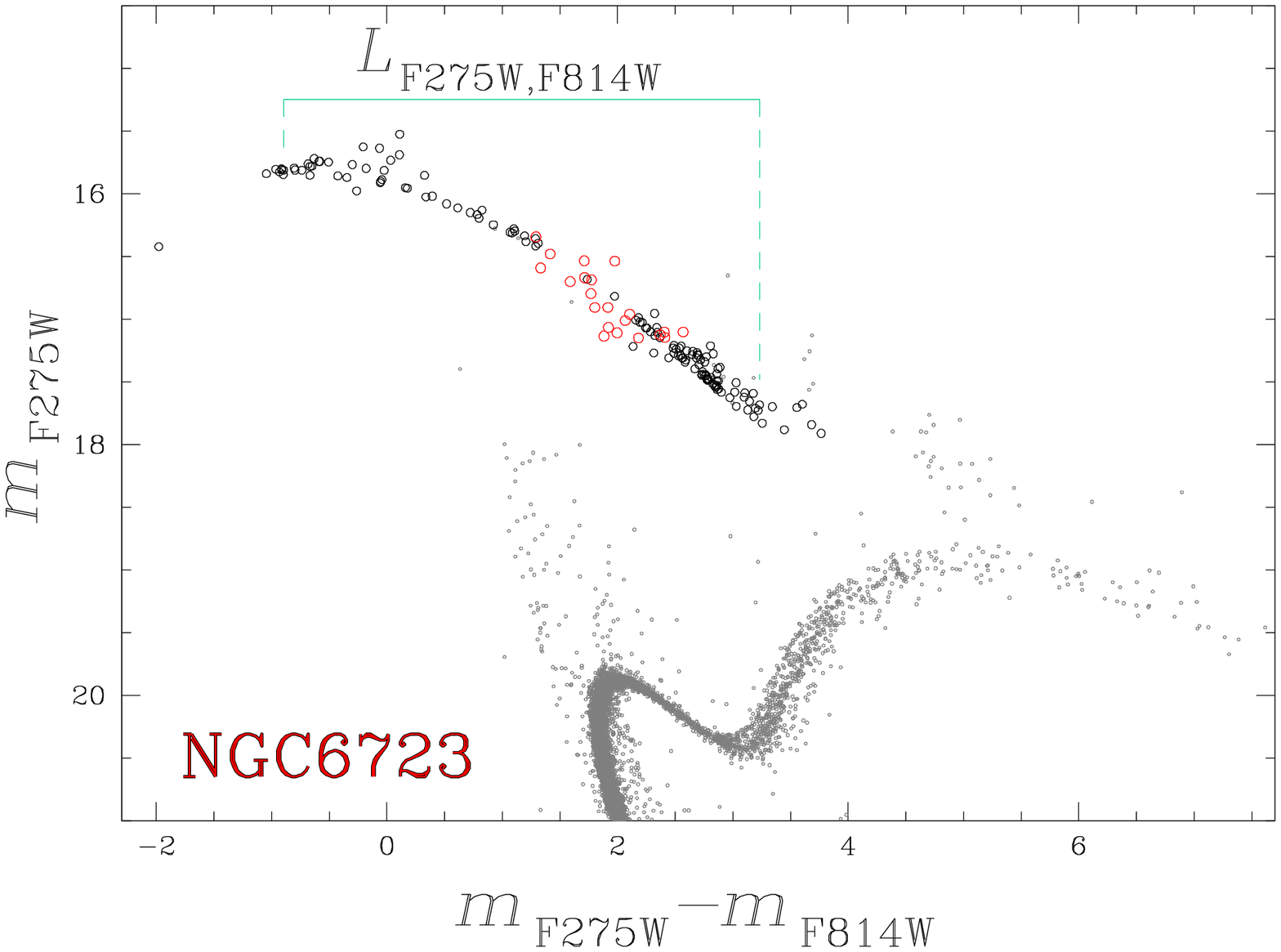} 
 %/home/milone/WORKS/treasury13/NGC6723/MATCH/hb.macro go
 \caption{$m_{\rm F275W}$ vs.\,$m_{\rm F275W}-m_{\rm F814W}$ CMD of NGC\,6723. The selected HB stars are represented with circles and the candidate RR\,Lyrae stars are colored red. The $m_{\rm F275W}-m_{\rm F814W}$ color extension of HB stars, $L_{\rm F275W-F814W}$, is indicated by the aqua segment.} 
 \label{fig:hb6723} 
\end{figure} 
\end{centering} 
%%%%%%%%%%%%%%%%%%%%%%%%%%%%%%%%%%%%%%%%%%%%%%%%%%%%%%%%%%%%%%%%%%%%%%%%%%%

The F275W and F814W bands provide a wide color which maximizes the sensitivity to the effective temperature of horizontal-branch (HB) stars. 
To investigate the relation between the helium abundance of stellar populations and the HB morphology, we estimated the $m_{\rm F275W}-m_{\rm F814W}$ color extension of the HB, $L_{\rm F275W-F814W}$. To do this we followed the recipe by Milone et al.\,(2014) that is illustrated in the upper panel of Fig.~\ref{fig:hb6723} for NGC\,6723 and selected by eye the sample of HB stars represented with large circles in Fig.~\ref{fig:hb6723}.
The red circles plotted in Fig.~\ref{fig:hb6723} are candidate RR-Lyrae stars and have been selected on the basis of their r.\,m.\,s.\, of the independent F275W, F336W, and F438W  magnitude measurements which are significantly larger than those of HB stars with similar luminosity. 
 The HB color extension, $L_{\rm F275W-F814W}$, is calculated as the difference between the 96$^{\rm th}$ and the 4$^{\rm th}$ percentile of the $m_{\rm F275W}-m_{\rm F814W}$ distribution of the selected HB stars. The corresponding uncertainty is estimated by means of bootstrapping statistics, with replacements over the sample of HB stars that we repeated 1,000 times. We assumed the 68.27$^{\rm th}$ percentile of the bootstrapped measurements as the uncertainty associated to $L_{\rm F275W-F814W}$.

\begin{centering} 
\begin{figure} 
 \includegraphics[width=8.0cm]{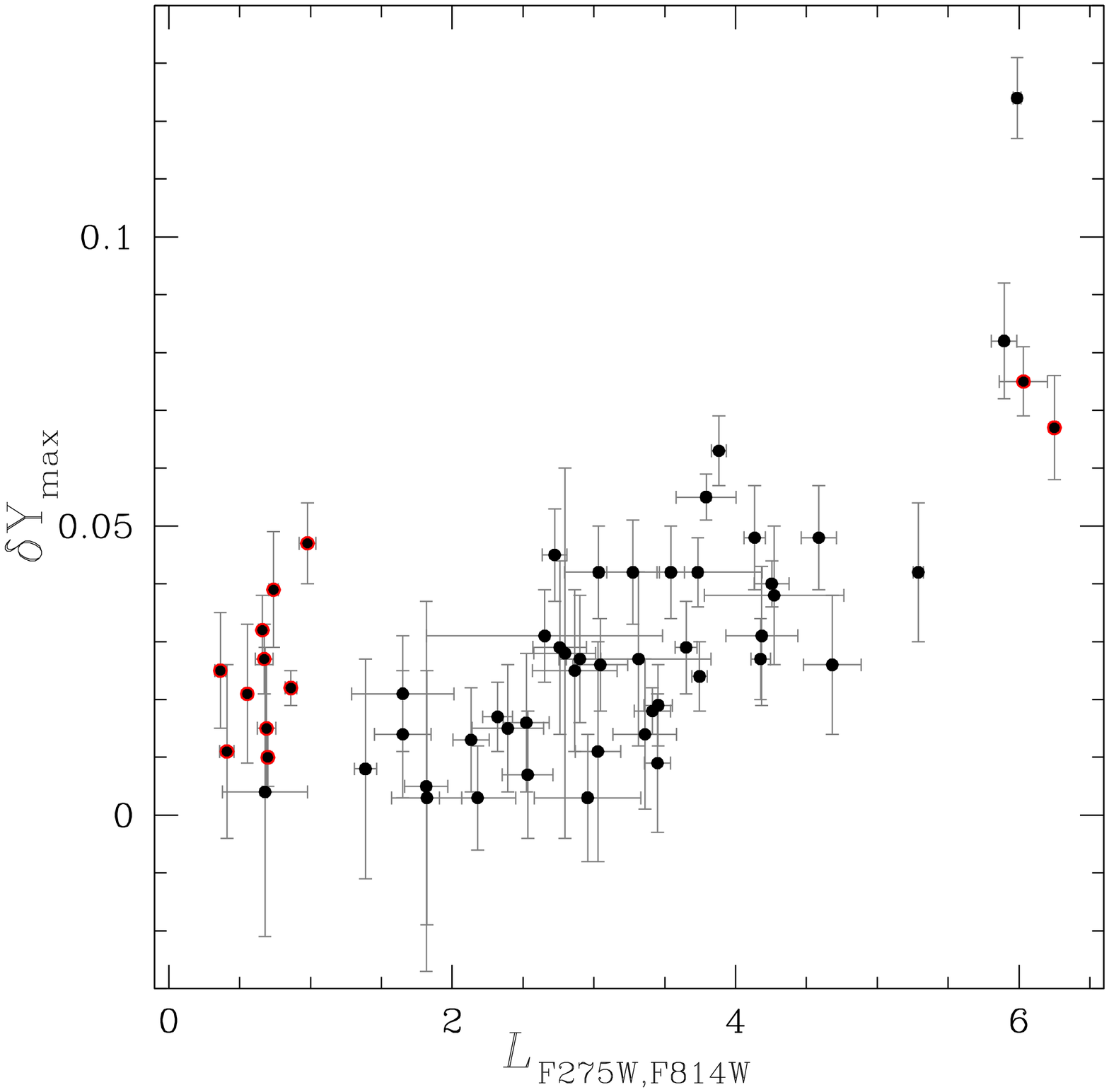} 
 \caption{Maximum internal helium variation, $\delta Y_{\rm max}$, against the F275W$-$F814W color extension of the HB. Metal rich clusters with [Fe/H]$>-0.99$ are marked with red circles.} 
 \label{fig:Yvshb} 
\end{figure} 
\end{centering}

 Fig.~\ref{fig:Yvshb} shows $L_{\rm F275W-F814W}$ as a function $\delta Y_{\rm max}$.
 When we consider all the clusters the Spearman's rank correlation coefficient is $r=0.60 \pm 0.09$.
 % We investigated the relation between the HB morphology and $\delta Y_{\rm max}$.
 Moreover, we exclude all the metal-rich clusters ([Fe/H]$> -0.99$), which display the red HB only (because the temperature distribution along the HB is dominated by the metallicity),  we find a significant correlation ($r=0.77\pm0.06$) between the maximum helium variation and the F275W$-$F814W color extension of the HB, where the clusters with extended HB having, on average, more-extreme helium variation, as expected.

%%%%%%%%%%%%%%%%%%%%%%%%%%%%%%%%%%%%%%%%%%%%%%%%%%%%%%%%%%%%%%%%%%%%%%%%%%%
%%%%%%%%%%%%%%%%%%%%%%%%%%%%%%%%%%%%%%%%%%%%%%%%%%%%%%%%%%%%%%%%%%%%%%%%%%%

\section{Comparison with the literature}
            As discussed in the Introduction, in addition to the method based on photometry of multiple sequences that we used in this paper, the internal helium variations in GCs is inferred from four additional techniques: i) multi-band photometry of the 1G and 2G RGB bumps (e.g.\,Papers III, XII; Bragaglia et al.\,2010), ii) spectroscopy of photospheric lines in HB stars (e.g.\,Villanova et al.\,2009; Marino et al.\,2014), spectroscopy of chromospheric lines of RGB stars (Pasquini et al.\,2011; Dupree et al.\,2011), iv) comparison between observations and models of HB (e.g.\,D'Antona et al.\,2002; Lee et al.\,2005; Salaris et al.\,2016). In the following we compare some estimates of the helium variations in GCs from literature with our results. 

\begin{centering} 
\begin{figure} 
\includegraphics[width=8.0cm]{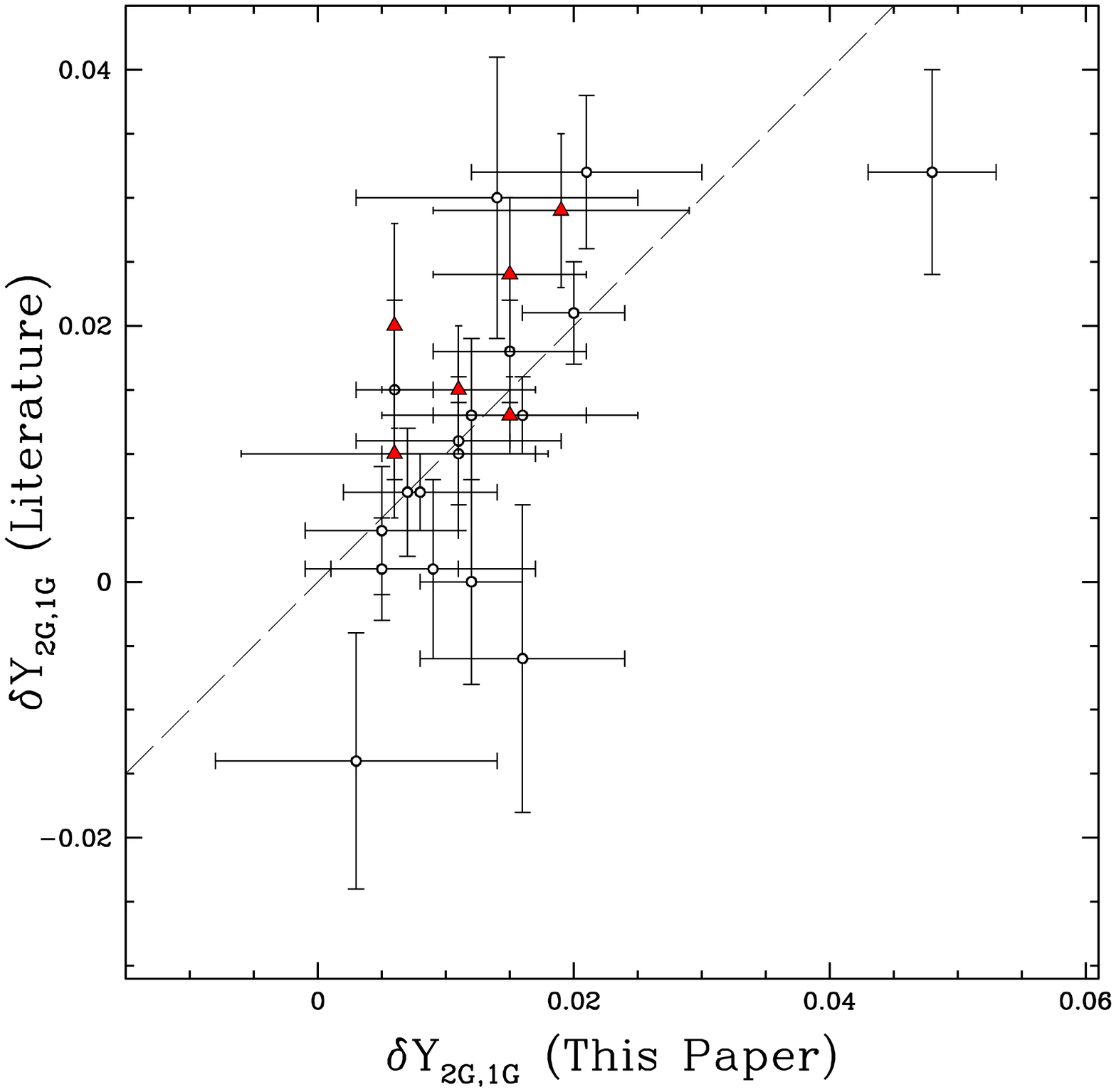} 
 %/home/milone/WORKS/treasury13/ELIO/test.macro 
 \caption{Comparison between the relative helium abundance of 2G and 1G stars derived in this paper and in the literature. Black dots represent results from Paper\,XII for 18 GCs. Red triangles refer to NGC\,104, NGC\,288, NGC\,6121, NGC\,6352, NGC\,6397, and NGC\,6752 (Milone et al.\,(2012a,b; Piotto et al.\,2013; Nardiello et al.\,2015a,b).} 
 \label{fig:confronto} 
\end{figure} 
\end{centering}

The finding that 2G stars are enhanced by $\delta_{\rm Y, 2G,1G} \sim$0.01 with respect to the 1G is consistent with the conclusion of Paper\,XII, where we estimated the helium content of 1G and 2G stars in 18 GCs from their RGB bumps, and found that 2G stars are enhanced by $\sim$0.011$\pm$0.002 in helium mass fraction with respect to 1G stars. Figure~\ref{fig:confronto} compares the results from this paper and from Paper\,XII (black points) and from other literature works (red triangles). 
%%%%%%%%%%%%%%%%%%%%%%%%%%%%%%%%%%%%%%%%%%%%%%%%%%%%%%%%%%%%%%%%%%%%%%%%%%%
%%%%%%%%%%%%%%%%%%%%%%%%%%%%%%%%%%%%%%%%%%%%%%%%%%%%%%%%%%%%%%%%%%%%%%%%%%%

  The helium lines at $\lambda \sim 5876 \AA$ and $\lambda 4471 \AA$ have been used to derive direct spectroscopic measurements of helium abundances in GC stars (e.g.\,Villanova et al.\,2009). Unfortunately, these lines are visible only in HB stars hotter than $\sim 8,500$K , with the 5876$\AA$ line significantly affected by departures from the local thermodynamic equilibrium (LTE) approximation (e.g.\,Marino et al.\,2014).
  Moreover, only stars cooler than $\sim 11,500 K$, which do not suffer from levitation of metals and sedimentation of He, provide atmospheric chemical
abundances that are representative of the pristine stellar chemical content. For these reasons these He lines can be analyzed in a small sample of HB stars of a few clusters only.

From LTE analysis of six stars, Villanova et al.\,(2012) concluded that blue HB stars of NGC\,6121 are enhanced by $\Delta Y \sim 0.04$ with respect to the primordial helium abundance. This value is significantly higher than the maximum helium variation derived in this paper ($\delta Y_{\rm max} \sim 0.013$).
However, appropriate non-LTE corrections applied to these stars are likely to decrease their He abundances possibly providing a better agreement with our results (see Marino et al.\,2014).
        Mucciarelli et al.\,(2014) show Y-[O/Fe] anti-correlation among HB stars of NGC\,6397 and NGC\,7099 and concluded that there is a small spread of the Y distributions, which would be qualitatively similar to that inferred in this paper. However, these authors detected for NGC\,6397 and NGC\,7099 very large oxygen variations of about 1.4 and 0.8 dex, respectively, in contrast with what is observed among RGB stars (e.g.\,Carretta et al.\,2009; Lind et al.\,2011). This fact confirms that accurate atmospheric parameters and NLTE corrections are needed to infer reliable spectroscopic helium and oxygen abundances in HB stars.
        Appropriate NLTE analysis of helium lines is provided by Marino et al.\,(2014) for blue-HB stars in NGC\,2808. These authors find that the analyzed HB segment in the effective temperature range $\sim 9,000-11,500$ K are enhanced by $\sim 0.09 \pm 0.01 \pm 0.05$ (internal plus systematic error) in helium mass fraction. This result provides direct spectroscopic evidence that NGC\,2808 hosts stellar populations with extreme helium abundances and agrees with what was inferred from photometry of multiple sequences in this work and in previous papers (D'Antona et al.\,2005; Piotto et al.\,2007; Milone et al.\,2012; Paper\,III). 
        
          Chromospheric spectral lines of RGB stars have been used to infer the helium content of stars in a few stars of three GCs.
Dupree \& Avrett\,(2013) estimated the helium abundance of two red giants of $\omega$\,Cen from the near-infrared transition of He\,I at 1.08$\mu$m.   The spectra suggest a helium abundance of Y$\leq$0.22 and Y$\ge$0.39--0.44 corresponding to a difference in the abundance $\delta Y \ge$0.17  (see also Dupree et al.\,2011). From a similar study on NGC\,2808, Pasquini et al.\,(2011) concluded that the helium abundance difference between the two analyzed RGB stars is larger than 0.17 in mass fraction. These values are significantly higher than those inferred from photometry of multiple sequences.
Strader et al.\,(2015) find no evidence for significant helium variation among M\,4 AGB and HB stars, concluding that a larger sample of data is needed to detect a subtle spectroscopic variations in helium, as inferred from multi-band photometry of multiple sequences of M\,4. %These authors concluded that a larger sample of data is needed to detect the subtle spectroscopic variations in He I inferred from multi-band photometry of multiple sequences of M\,4.
If these pioneer studies based on chromospheric abundances qualitatively confirm the presence of He variations in GCs, they are hardly to provide a quantitative estimate of these variations until proper models of the chromosphere will be available. 

\begin{centering} 
\begin{figure} 
\includegraphics[width=8.5cm]{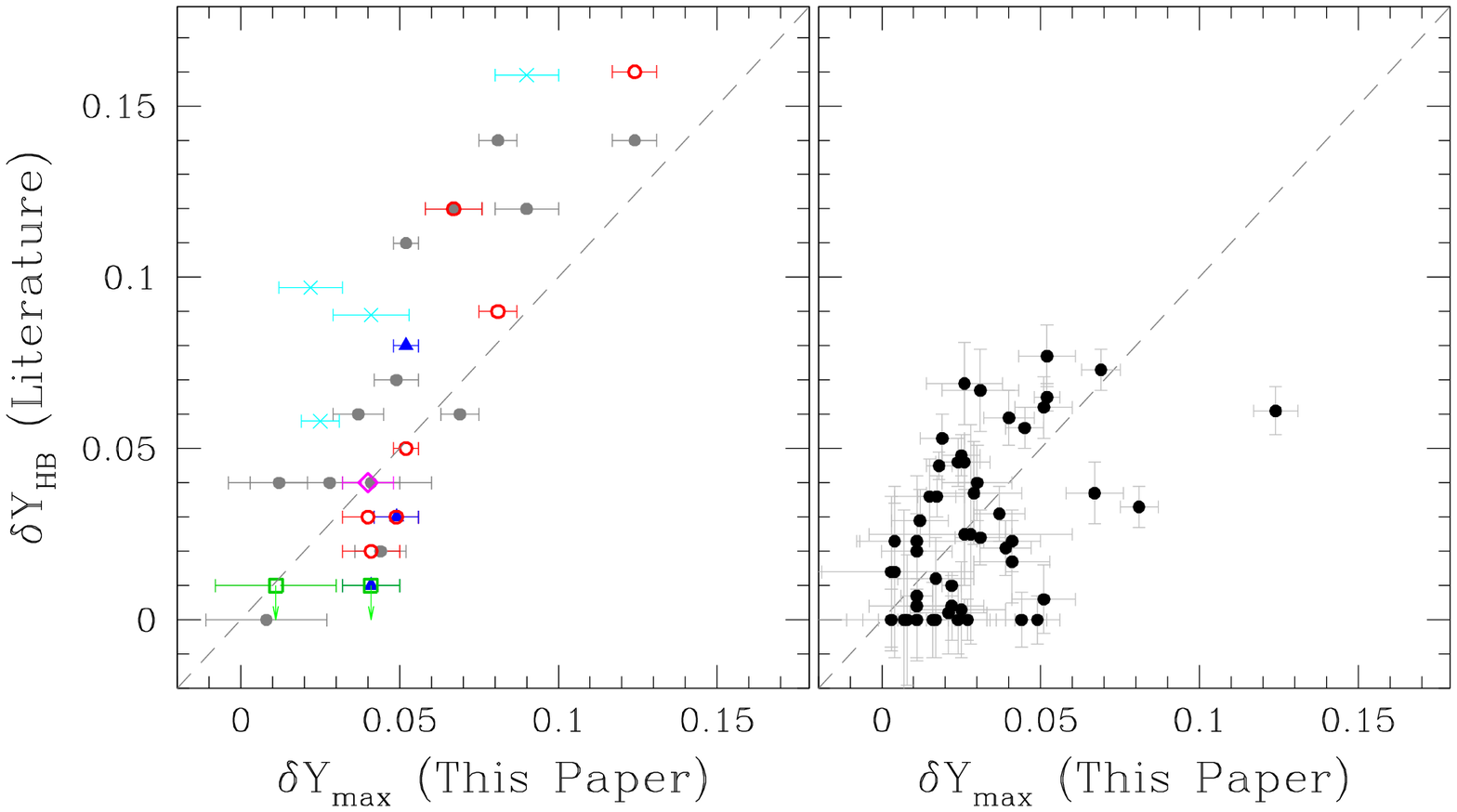} 
 %/home/milone/WORKS/treasury13/ELIO/test.macro 
\caption{ Comparison between the maximum helium abundance derived in this paper and the helium spread used in literature papers to reproduce the HB ($\delta Y_{\rm HB}$). In the left panel we show the values of $\delta Y_{\rm HB}$ adopted by D'Antona \& Caloi (2008), Caloi \& D'Antona (2011) and Tailo et al.\,(2017), which are represented as grey dots, while those from Busso et al.\,(2007), Dalessandro et al.\,(2011), Cassisi et al.\,(2014), and Salaris et al.\,(2016) are indicated by red circles. The helium spread used by Catelan et al.\,(2009) and Valcarce et al.\,(2016) are represented by green squares, while blue triangles and magenta diamonds indicate results by Denissenkov et al.\,(2017) and Campbell et al.\,(2013), respectively. Cyan crosses refer to the helium-difference estimates provided by Jang et al.\,(2014), Joo \& Lee (2013) and Lee et al.\,(2009).
     In the right panel we show the comparison with Gratton et al.\,(2010). } 
 \label{fig:confronto2} 
\end{figure} 
\end{centering}

The most-commonly used method to estimate the intrinsic helium variation in GCs is based on the comparison between observations and theoretical distribution of HB stars (e.g.\,D'Antona et al.\,2002). The comparison between the values of $\delta Y_{\rm max}$ derived in our work and the helium variations used in various literature papers to reproduce the HB is provided in Fig.~\ref{fig:confronto2}.
 Despite the overall correlation between the helium variations derived from RGB and HB, there are significant differences in the amount of helium variations derived  from these two methods.
 The $\delta Y_{\rm max}$ come from various authors who used different photometry, HB models and assumptions on the cluster properties, including age and mass loss.  Although the comparison between the results inferred in the various papers is beyond the purposes of the present work, we emphasize that the incomplete knowledge of the second parameters governing the HB morphology of star clusters is possibly the major challenge of the method to derive helium from the HB morphology. In particular, the results inferred from the HB strongly depend on the adopted cluster age and on the mass-loss law, which is a poorly-constrained quantity for GC stars (see, e.g.\,D'Antona et al.\,2002 and Salaris et al.\,2010 for discussion on the degeneracy among age, mass-loss, and helium).

%%%%%%%%%%%%%%%%%%%%%%%%%%%%%%%%%%%%%%%%%%%%%%%%%%%%%%%%%%%%%%%%%%%%%%%%%%%
%%%%%%%%%%%%%%%%%%%%%%%%%%%%%%%%%%%%%%%%%%%%%%%%%%%%%%%%%%%%%%%%%%%%%%%%%%%

\section{A possible pure-helium spread among 1G stars}
\label{sec:He1G}
In Sect.~\ref{sec:read} we show that the position of 1G stars in the chromosome map is consistent with a sequence of stars with the same C, N, O abundance but different helium content, as earlier discussed in Papers\,III and IX. In the following we provide some speculative scenarios to make a pure helium spread among 1G stars. 

\begin{centering} 
\begin{figure} 
 \includegraphics[width=9.cm]{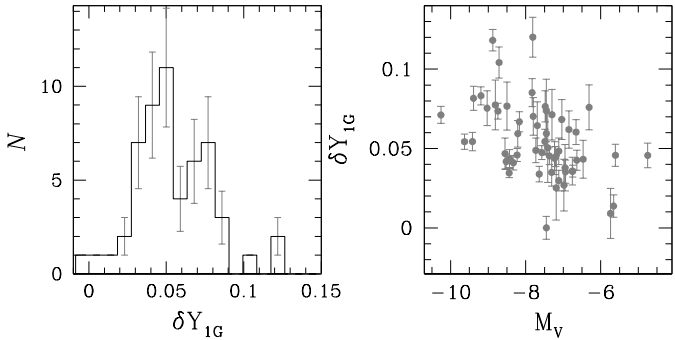} 
 \caption{\textit{Left Panel.} Histogram distribution of $\delta  Y_{\rm 1G}$ for the clusters studied in this paper. \textit{Left Panel.} $\delta  Y_{\rm 1G}$ as a function of the absolute magnitude of the host GC.} 
 \label{fig:elio1G} 
\end{figure} 
\end{centering}

In Paper\,IX we already estimated the $m_{\rm F275W}-m_{\rm F814W}$ color width of 1G RGB stars of the analyzed GCs ($W_{\rm mF275W-m814W}^{\rm 1G}$) and found that in the majority of GCs it is not consistent with a simple population. By assuming that the derived color spreads are entirely due to helium variation (Paper\,III), we used the relation between the $m_{\rm F275W}-m_{\rm F814W}$ color and the helium abundance from Dotter et al.\,(2008) to derive the internal helium variation among 1G stars, $\delta Y_{\rm 1G}$, from the values $W_{\rm mF275W-m814W}^{\rm 1G}$ of Paper\,IX.
Results are illustrated in Figure~\ref{fig:elio1G}, where we plot the histogram distribution of $\delta Y_{\rm 1G}$.
We find that, if the extension of 1G stars is entirely due to helium spread, the internal helium variation among 1G stars dramatically changes from one cluster to another and ranges from $\delta Y^{\rm 1G} \sim 0.00$ to $\sim$0.12. 
The average helium spread is $\delta  Y^{\rm 1G} \sim 0.05$ and  $\delta_{Y}^{\rm 1G}$ is characterized by a bimodal distribution with two main peaks around $\delta Y^{\rm 1G} \sim 0.04$ and $\sim 0.08$ with a tail of few clusters, including NGC\,5024, NGC\,5927 and NGC\,5272 with $\delta Y \gtrsim 0.10$.
 As shown in the right panel of Fig.~\ref{fig:elio1G}, we find a mild correlation between $\delta Y_{\rm 1G}$ (r=$-$0.51) and the cluster absolute luminosity but there are GCs with similar mass but different values of $\delta Y_{\rm 1G}$. 

As well known, the {\it pp}-chain is the dominant energy producer during the MS phase of low-mass stars, such as those in globular clusters. In canonical models, these stars are almost completely in radiative equilibrium, so one should appeal to an {\it ad hoc}
mixing process in order to mix in the stellar envelope and bring to  the surface the required helium of pure {\it pp}-chain origin, so to produce a  $\delta Y$ of the order of $\sim$0.03-0.05 in the whole envelope. Fig.~9 in one of the classical papers of Icko Iben (Iben\,1967) shows that in a $1\; \mathcal{M}_\odot$ star one should mix the envelope down to a mass fraction $\sim 0.25$ in order to increase helium in the whole envelope by the just mentioned amount, something that perhaps rotation-induced meridional circulation could accomplish. Even so, the problem, unfortunately, would not be solved. From the same figure one can indeed see that below a mass fraction $\sim 0.4$ carbon is severely depleted leading to a factor up to $\sim  5$ increase in nitrogen. If the envelope would have been kept mixed down to a mass fraction $\sim 0.3$ then the whole carbon in the star would have been processed at sufficiently high temperatures to be fully converted into nitrogen.  Even if energetically sub-dominant, the CNO cycle would have dominated the surface chemistry, hence failing to align stars along the 1G sequence.

A closer look to Fig.~9 of Iben\,1967 reveals that down to a mass fraction $\sim 0.4$ the CNO cycle did not operate much, while at precisely this mass fraction at the end of the main sequence helium was increased by only $\delta Y\sim 0.04$, too little for producing a global helium increase of the same size having to mix the whole envelope. We can also mention that mixing down to a mass fraction 0.6 would bring to the surface a large among of $^3$He,  {\it relative}  to its pristine abundance, but still more than one order of magnitude too little with respect to what would be required to account for the $\delta_{\rm F275W,F814W}$ spread of 1G stars in very many GCs.

So, this kind of mixing, either would produce too little helium, or it would still be accompanied by substantial nitrogen enhancement, leading to something resembling the 2G stars rather than the 1G ones. One may argue that those of Iben are models half a century old and that modern calculations may open up this opportunity. Yet, we believe that those models belong to an excellent vintage and indeed it does not appear that the critical cross section of the reaction $^{12}$C($p,\gamma)^{13}$N has changed much since that used by Iben (Parker et al.\,1964) to the latest determinations (Burtebaev et al.\,2008; Li et al.\,2010). We conclude that an hypothetical deep extra-mixing during the MS stage does not offer a viable solution for the putative helium spread among 1G stars.

Perhaps a less-conjectural alternative would be offered by a variable first dredge up, i.e., when the envelope convection penetrates deeply inside the star and pp-chain products are brought to the surface. Canonical models of near-solar mass stars predict an increase of the helium abundance in the envelope of $\delta Y\simeq 0.02$ accompanied by a modest decrease of carbon to the advantage of nitrogen (e.g., Renzini \& Voli 1981). If the penetration of convection or any additional form of mixing were deeper in some stars than others, then a spread of helium abundances would be generated, again with relatively modest increase in nitrogen. However, in order to nearly double $\delta Y$ the extra-mixing should penetrate $\sim 0.05\, M_\odot$ more than in models without such extra-mixing (cf. Fig.~11 in Iben 1967) and this would have an undesired side effect. Indeed, the luminosity of the RGB bump is directly controlled by the mass coordinate marked by the deepest penetration of envelope mixing and a dispersion of $\sim 0.05\, M_\odot$ in such mass coordinate would produce a broadening of the RGB bump by almost 3/4 of a magnitude (cf. Table~3 in Sweigart \& Gross 1978). Such a broadening of the RGB Bump is not observed in clusters with a broad 1G locus in the chromosome map, which instead appears to be narrower than $\sim 0.2$ mag (Paper XII). Actually, this not the whole story, because the RGB Bump luminosity has also a direct dependence on helium, with the Bump getting brighter with increasing helium as $\delta m_{\rm F814W}^{\rm Bump}/\delta Y\simeq -2$ (cf. Fig. 9 in Paper XII). Even a  $\delta Y=0.05$ will produce a brightening of the Bump by only $\sim 0.1$ magnitudes, too little to compensate the large, opposite effect due to a variable depth of extra-mixing. Actually, further investigation of the Bump luminosities in 1G stars of various GCs (Lagioia et al.\, in preparation) reveals that there is indeed a correlation between the $\delta$ $m_{\rm F275W}-m_{\rm F814W}$ color extension of 1G stars and the RGB Bump luminosity, with bluer stars having a brighter Bump, qualitatively consistent with having higher helium, but opposite to what expected  if the higher helium was the result of deeper first dredge-up mixing. We conclude that neither this hypothetical form of extra-mixing offers a plausible source of  $pp$-chain-only helium for 1G stars.

We have also considered accretion from the ISM or from binary companions but in all cases we found that significant helium enrichment is always accompanied by extensive CNO processing.

Thus, although an interpretation of the 1G as $pp$-chain sequence in the chromosome map is quite tantalizing, we are left without a concrete astrophysical environment that could produce a pure pp-chain helium enrichment.
Yet another option to consider is offered by Population III stars, i.e., stars with Big Bang pristine, zero metals, composition. Massive Pop.\,III stars begin burning hydrogen via the $pp$-chain but do so at such high temperature that some triple-$alpha$ reactions also take place, so producing a tiny amount ($\sim 10^{-8}$ by mass) of $^{12}$C, yet sufficient to have then stars running on CNO cycle (e.g., Limongi \& Chieffi 2012, and references therein). In this way, during the main sequence phase of massive Pop. III stars helium is produced thanks only to  an insignificant amount of CNO elements. Objects of this kind would then be a potential source of helium without a concomitant enrichment in CNO elements, and this is why we mention them here. However, it is a long way to go from them to a possible star-by-star variation of helium in 1G stars of globular clusters. Limongi \& Chieffi (2012) models do not lose mass, because the lack of metals deprives them of the radiation force driving winds in hot stars. So, to extract helium from them one should invoke some sort of rotational mixing accompanied by an equatorial extrusion disk. But then the problem remains of how to have a helium spread in the 1G forming cloud while the cloud being homogeneous in other elements such as iron. Moreover, Limongi \& Chieffi models eventually explode as supernovae, and besides helium produce vast amounts of carbon, oxygen, neon, magnesium and silicon. Definitely, neither Pop. III stars appear to offer a plausible source of helium for the spread of 1G stars.

In conclusion, we consider still unproven that the 1G spread is due to helium and believe that more spectroscopic observations may be  needed to solve  this additional puzzle of GC multiple populations. For this reason, the possible helium variation within 1G stars  will be further studied in a separate paper.

%%%%%%%%%%%%%%%%%%%%%%%%%%%%%%%%%%%%%%%%%%%%%%%%%%%%%%%%%%%%%%%%%%%%%%%%%%%
%%%%%%%%%%%%%%%%%%%%%%%%%%%%%%%%%%%%%%%%%%%%%%%%%%%%%%%%%%%%%%%%%%%%%%%%%%%

\section{Summary and Conclusions}
\label{sec:discussion} 
We exploited multi-wavelength UVIS/WFC3 and WFC/ACS photometry of 57 Galactic GCs  from the {\it HST} UV legacy survey of Galactic GCs and from the {\it HST} archive to determine the relative helium abundance of 2G and 1G stars and the maximum internal variation.
 This work is based on all the RGB stars of type-I GCs and the blue-RGB stars of type-II GCs. We excluded from the analysis the red RGB stars of type-II GCs, which are enhanced in metallicity with respect to the remaining GC stars (see Marino et al.\,2015 and references in their Table~10).

 For each cluster we compare synthetic spectra and observed colors of multiple populations to derive the helium difference between the two groups of 2G and 1G stars identified in paper IX and to estimate the maximum internal helium variation.
 
 We find that 2G stars are consistent with having higher helium abundance than 1G stars in all the analyzed clusters, with an average helium difference of $\sim$0.01  in mass fraction. This finding is in agreement with the conclusion from Paper\,XII, where we used the RGB-bump observations to infer the relative helium abundance of 1G and 2G stars in 18 GCs, finding an average helium enhancement of 0.011$\pm$0.002 of 2G stars with respect to 1G stars.
 
The maximum variation in helium mass fraction ranges from less than 0.01 in low-mass clusters to more than 0.1 in the most-massive clusters and exhibits a significant correlation with the cluster mass and an anticorrelation with the cluster luminosity.  These results confirm the conclusion of Paper\,IX that the complexity of the multiple stellar population phenomenon increases with the cluster mass.

In the analyzed sample of metal-intermediate and metal-poor GCs ([Fe/H]$\lesssim -1.0$), the maximum helium variation correlates with the F275W$-$F814W extension of the HB, thus confirming previous results based on a smaller number of clusters (Milone et al.\,2014).
This fact is consistent with spectroscopic studies that stars with different light-element (hence helium) abundance populate distinct segment of the HB (e.g.\,Marino et al.\,2011, 2013, 2014; Gratton et al.\,2011, 2012, 2013). 
 These findings strongly support the idea, suggested previously by D'Antona et al.\,(2002), that star-to-star helium variations are one of the main second parameters that determine the HB morphology of GCs. 

\section*{acknowledgments} 
\small
 Based on observations with the NASA/ESA Hubble Space Telescope, obtained at the Space Telescope Science Institute, which is operated by AURA, Inc., under NASA contract NAS 5-26555. J.\,A.\, and A.\,B.\, acknowledge support from STScI grants GO-13297.
 This work has been supported by the European Research Council through the ERC-StG 2016, project 716082 `GALFOR' and by the MIUR through the the FARE project R164RM93XW `SEMPLICE'.
AFM acknowledges support by the Australian Research Council through Discovery Early Career Researcher Award DE160100851.

\bibliographystyle{aa} 
 
\onecolumn
\begin{longtable}{cccccc}
\caption{Decription of the archive data used in this paper.}\label{tab:data} \\ \hline
ID        & date      &  camera        & Filter & N$\times$ exposure time            & GO    \\
\hline
\endfirsthead
\caption{continued.}\\\hline
\endhead
\endfoot
\hline
IC\,4499 & May 29-30 2017& WFC3/UVIS & F275W & 985s$+$1078s$+$6$\times$1087s  & 14723 \\
         & May 30 2017   & WFC3/UVIS & F336W & 4$\times$650s  & 14723 \\
         & May 30 2017   & WFC3/UVIS & F438W & 4$\times$97s   & 14723 \\
         & Jul 01 2010   & ACS/WFC   & F606W & 60s$+$4$\times$603s  & 11586 \\
         & Jul 24 2016   & ACS/WFC   & F606W & 65s$+$3$\times$907s$+$911s$+$2$\times$912s  & 14235 \\
         & Jul 01 2010   & ACS/WFC   & F814W & 65s$+$3$\times$636s$+$637s  & 11586 \\
\hline
NGC\,104 & Nov 14 2012 --- Sep 9 2013   & WFC3/UVIS & F225W & 10$\times$380s$+$10$\times$700s  & 12971 \\
         & Sep 28 2010   & WFC3/UVIS & F390M & 50s$+$2$\times$700s  & 11729 \\
         & Sep 28 2010   & WFC3/UVIS & F390W & 2$\times$10s$+$2$\times$348s$+$2$\times$940s & 11644 \\
         & Sep 28 2010   & WFC3/UVIS & F390W & 10s                  & 11729 \\
         & Aug 13 2013   & WFC3/UVIS & F390W & 567s$+$573s$+$2$\times$577s$+$2$\times$578s$+$584s$+$590s & 11729 \\
         & Sep 28 2010   & WFC3/UVIS & F390M & 50s$+$2$\times$700s  & 11729 \\
         & Sep 28 2010   & WFC3/UVIS & F395N & 90s$+$2$\times$1120s & 11729 \\
         & Sep 29 2010   & WFC3/UVIS & F410M & 40s$+$2$\times$800s  & 11729 \\
         & Sep 29 2010   & WFC3/UVIS & F467M & 40s$+$2$\times$450s  & 11729 \\
         & Apr 05 2002   & ACS/WFC   & F475W & 20$\times$60s        & 9028  \\
         & Jul 07 2002   & ACS/WFC   & F475W & 5$\times$60s$+$150s  & 9443  \\
         & Jan 08 2003   & ACS/WFC   & F475W & 60s                  & 9503  \\
         & Jul 29 2012   & ACS/WFC   & F475W & 2$\times$7s$+$4$\times$475s & 12116  \\
         & Jan 08 2003   & ACS/WFC   & F502N & 340s                 & 9503  \\
         & Sep 28 2010   & WFC3/UVIS & F547M & 5s$+$40s$+$400s      & 11729 \\
         & Jan 08 2003   & ACS/WFC   & F550M & 3$\times$90s         & 9503  \\
         & Sep 28 2010   & WFC3/UVIS & F555W & 30s$+$2$\times$665s  & 11664 \\  %+0s
         & Sep 30 2002 --- Oct 11 2002  & ACS/WFC   & F625W & 2$\times$10s$+$20$\times$65s & 9281  \\
         & Sep 30 2002 --- Oct 11 2002  & ACS/WFC   & F658N & 6$\times$350s$+$6$\times$370s$+$8$\times$390s & 9281  \\
\hline
NGC\,288 & Nov 10 2010   & WFC3/UVIS & F395N & 1260s$+$1300s  & 12193 \\
         & Sep 20 2004   & ACS/WFC   & F625W &  10s$+$75s$+$115s$+$120s  & 10120 \\
         & Nov 10 2010   & WFC3/UVIS & F467M &  964s$+$1055s  & 12193 \\
         & Nov 10 2010   & WFC3/UVIS & F547M &  2$\times$360s & 12193 \\
         & Sep 20 2004   & ACS/WFC   & F658N &  2$\times$340s$+$2$\times$540s  & 10120 \\
\hline
NGC\,362 & Apr 13 2012   & WFC3/UVIS & F390W & 14$\times$348s  & 12516 \\
         & Apr 13 2012   & WFC3/UVIS & F390W & 144s$+$145s$+$6$\times$150s$+$160s$+$200s  & 12516 \\  %ACS F625W f658N
\hline
NGC\,5904 & Jul 5 2010   & WFC3/UVIS & F390W & 6$\times$500s  & 11615 \\
          & Jun 6-9 2012 & WFC3/UVIS & F390W & 4$\times$735s  & 12517 \\
          & Jul 5 2010   & WFC3/UVIS & F656N & 4$\times$800s$+$950s$+$1100s  & 11615 \\
          & Aug 1 2004   & ACS/WFC   & F625W & 10s$+$70s$+$2$\times$110s  & 10120 \\
          & Aug 1 2004   & ACS/WFC   & F658N & 2$\times$340s$+$2$\times$540s  & 10120 \\
\hline
NGC\,5927 & Sep 1  2010   & WFC3/UVIS & F390M & 50s$+$2$\times$700s  & 11729 \\ % 9 nel file?
          & Sep 1  2010   & WFC3/UVIS & F390W & 10s  & 11729 \\
          & Aug 28 2010   & WFC3/UVIS & F390W & 2$\times$40s$+$2$\times$348s$+$2$\times$800s  & 11664 \\
          & Sep 1  2010   & WFC3/UVIS & F395N & 90s$+$1015s  & 11729 \\
          & Sep 1  2010   & WFC3/UVIS & F410M & 40s$+$2$\times$800s  & 11729 \\
          & Sep 1  2010   & WFC3/UVIS & F467M & 365s  & 11729 \\
          & Jul 31 2012   &   ACS/WFC & F475W & 2$\times$7s$+$2$\times$425s  & 12116 \\
          & Sep 1  2010   & WFC3/UVIS & F547M & 5s$+$40s$+$400s  & 11729 \\
          & Aug 28 2010   & WFC3/UVIS & F555W & 1s$+$50s$+$2$\times$665s  & 11664 \\
\hline
NGC\,6093 & Feb 2  2006   &   ACS/WFC & F555W & 180s  & 10573 \\ 
\hline
NGC\,6121 & Jul 7 2011   & WFC3/UVIS & F395N &  2$\times$646s & 12193 \\
          & Jul 7 2011   & WFC3/UVIS & F467M &  2$\times$350s & 12193 \\
          & Jul 7 2011   & WFC3/UVIS & F547M &  2$\times$75s  & 12193 \\
          & Jul 26 2004   &   ACS/WFC & F606W &  1s$+$10s$+$2$\times$1170s$+$2$\times$1175s$+$2$\times$1210s$+$12$\times$1218s$+$2$\times$1259s & 10146 \\ 
          & Jul 26 2004   &   ACS/WFC & F625W & 15s$+$30s  & 10120 \\ 
          & Jul 26 2004   &   ACS/WFC & F658N & 50s$+$340s & 10120 \\ 
\hline
NGC\,6205 & Aug 15 2005   & ACS/WFC & F625W &  2$\times$10s$+$4$\times$90s & 10349 \\
          & Aug 15 2005   & ACS/WFC & F658N &  2$\times$60s$+$2$\times$345s$+$2$\times$400s & 10349 \\
\hline
NGC\,6218 & Feb  1 2006   & ACS/WFC & F555W &  50s & 10573 \\
          & Jun 14 2004   & ACS/WFC & F625W &  2$\times$40s$+$2$\times$60s & 10005 \\
          & Jun 14 2004   & ACS/WFC & F658N &  4$\times$340s & 10005 \\
\hline

NGC\,6341 & Oct 11 2009   & WFC3/UVIS & F390M & 50s$+$2$\times$700s  & 11729 \\ 
          & Oct 11 2009   & WFC3/UVIS & F390W & 10s  & 11729 \\
          & Oct 10 2009   & WFC3/UVIS & F390W & 2$\times$2s$+$2$\times$348s$+$2$\times$795s  & 11664 \\
          & Oct 10-11 2009   & WFC3/UVIS & F395N & 90s$+$2$\times$965s  & 11729 \\
          & Oct 11 2009   & WFC3/UVIS & F410M & 40s$+$2$\times$765s  & 11729 \\
          & Oct 11 2009   & WFC3/UVIS & F467M & 40s$+$2$\times$350s  & 11729 \\
          & Aug 21 2012   & ACS/WFC & F475W &  4$\times$400s & 12116 \\
          & Oct 11 2009   & WFC3/UVIS & F547M & 5s$+$40s$+$400s  & 11729 \\
          & Oct 10 2009   & WFC3/UVIS & F555W & 1s$+$30s$+$2$\times$665s  & 11664 \\
          & Aug 07 2004   & ACS/WFC & F625W &  10s$+$3$\times$120s & 10120 \\
          & Aug 07 2004   & ACS/WFC & F658N &  2$\times$350s$+$2$\times$555s & 10120 \\
\hline
NGC\,6352 & Feb 02 2012   & ACS/WFC & F625W &  2$\times$150s & 12746 \\
          & Feb 02 2012   & ACS/WFC & F658N &  643s$+$645s   & 12746 \\
\hline
NGC\,6362 & Mar 30 2011   & ACS/WFC & F625W &  140s$+$145s & 12008 \\
          & Mar 30 2011   & ACS/WFC & F658N &  750s$+$766s   & 12008 \\
\hline
NGC\,6388 & Sep 4 30 2003 --- Jun 23 2004   & ACS/WFC & F555W &  7$\times$7s & 9821 \\
\hline
NGC\,6397 & Mar 9-11 2010   & WFC3/UVIS & F225W & 24$\times$680  & 11633 \\
          & Mar 21 2011   & WFC3/UVIS & F395N &  2$\times$200s  & 12193 \\
          & Mar 21 2011   & WFC3/UVIS & F467M &  2$\times$140s  & 12193 \\
          & Mar 21 2011   & WFC3/UVIS & F547M &  2$\times$40s   & 12193 \\
          & Jul 16 2004 --- Jun 19 2005   & ACS/WFC & F625W &  5$\times$10s$+$5$\times$340s   & 12746 \\
          & Jul 16 2004 --- Jun 28 2005   & ACS/WFC & F658N &  20$\times$390s$+$20$\times$395s   & 12746 \\
\hline
NGC\,6535 & Apr 9 2010   & ACS/WFC & F625W &  100s$+$148s & 12008 \\
          & Apr 9 2010   & ACS/WFC & F658N &  588s$+$600s & 12008 \\
\hline
NGC\,6541 & Feb 24 2012   & WFC3/UVIS & F390W & 12$\times$348s  & 12516 \\
          & Feb 24 2012   & WFC3/UVIS & F555W & 2$\times$145s$+$8$\times$150s & 12516 \\
          & Aug 3 2004 - Jun 28 2006  & ACS/WFC   & F625W &  10s$+$5$\times$120s  & 10120 \\
          & Aug 3 2004 - Jun 28 2006  & ACS/WFC   & F658N &  2$\times$350s$+$2$\times$520s  & 10120 \\
\hline
NGC\,6624 & Jun 5  2006   &   ACS/WFC & F555W & 160s  & 10573 \\ 
\hline
NGC\,6637 & Jun 6  2006   &   ACS/WFC & F555W & 120s  & 10573 \\ 
\hline
NGC\,6656 & May 18 2011   & WFC3/UVIS & F395N &  2$\times$631s$+$2$\times$697s & 12193 \\
          & May 18 2011   & WFC3/UVIS & F467M &  2$\times$361s$+$2$\times$367s & 12193 \\
          & Mar 2  2010   &   ACS/WFC & F502N &  2$\times$441s$+$2102s$+$2322s & 11558 \\
          & May 18 2011   & WFC3/UVIS & F547M &  74s$+$3$\times$75s & 12193 \\
\hline
NGC\,6681 & Nov  5 2011   & WFC3/UVIS & F390W & 12$\times$348s  & 12516 \\
          & Nov  5 2011   & WFC3/UVIS & F555W & 2$\times$127s$+$8$\times$150s & 12516 \\
\hline
NGC\,6752 & Jul 31 --- Aug 21 2010   & WFC3/UVIS & F225W & 18$\times$120  & 11904 \\
          & May 5 2010   & WFC3/UVIS & F390M & 50s$+$2$\times$700s  & 11729 \\ 
          & May 5 2010   & WFC3/UVIS & F390W & 10s  & 11729 \\
          & May 1 2010   & WFC3/UVIS & F390W & 2$\times$2s$+$2$\times$348s$+$2$\times$880s  & 11664 \\
          & May 5 2010   & WFC3/UVIS & F395N & 90s$+$2$\times$1015s  & 11729 \\
          & May 21 2011  & WFC3/UVIS & F395N & 2$\times$748s  & 12193 \\
          & May 5  2010   & WFC3/UVIS & F410M & 40s$+$2$\times$800s  & 11729 \\
          & May 5  2010   & WFC3/UVIS & F467M & 40s$+$2$\times$400s  & 11729 \\
          & Jul 18 2004   & ACS/WFC & F475W &  6$\times$340s & 9899 \\
          & May 21 2011  & WFC3/UVIS & F467M & 2$\times$350s  & 12193 \\
          & Jul 31 --- Aug 21 2010  & WFC3/UVIS & F502N & 18$\times$670s  & 12193 \\
          & May 5  2010   & WFC3/UVIS & F547M & 5s$+$40s$+$400s  & 11729 \\
          & May 21 2011  & WFC3/UVIS & F547M & 2$\times$100s  & 12193 \\
          & May 1  2010   & WFC3/UVIS & F555W & 30s$+$2$\times$665s  & 11664 \\  %+0s
          & Jul 31 --- Aug 21 2010   & WFC3/UVIS & F555W & 15$\times$550s  & 11904 \\
          & May 19 --- Aug 31 2011   & ACS/WFC & F625W &   6$\times$10s$+$12$\times$360s & 12254 \\
          & May 19 --- Aug 31 2011   & ACS/WFC & F658N &  12$\times$724s$+$12$\times$820s & 12254 \\
\hline
NGC\,7078 & Sep 1 2013   & WFC3/UVIS & F343N &  2$\times$350s & 13295 \\
          & May 19-20 2010   & WFC3/UVIS & F390W &  6$\times$827s & 11233 \\
%          & Sep 1 2013   & WFC3/UVIS & F555W &  2$\times$10s & 13295 \\
%KRON\,34   & 2017, Jan, 13    & UVIS/WFC3 & F814W  & 90s$+$680s                         & 14710 & A.\,P.\,Milone \\

\hline
\end{longtable}

% %%%%%%%%%%%%%%%%%%%%%%%%%%%%%%%%%%%%%%%%%%%%%%%%%%%%%%%%%%%%%%%%%%%%%%%%%%    
 
\onecolumn
\begin{longtable}{lcccccccl}
\caption{Abundance difference between 2G and 1G stars adopted to infer the average helium difference between 2G and 1G stars $\delta Y_{\rm 2G,1G}$. For the clusters without literature determination of Mg, Al, and Si we adopted two sets of abundance differences corresponding to different choices of $\Delta$[Mg/Fe], $\Delta$[Al/Fe] and $\Delta$[Si/Fe].}\label{tab:CNO} \\ \hline %\hline
ID        &  $\Delta$[C/Fe]     &   $\Delta$[N/Fe]  &   $\Delta$[O/Fe]  &   $\Delta$[Mg/Fe]  &   $\Delta$[Al/Fe]    &   $\Delta$[Si/Fe]  & Reference\\
\hline
\endfirsthead
\caption{continued.}\\\hline %\hline
ID        &  $\Delta$[C/Fe]     &   $\Delta$[N/Fe]  &   $\Delta$[O/Fe]  &   $\Delta$[Mg/Fe]  &   $\Delta$[Al/Fe]    &   $\Delta$[Si/Fe]  & Reference \\
\hline
\endhead
\endfoot
%          &                    &                    & 2G-1G&            &                  &                    &              &   \\
\hline
IC\,4499  & $-$0.05            &  0.65              & $-$0.20           &   0.00           &   0.00             &    0.00      &  --  \\
          & $-$0.05            &  0.60              & $-$0.00           &$-$0.10           &   0.80             &    0.05      &  --  \\
NGC\,104  & $-$0.30            &  0.70              & $-$0.30           &   0.00           &   0.30             &    0.00      &  Pancino et al.\,(2017) \\
NGC\,288  & $-$0.20            &  0.70              & $-$0.45           &   0.00           &   0.20             &    0.00      &  Carretta et al.\,(2009) \\
NGC\,362  & $-$0.25            &  0.85              & $-$0.40           &$-$0.00           &   0.40             &    0.00      &  Pancino et al.\,(2017)   \\
NGC\,1261 & $-$0.20            &  0.75              & $-$0.25           &   0.00           &   0.00             &    0.00      &  --  \\
          & $-$0.15            &  0.70              & $-$0.10           &$-$0.10           &   0.80             &    0.05      &  --  \\
NGC\,1851 & $-$0.25            &  0.85              & $-$0.35           &   0.00           &   0.40             &    0.00      &  Pancino et al.\,(2017)   \\
NGC\,2298 & $-$0.30            &  0.75              & $-$0.80           &   0.00           &   0.00             &    0.00      &  --  \\
          & $-$0.15            &  0.75              & $-$0.70           &$-$0.10           &   0.80             &    0.05      &  --  \\
NGC\,2808 & $-$0.75            &  1.00              & $-$0.60           &$-$0.25           &   1.00             &    0.10      &  Carretta et al.\,(2018)  \\
NGC\,3201 & $-$0.15            &  0.90              & $-$0.45           &$-$0.00           &   0.50             &    0.00      &  Mu{\~n}oz et al.(2013)  \\ %
NGC\,4590 & $-$0.05            &  0.80              & $-$0.15           &   0.00           &   0.30             &    0.00      &  Carretta et al.\,(2009)  \\ %
NGC\,4833 & $-$0.10            &  0.85              & $-$0.30           &$-$0.15           &   0.50             &    0.05      &  Pancino et al.\,(2017)  \\ %
NGC\,5024 & $-$0.05            &  0.65              & $-$0.45           &   0.00           &   0.50             &    0.00      &  M{\'e}sz{\'a}ros et al.\,(2015)  \\ %
NGC\,5053 &    0.20            &  0.50              &    0.00           &$-$0.10           &   1.00             &    0.10      &  Tang et al.\,(2009)  \\ %
NGC\,5139 & $-$0.30            &  0.75              & $-$0.50           &$-$0.25           &   0.60             &    0.10      &  Johnson \& Pilachowski (2010)  \\
NGC\,5272 & $-$0.15            &  0.70              & $-$0.25           &$-$0.05           &   0.50             &    0.10      &  Sneden et al.\,(2004)  \\ %
NGC\,5286 & $-$0.25            &  0.90              & $-$0.55           &   0.00           &   0.00             &    0.00      &   -- \\ %
          & $-$0.25            &  0.80              & $-$0.20           &$-$0.10           &   0.80             &    0.05      &   -- \\ %
NGC\,5466 & $-$0.15            &  0.45              & $-$0.35           &   0.00           &   0.50             &    0.00      &  M{\'e}sz{\'a}ros et al.\,(2015)  \\ %
NGC\,5904 & $-$0.20            &  0.80              & $-$0.35           &$-$0.05           &   0.55             &    0.00      &  M{\'e}sz{\'a}ros et al.\,(2015)  \\ %
NGC\,5927 & $-$0.25            &  0.30              & $-$0.10           &  0.00            &   0.10             &    0.00      &  Pancino et al.\,(2017)  \\ %
NGC\,5986 & $-$0.25            &  0.80              & $-$0.40           &$-$0.15           &   0.50             &    0.00      &   Johnson et al.\,(2017)   \\ %
NGC\,6093 &    0.00            &  0.75              & $-$0.20           &   0.00           &   0.50             &    0.00      &  Cavallo et al.\,(2004); Carretta et al.\,(2015)  \\ %
NGC\,6101 & $-$0.25            &  0.85              & $-$0.60           &   0.00           &   0.00             &    0.00      &   -- \\ %
          & $-$0.25            &  0.70              & $-$0.25           &$-$0.10           &   0.80             &    0.05      &   -- \\ %
NGC\,6121 & $-$0.15            &  0.70              & $-$0.10           &   0.00           &   0.00             &    0.00      &  Marino et al.\,(2008)  \\ %
NGC\,6144 &    0.00            &  0.75              & $-$0.30           &   0.00           &   0.00             &    0.00      &  -- \\ %
          & $-$0.00            &  0.70              & $-$0.05           &$-$0.10           &   0.80             &    0.05      &  -- \\ %
NGC\,6171 & $-$0.20            &  0.60              & $-$0.10           &   0.00           &   0.00             &    0.00      &  M{\'e}sz{\'a}ros et al.(2015)  \\ %
NGC\,6205 & $-$0.15            &  0.75              & $-$0.35           &$-$0.10           &   0.80             &    0.00      &  Johnson et al.\,(2005); Johnson \& Pilachowski (2012)  \\ %
NGC\,6218 & $-$0.10            &  0.70              & $-$0.20           &   0.00           &   0.30             &    0.00      &  Carretta et al.\,(2009)  \\ %
NGC\,6254 & $-$0.30            &  0.95              & $-$0.55           &   0.00           &   0.60             &    0.00      &  Carretta et al.\,(2009)  \\ %
NGC\,6304 & $-$0.50            &  0.55              & $-$0.25           &   0.00           &   0.00             &    0.00      &  --  \\ %
          & $-$0.55            &  0.55              & $-$0.20           &$-$0.10           &   0.80             &    0.05      &  --  \\ %
NGC\,6341 & $-$0.40            &  0.70              & $-$0.45           &$-$0.15           &   0.75             &    0.00      &  M{\'e}sz{\'a}ros et al.\,(2015)  \\ %
NGC\,6352 & $-$0.50            &  0.60              & $-$0.20           &   0.00           &   0.00             &    0.00      &  --  \\ %
          & $-$0.45            &  0.65              & $-$0.15           &$-$0.10           &   0.80             &    0.05      &  --  \\ %
NGC\,6362 & $-$0.35            &  0.70              & $-$0.30           &$-$0.00           &   0.00             &    0.00      & Massari et al.\,(2017)   \\ %
NGC\,6366 & $-$0.35            &  0.50              & $-$0.15           &$-$0.05           &   0.10             &    0.05      & Johnson et al.\,(2016); Puls et al.\,(2018)   \\ %
NGC\,6388 & $-$0.40            &  0.65              & $-$0.15           &$-$0.05           &   0.50             &    0.05      & Carretta \& Bragaglia (2018)   \\
NGC\,6397 & $-$0.20            &  0.65              & $-$0.20           &   0.00           &   0.10             &    0.00      & Lind et al.\,(2011)   \\ %
NGC\,6441 & $-$0.70            &  0.70              & $-$0.30           &   0.10           &   0.30             &    0.10      &  Gratton et al.\,(2006); Carretta et al.\,(2009)  \\ %
NGC\,6496 & $-$0.35            &  0.45              & $-$0.15           &   0.00           &   0.00             &    0.00      &  --  \\ %
          & $-$0.35            &  0.50              & $-$0.10           &$-$0.10           &   0.80             &    0.05      &  --  \\ %
NGC\,6535 & $-$0.15            &  0.75              & $-$0.40           &   0.00           &   0.40             &    0.00      &  Bragaglia et al.\,(2017)   \\ %
NGC\,6541 & $-$0.25            &  1.00              & $-$0.75           &   0.00           &   0.00             &    0.00      &  --  \\ %
          & $-$0.20            &  0.90              & $-$0.50           &$-$0.10           &   0.80             &    0.05      &  --  \\ %
NGC\,6584 & $-$0.10            &  0.70              & $-$0.05           &   0.00           &   0.00             &    0.00      &  --  \\ %
          & $-$0.05            &  0.65              &    0.10           &$-$0.10           &   0.80             &    0.05      &  --  \\ %
NGC\,6624 & $-$0.60            &  0.60              & $-$0.40           &   0.00           &   0.00             &    0.00      &   -- \\
          & $-$0.65            &  0.65              & $-$0.35           &$-$0.10           &   0.80             &    0.05      &   -- \\ %
NGC\,6637 & $-$0.30            &  0.65              & $-$0.20           &   0.00           &   0.00             &    0.00      &   -- \\
          & $-$0.30            &  0.65              & $-$0.10           &$-$0.10           &   0.80             &    0.05      &   -- \\ %
NGC\,6652 & $-$0.30            &  0.65              & $-$0.15           &   0.00           &   0.00             &    0.00      &   -- \\
          & $-$0.25            &  0.65              & $-$0.05           &$-$0.10           &   0.80             &    0.05      &   -- \\ %
NGC\,6656 & $-$0.10            &  0.70              & $-$0.20           &$-$0.05           &   0.40             &    0.00      &  Marino et al.\,(2011)  \\ %
NGC\,6681 & $-$0.30            &  1.05              & $-$0.85           &   0.00           &   0.20             &    0.00      &  O'Malley et al.\,(2017)  \\ %
NGC\,6715 & $-$0.15            &  0.75              & $-$0.20           &$-$0.10           &   0.50             &    0.05      &  Carretta et al.\,(2010)  \\ %
NGC\,6717 & $-$0.20            &  0.80              & $-$0.25           &   0.00           &   0.00             &    0.00      &   -- \\ %
          & $-$0.20            &  0.80              & $-$0.10           &$-$0.10           &   0.80             &    0.05      &   -- \\ %
NGC\,6723 & $-$0.35            &  0.80              & $-$0.30           &   0.00           &   0.00             &    0.00      &   -- \\ %
          & $-$0.30            &  0.75              & $-$0.20           &$-$0.10           &   0.80             &    0.05      &   -- \\ %
NGC\,6752 & $-$0.10            &  0.95              & $-$0.40           &$-$0.05           &   0.80             &    0.05      &  Yong et al.\,(2005)  \\ %
NGC\,6779 & $-$0.35            &  1.10              & $-$1.00           &   0.00           &   0.00             &    0.00      &   -- \\
          & $-$0.30            &  0.95              & $-$0.60           &$-$0.10           &   0.80             &    0.05      &   -- \\ %
NGC\,6809 & $-$0.05            &  0.80              & $-$0.20           &   0.00           &   0.50             &    0.00      &  M{\'e}sz{\'a}ros et al.\,(2015)  \\ %
NGC\,6838 & $-$0.40            &  0.60              & $-$0.25           &   0.00           &   0.00             &    0.00      &  M{\'e}sz{\'a}ros et al.(2015)  \\ %
NGC\,6934 & $-$0.40            &  0.90              & $-$0.65           &   0.00           &   0.00             &    0.00      &   -- \\
          & $-$0.30            &  0.85              & $-$0.35           &$-$0.10           &   0.80             &    0.05      &   -- \\ %
NGC\,6981 & $-$0.15            &  0.80              & $-$0.35           &   0.00           &   0.00             &    0.00      &   -- \\
          & $-$0.10            &  0.75              & $-$0.10           &$-$0.10           &   0.80             &    0.05      &   -- \\ %
NGC\,7078 & $-$0.20            &  0.85              & $-$0.50           &$-$0.20           &   0.50             &    0.10      &  M{\'e}sz{\'a}ros et al.\,(2015)  \\ %
NGC\,7089 & $-$0.35            &  0.70              & $-$0.60           &$-$0.10           &   0.40             &    0.00      &  Pancino et al.\,(2017)  \\
NGC\,7099 &    0.00            &  0.80              & $-$0.40           &   0.00           &   0.30             &    0.00      &  Carretta et al.\,(2009)  \\ %
\hline
\end{longtable}
% 5139 2808

\onecolumn
\begin{longtable}{lcccccccl}
\caption{Abundance difference between 2Ge and 1Ge stars adopted to infer the maximum internal variation of helium, $\delta Y_{\rm 2G,1G}$. For the clusters without literature determination of Mg, Al, and Si we adopted two sets of abundance differences corresponding to different choices of $\Delta$[Mg/Fe], $\Delta$[Al/Fe] and $\Delta$[Si/Fe].}\label{tab:CNO} \\ \hline %\hline
ID        &  $\Delta$[C/Fe]     &   $\Delta$[N/Fe]  &   $\Delta$[O/Fe]  &   $\Delta$[Mg/Fe]  &   $\Delta$[Al/Fe]    &   $\Delta$[Si/Fe]  & Reference\\
\hline
\endfirsthead
\caption{continued.}\\\hline\hline
ID        &  $\Delta$[C/Fe]     &   $\Delta$[N/Fe]  &   $\Delta$[O/Fe]  &   $\Delta$[Mg/Fe]  &   $\Delta$[Al/Fe]    &   $\Delta$[Si/Fe]  & Reference \\
\hline
\endhead
\endfoot
%          &                    &                    & 2G-1G&            &                  &                    &              &   \\
\hline   
 IC\,4499 & $-$0.10            &  0.85              & $-$0.55           &   0.00           &   0.00             &    0.00      &  --  \\
          & $-$0.05            &  0.55              & $-$0.00           &$-$0.30           &   1.10             &    0.10      &  --  \\
NGC\,104  & $-$0.55            &  0.90              & $-$0.50           &   0.00           &   0.45             &    0.00      &  Pancino et al.\,(2017)  \\
NGC\,288  & $-$0.20            &  0.85              & $-$0.45           &   0.00           &   0.30             &    0.00      &  Carretta et al.\,(2009)  \\
NGC\,362  & $-$0.40            &  0.95              & $-$0.65           &$-$0.10           &   0.50             &    0.05      &  Pancino et al.\,(2017)  \\
NGC\,1261 & $-$0.40            &  1.00              & $-$0.75           &   0.00           &   0.00             &    0.00      &  --  \\
          & $-$0.45            &  0.85              & $-$0.50           &$-$0.30           &   1.10             &    0.10      &  --  \\
NGC\,1851 & $-$0.35            &  1.12              & $-$0.55           &$-$0.10           &   0.50             &    0.05      &  Pancino et al.\,(2017)  \\
NGC\,2298 & $-$0.40            &  1.05              & $-$1.50           &   0.00           &   0.00             &    0.00      &  --  \\
          & $-$0.45            &  0.85              & $-$0.85           &$-$0.30           &   1.10             &    0.10      &  --  \\
NGC\,2808 & $-$0.90            &  1.20              & $-$0.95           &$-$0.50           &   1.20             &    0.20      &  Carretta et al.\,(2018) \\ %
NGC\,3201 & $-$0.45            &  1.10              & $-$1.00           &$-$0.05           &   0.90             &    0.00      &  Mu{\~n}oz et al.(2013)  \\ %
NGC\,4590 & $-$0.15            &  0.85              & $-$0.45           &   0.00           &   0.40             &    0.00      &  Carretta et al.\,(2009)  \\ %
NGC\,4833 & $-$0.15            &  1.05              & $-$0.70           &$-$0.40           &   0.80             &    0.10      &  Pancino et al.\,(2017)  \\ %
NGC\,5024 & $-$0.30            &  0.75              & $-$1.25           &$-$0.10           &   1.00             &    0.00      &  M{\'e}sz{\'a}ros et al.\,(2015)  \\ %
NGC\,5053 & $-$0.00            &  0.70              & $-$0.40           &$-$0.10           &   1.00             &    0.10      &  Tang et al.\,(2009)  \\ %
NGC\,5139 & $-$0.45            &  1.00              & $-$0.60           &$-$0.55           &   1.00             &    0.20      &  Johnson \& Pilachowski (2010) \\ %
NGC\,5272 & $-$0.30            &  0.85              & $-$0.65           &$-$0.30           &   0.90             &    0.15      &  Sneden et al.\,(2004); M{\'e}sz{\'a}ros et al.\,(2015)  \\ %
NGC\,5286 & $-$0.60            &  1.20              & $-$1.50           &   0.00           &   0.00             &    0.00      &  --  \\ %
          & $-$0.60            &  0.95              & $-$0.90           &$-$0.30           &   1.10             &    0.10      &  --  \\ %
NGC\,5466 & $-$0.05            &  0.50              & $-$0.30           &   0.00           &   0.70             &    0.00      &  M{\'e}sz{\'a}ros et al.\,(2015)  \\ %
NGC\,5904 & $-$0.45            &  1.10              & $-$0.80           &$-$0.10           &   1.10             &    0.10      &  M{\'e}sz{\'a}ros et al.\,(2015)  \\ %
NGC\,5927 & $-$0.80            &  0.35              & $-$0.50           &   0.00           &   0.20             &    0.00      &  Pancino et al.\,(2017)  \\ %
NGC\,5986 & $-$0.50            &  1.10              & $-$0.90           &$-$0.40           &   1.10             &    0.05      &  Johnson et al.\,(2017)  \\ %
NGC\,6093 & $-$0.15            &  1.15              & $-$0.65           &$-$0.10           &   1.00             &    0.00      &  Cavallo et al.\,(2004); Carretta et al.\,(2015)  \\ %
NGC\,6101 & $-$0.30            &  0.90              & $-$0.65           &   0.00           &   0.00             &    0.00      &  --  \\ %
          & $-$0.30            &  0.65              & $-$0.05           &$-$0.30           &   1.10             &    0.10      &  --  \\ %
NGC\,6121 & $-$0.15            &  0.80              & $-$0.20           &   0.00           &   0.05             &    0.00      &  Marino et al.\,(2008)  \\ %
NGC\,6144 & $-$0.10            &  0.85              & $-$0.55           &   0.00           &   0.00             &    0.00      &  --  \\ %
          & $-$0.10            &  0.65              &    0.00           &$-$0.30           &   1.10             &    0.10      &  --  \\ %
NGC\,6171 & $-$0.30            &  0.70              & $-$0.30           &   0.00           &   0.00             &    0.00      &  M{\'e}sz{\'a}ros et al.\,(2015)  \\ %
NGC\,6205 & $-$0.25            &  1.10              & $-$0.90           &$-$0.30           &   1.10             &    0.00      &  Johnson et al.\,(2005); Johnson \& Pilachowski (2012)  \\ %
NGC\,6218 & $-$0.35            &  0.95              & $-$0.60           &   0.00           &   0.50             &    0.00      &  Carretta et al.\,(2009)  \\ %
NGC\,6254 & $-$0.55            &  1.10              & $-$1.05           &$-$0.20           &   1.00             &    0.00      &  Carretta et al.\,(2009)  \\ %
NGC\,6304 & $-$0.80            &  0.65              & $-$0.45           &   0.00           &   0.00             &    0.00      &  --  \\ %
          & $-$0.95            &  0.60              & $-$0.50           &$-$0.30           &   1.10             &    0.10      &  --  \\ %
NGC\,6341 & $-$0.50            &  0.85              & $-$0.50           &$-$0.40           &   1.30             &    0.00      &  M{\'e}sz{\'a}ros et al.\,(2015)  \\ %
NGC\,6352 & $-$0.60            &  0.70              & $-$0.35           &   0.00           &   0.00             &    0.00      &  --  \\ %
          & $-$0.75            &  0.60              & $-$0.35           &$-$0.30           &   1.10             &    0.10      &  --  \\ %
NGC\,6362 & $-$0.55            &  0.75              & $-$0.70           &$-$0.00           &   0.00             &    0.00      & Massari et al.\,(2017)   \\ %
NGC\,6366 & $-$0.60            &  0.70              & $-$0.30           &$-$0.10           &   0.20             &    0.10      & Johnson et al.\,(2016); Puls et al.\,(2018)   \\ %
NGC\,6388 & $-$0.55            &  0.75              & $-$0.40           &$-$0.10           &   1.00             &    0.10      & Carretta \& Bragaglia (2018)   \\ %
NGC\,6397 & $-$0.20            &  0.75              & $-$0.20           &$-$0.05           &   0.20             &    0.00      & Lind et al.\,(2011)   \\ %
NGC\,6441 & $-$1.05            &  0.85              & $-$0.50           &   0.25           &   0.50             &    0.20      & Gratton et al.\,(2006); Carretta et al.\,(2009)  \\ %
NGC\,6496 & $-$0.45            &  0.50              & $-$0.25           &   0.00           &   0.00             &    0.00      &  --  \\ %
          & $-$0.60            &  0.40              & $-$0.25           &$-$0.30           &   1.10             &    0.10      &  --  \\ %
NGC\,6535 & $-$0.15            &  0.75              & $-$0.40           &   0.00           &   0.40             &    0.00      &  Bragaglia et al.\,(2017)  \\ %
NGC\,6541 & $-$0.50            &  1.30              & $-$1.40           &   0.00           &   0.00             &    0.00      &  --  \\ %
          & $-$0.40            &  1.05              & $-$0.75           &$-$0.30           &   1.10             &    0.10      &  --  \\ %
NGC\,6584 & $-$0.10            &  0.85              & $-$0.35           &   0.00           &   0.00             &    0.00      &  --  \\ %
          & $-$0.10            &  0.60              &    0.10           &$-$0.30           &   1.10             &    0.10      &  --  \\ %
NGC\,6624 & $-$0.70            &  0.70              & $-$0.55           &   0.00           &   0.00             &    0.00      &  --  \\
          & $-$0.85            &  0.60              & $-$0.60           &$-$0.30           &   1.10             &    0.10      &  --  \\
NGC\,6637 & $-$0.45            &  0.75              & $-$0.40           &   0.00           &   0.00             &    0.00      &  --  \\
          & $-$0.60            &  0.70              & $-$0.40           &$-$0.30           &   1.10             &    0.10      &  --  \\
NGC\,6652 & $-$0.45            &  0.75              & $-$0.30           &   0.00           &   0.00             &    0.00      &  --  \\ %
          & $-$0.50            &  0.65              & $-$0.50           &$-$0.30           &   1.10             &    0.10      &  --  \\ %
NGC\,6656 & $-$0.35            &  0.95              & $-$0.55           &$-$0.10           &   0.70             &    0.10      &  Marino et al.\,(2011)  \\ %
NGC\,6681 & $-$0.30            &  1.25              & $-$1.10           &$-$0.10           &   0.60             &    0.00      &  O'Malley et al.\,(2017)  \\ %
NGC\,6715 & $-$0.45            &  1.05              & $-$1.10           &$-$0.40           &   1.20             &    0.10      &  Carretta et al.\,(2010)  \\ %
NGC\,6717 & $-$0.40            &  0.95              & $-$0.40           &   0.00           &   0.00             &    0.00      &   -- \\
          & $-$0.40            &  0.80              & $-$0.15           &$-$0.30           &   1.10             &    0.10      &   -- \\
NGC\,6723 & $-$0.80            &  1.05              & $-$0.95           &   0.00           &   0.00             &    0.00      &   -- \\
          & $-$0.85            &  0.95              & $-$0.80           &$-$0.30           &   1.10             &    0.10      &   -- \\
NGC\,6752 & $-$0.40            &  1.25              & $-$0.80           &$-$0.25           &   1.10             &    0.10      &  Yong et al.\,(2005)  \\ %
NGC\,6779 & $-$0.50            &  1.40              & $-$1.50           &   0.00           &   0.00             &    0.00      &   -- \\
          & $-$0.45            &  1.10              & $-$0.80           &$-$0.30           &   1.10             &    0.10      &   -- \\
NGC\,6809 & $-$0.20            &  1.10              & $-$0.70           &$-$0.10           &   1.00             &    0.00      &  M{\'e}sz{\'a}ros et al.\,(2015)  \\ %
NGC\,6838 & $-$0.45            &  0.60              & $-$0.40           &   0.00           &   0.00             &    0.00      &  M{\'e}sz{\'a}ros et al.\,(2015)  \\ %
NGC\,6934 & $-$0.50            &  1.10              & $-$0.90           &   0.00           &   0.00             &    0.00      &   -- \\
          & $-$0.50            &  0.85              & $-$0.35           &$-$0.30           &   1.10             &    0.10      &   -- \\
NGC\,6981 & $-$0.60            &  1.15              & $-$1.50           &   0.00           &   0.00             &    0.00      &   -- \\
          & $-$0.60            &  0.90              & $-$0.90           &$-$0.30           &   1.10             &    0.10      &   -- \\
NGC\,7078 & $-$0.50            &  1.00              & $-$0.80           &$-$0.50           &   1.10             &    0.30      &   M{\'e}sz{\'a}ros et al.\,(2015) \\ %
NGC\,7089 & $-$0.65            &  0.75              & $-$1.20           &$-$0.40           &   1.10             &    0.10      &   Pancino et al.\,(2017) \\
NGC\,7099 & $-$0.15            &  1.00              & $-$0.80           &   0.00           &   0.50             &    0.00      &   Carretta et al.\,(2009) \\ %
\hline
\end{longtable}
% 5139 2808

\begin{table*}
\begin{centering} 
  \caption{Average helium difference between 2G and 1G stars and maximum internal helium variation. For the clusters without literature determination of Mg, Al, and Si we derived two
    estimates of $\delta Y_{\rm 2G,1G}$ and $\delta Y_{\rm max}$ corresponding to different choices of $\Delta$[Mg/Fe], $\Delta$[Al/Fe] and $\Delta$[Si/Fe].}
 \begin{tabular}{lcclcc}
   \hline
   \hline
ID        &  $\delta Y_{\rm 2G,1G}$     &  $\delta Y_{\rm max}$ & ID        &  $\delta Y_{\rm 2G,1G}$     &  $\delta Y_{\rm max}$ \\
\hline
 IC\,4499 &   0.004$\pm$0.006 &  0.017$\pm$0.008 &  NGC\,6362 &   0.003$\pm$0.011 &  0.004$\pm$0.011      \\
          &   0.004$\pm$0.006 &  0.017$\pm$0.008 &  NGC\,6366 &   0.011$\pm$0.010 &  0.011$\pm$0.015      \\
NGC\,104  &   0.011$\pm$0.005 &  0.049$\pm$0.005 &  NGC\,6388 &   0.019$\pm$0.007 &  0.067$\pm$0.009      \\
NGC\,288  &   0.015$\pm$0.010 &  0.016$\pm$0.012 &  NGC\,6397 &   0.006$\pm$0.009 &  0.008$\pm$0.011      \\
NGC\,362  &   0.008$\pm$0.006 &  0.026$\pm$0.008 &  NGC\,6441 &   0.029$\pm$0.006 &  0.081$\pm$0.022      \\
NGC\,1261 &   0.004$\pm$0.004 &  0.019$\pm$0.007 &  NGC\,6496 &   0.009$\pm$0.011 &  0.021$\pm$0.006      \\
          &   0.004$\pm$0.004 &  0.019$\pm$0.007 &            &   0.009$\pm$0.011 &  0.025$\pm$0.007      \\
NGC\,1851 &   0.007$\pm$0.005 &  0.025$\pm$0.006 &  NGC\,6535 &   0.003$\pm$0.021 &  0.003$\pm$0.022      \\
NGC\,2298 &$-$0.003$\pm$0.009 &  0.011$\pm$0.012 &  NGC\,6541 &   0.024$\pm$0.005 &  0.045$\pm$0.006      \\
          &$-$0.003$\pm$0.008 &  0.011$\pm$0.012 &            &   0.024$\pm$0.005 &  0.044$\pm$0.006      \\
NGC\,2808 &   0.048$\pm$0.005 &  0.124$\pm$0.007 &  NGC\,6584 &   0.000$\pm$0.007 &  0.015$\pm$0.011      \\
NGC\,3201 &$-$0.001$\pm$0.013 &  0.028$\pm$0.032 &            &   0.000$\pm$0.007 &  0.015$\pm$0.011      \\
NGC\,4590 &   0.007$\pm$0.009 &  0.012$\pm$0.009 &  NGC\,6624 &   0.010$\pm$0.004 &  0.022$\pm$0.003      \\
NGC\,4833 &   0.016$\pm$0.008 &  0.051$\pm$0.009 &            &   0.010$\pm$0.004 &  0.024$\pm$0.004      \\
NGC\,5024 &   0.013$\pm$0.007 &  0.044$\pm$0.008 &  NGC\,6637 &   0.004$\pm$0.006 &  0.011$\pm$0.005      \\
NGC\,5053 &$-$0.002$\pm$0.013 &  0.004$\pm$0.025 &            &   0.004$\pm$0.006 &  0.013$\pm$0.005      \\
NGC\,5139 &   0.033$\pm$0.006 &  0.090$\pm$0.010 &  NGC\,6652 &   0.008$\pm$0.007 &  0.017$\pm$0.011      \\
NGC\,5272 &   0.016$\pm$0.005 &  0.041$\pm$0.009 &            &   0.010$\pm$0.007 &  0.017$\pm$0.011      \\
NGC\,5286 &   0.007$\pm$0.006 &  0.044$\pm$0.004 &  NGC\,6656 &   0.005$\pm$0.008 &  0.041$\pm$0.012      \\
          &   0.007$\pm$0.006 &  0.044$\pm$0.004 &  NGC\,6681 &   0.009$\pm$0.008 &  0.029$\pm$0.015      \\
NGC\,5466 &   0.002$\pm$0.017 &  0.007$\pm$0.024 &  NGC\,6715 &   0.012$\pm$0.003 &  0.052$\pm$0.012      \\
NGC\,5904 &   0.012$\pm$0.004 &  0.037$\pm$0.007 &  NGC\,6717 &   0.003$\pm$0.006 &  0.003$\pm$0.009      \\
NGC\,5927 &   0.011$\pm$0.004 &  0.055$\pm$0.015 &            &   0.003$\pm$0.006 &  0.003$\pm$0.009      \\
NGC\,5986 &   0.005$\pm$0.006 &  0.031$\pm$0.012 &  NGC\,6723 &   0.005$\pm$0.006 &  0.024$\pm$0.007      \\
NGC\,6093 &   0.011$\pm$0.008 &  0.027$\pm$0.012 &            &   0.005$\pm$0.006 &  0.026$\pm$0.007      \\
NGC\,6101 &   0.005$\pm$0.010 &  0.017$\pm$0.011 &  NGC\,6752 &   0.015$\pm$0.005 &  0.042$\pm$0.004      \\
          &   0.004$\pm$0.010 &  0.019$\pm$0.011 &  NGC\,6779 &   0.011$\pm$0.007 &  0.031$\pm$0.008      \\
NGC\,6121 &   0.009$\pm$0.006 &  0.014$\pm$0.006 &            &   0.012$\pm$0.007 &  0.031$\pm$0.008      \\ %%%
NGC\,6144 &   0.009$\pm$0.011 &  0.017$\pm$0.013 &  NGC\,6809 &   0.014$\pm$0.008 &  0.026$\pm$0.015      \\
NGC\,6171 &   0.019$\pm$0.011 &  0.024$\pm$0.014 &  NGC\,6838 &   0.005$\pm$0.009 &  0.024$\pm$0.010      \\ %%%
          &   0.009$\pm$0.011 &  0.017$\pm$0.014 &  NGC\,6934 &   0.006$\pm$0.003 &  0.018$\pm$0.004      \\
NGC\,6205 &   0.020$\pm$0.004 &  0.052$\pm$0.004 &  NGC\,6981 &   0.011$\pm$0.006 &  0.017$\pm$0.006      \\
NGC\,6218 &   0.009$\pm$0.007 &  0.011$\pm$0.011 &            &   0.010$\pm$0.006 &  0.017$\pm$0.006      \\
NGC\,6254 &   0.006$\pm$0.008 &  0.029$\pm$0.011 &  NGC\,7078 &   0.021$\pm$0.009 &  0.069$\pm$0.006      \\
NGC\,6304 &   0.008$\pm$0.005 &  0.025$\pm$0.006 &  NGC\,7089 &   0.013$\pm$0.005 &  0.052$\pm$0.009      \\
          &   0.010$\pm$0.005 &  0.028$\pm$0.007 &  NGC\,7099 &   0.015$\pm$0.010 &  0.022$\pm$0.010      \\
NGC\,6341 &   0.022$\pm$0.004 &  0.039$\pm$0.006 &            &                   &                       \\
NGC\,6352 &   0.019$\pm$0.014 &  0.027$\pm$0.006 &            &                   &                       \\
          &   0.019$\pm$0.014 &  0.028$\pm$0.006 &            &                   &                       \\
\hline
\hline
 \end{tabular}\\
 \label{tab:He}
\end{centering} 
\end{table*}

% %%%%%%%%%%%%%%%%%%%%%%%%%%%%%%%%%%%%%%%%%%%%%%%%%%%%%%%%%%%%%%%%%%%%%%%%%%    
 
\begin{table}
 \caption{Spearman's rank correlation coefficient indicating the dependence between $\delta Y_{\rm 2G,1G}$ (Column 2), $\delta Y_{\rm max}$ (Column 3) and several parameters of the host GC. For each couple of parameters we provide the number of analyzed GCs. In the case of the $L_{\rm F275W-F814W}$ parameter we provide the correlation coefficients for all the analyzed clusters and by excluding those clusters with [Fe/H]$> -0.99$ that display only the red HB.}
 \begin{tabular}{ccc}
   \hline
   \hline
   Parameter        &  $\delta Y_{\rm 2G,1G}$     &  $\delta Y_{\rm max}$ \\
   \hline
[Fe/H] 	                              & $0.05 \pm 0.13$, 57    & $0.02 \pm 0.14$, 57     \\
$M_{\rm V}$ 	                      & $-0.50 \pm 0.11$, 57   & $-0.83 \pm 0.05$, 57     \\
$log(\mathcal{M}/\mathcal{M}_{\odot})$  & $0.54\pm 0.11$, 44     & $0.87\pm 0.05$, 44      \\
E(B$-$V) 	                      & $-0.01\pm 0.13$, 57     & $0.07\pm 0.14$, 57      \\
$\epsilon$                            & $0.01\pm 0.14$, 55     & $0.04 \pm 0.13$, 55     \\
$R_{\rm GC}$                            & $-0.08\pm 0.14$, 57    & $0.04\pm 0.14$, 57      \\
$v_{\rm r}$ 	                      & $-0.27\pm 0.14$, 56    & $0.01\pm 0.13$, 56      \\
%$V_{lsr}$ 	                      & $-0.26 \pm 0.13$, 56   & $0.02\pm 0.12$, 56      \\
$\sigma_{\rm V}$ 	              & $0.53\pm 0.12$, 40     & $0.81 \pm 0.09$, 40     \\
$c$ 	                              & $0.35\pm 0.12$, 57     & $0.35\pm 0.12$, 57      \\
$\mu_{\rm v}$	                      & $-0.46\pm 0.11$, 57    & $-0.60\pm 0.09$, 57     \\
$\rho_{0}$	                      & $0.38\pm 0.12$, 57     & $0.47\pm 0.11$, 57      \\
$log(\tau_{\rm c})$ 	              & $-0.10 \pm 0.13$, 57   & $-0.03\pm 0.14$, 57     \\
$log(\tau_{\rm h})$ 	              & $0.18\pm 0.14$, 57     & $0.36\pm 0.13$, 57      \\
age (D10)                             & $0.06\pm 0.14$, 57     & $-0.04\pm 0.14$, 57     \\
age (MF09)                            & $-0.03\pm 0.14$, 56    & $-0.10\pm 0.13$, 56     \\
age (V13)                             & $0.05\pm 0.14$, 52     & $-0.04\pm 0.14$, 52     \\
HBR 	                              & $0.09\pm 0.13$, 54     & $0.05\pm 0.14$, 54      \\
$\Delta$(V-I)                         & $0.20\pm 0.13$, 57     & $0.19\pm 0.14$, 57   \\
$L_{\rm F275W-F814W}$ (all)               & $0.60\pm 0.09$, 57     & $0.36\pm0.12$, 57   \\
$L_{\rm F275W-F814W}$ (blue-HB GCs)       & $0.77\pm 0.06$, 47     & $0.48\pm0.12$, 47   \\
$f_{\rm bin}^{\rm C}$                     & $-0.19\pm 0.18$, 35    & $-0.41\pm 0.15$, 35   \\
$f_{\rm bin}^{\rm C-HM}$                   & $-0.41\pm 0.14$, 46    & $-0.55\pm 0.10$, 46    \\
$f_{\rm bin}^{\rm oHM}$                    & $-0.29\pm 0.14$, 43   & $-0.45\pm 0.12$, 43    \\
  \hline 
 \hline\hline
 \end{tabular}\\
 \label{tab:rel}
 \end{table}
% %%%%%%%%%%%%%%%%%%%%%%%%%%%%%%%%%%%%%%%%%%%%%%%%%%%%%%%%%%%%%%%%%%%%%%%%%%    

\end{document}